\documentclass[useAMS,usenatbib]{mn2e}

\usepackage{graphicx}
\def\lesssim{\la}
\def\gtrsim{\ga}

\title[]{A Vertical Resonance Heating Model for X- or Peanut-Shaped Galactic Bulges}
\author[Quillen et al.]{
Alice C. Quillen$^1$, Ivan Minchev$^2$,  Sanjib Sharma$^3$,
\newauthor
Yu-Jing Qin$^4$ , \& Paola Di Matteo$^5$  \\
$1$Department of Physics and Astronomy, University of Rochester, Rochester, NY 14627, USA \\
$2$Leibniz-Institut fur Astrophysik Potsdam (AIP), An der Sternwarte 16, 14482, Potsdam, Germany  \\ 
$3$Sydney Institute for Astronomy, School of Physics, University of Sydney, NSW 2006, Australia \\
$4$Shanghai Astronomical Observatory, 80 Nandan Road, Shanghai, China \\
$5$Observatoire de Paris-Meudon, GEPI, CNRS UMR 8111, 5 pl. Jules Janssen, Meudon, 92195, France 
}

\begin{document}
\maketitle

\begin{abstract}
We explore a second order Hamiltonian vertical resonance model for X-shaped or peanut-shaped galactic bulges. The X- or peanut-shape is caused by the 2:1 vertical Lindblad resonance with the bar,   with two vertical oscillation periods per orbital period in the bar frame. We examine N-body simulations and find that due to the bar slowing down and disk thickening during bar buckling, the resonance and associated peanut-shape moves outward.  The peanut-shape is consistent with the  location of the 2:1 vertical  resonance, independent of whether the bar buckled or not. We estimate the resonance width from the potential $m=4$ Fourier component and find that the resonance is narrow, affecting orbits  over a narrow range in the angular momentum distribution, $dL/L \sim 0.05$. As the resonance moves outward, stars originally in the mid plane  are forced out of the mid plane and into orbits just within the resonance separatrix.  The height of the separatrix orbits, estimated from the Hamiltonian model,  is approximately consistent with the peanut-shape height. The peanut- or X-shape is comprised of stars in the vicinity of the resonance separatrix.  The velocity distributions from the simulations illustrate that low inclination orbits are depleted within resonance.  Within resonance, the vertical velocity distribution is broad,  consistent with resonant heating caused by the passage of the resonance through the disk. 
In the Milky Way bulge  we relate the azimuthally averaged mid-plane mass density near the vertical resonance to the rotation curve and bar pattern speed.  At an estimated vertical resonance galactocentric radius  of $\sim 1.3$ kpc,  we confirm  a mid-plane density of $\sim 5\times 10^8 M_\odot {\rm kpc}^{-3}$, consistent with recently estimated mass distributions. We find that the rotation curve, bar pattern speed, 2:1 vertical resonance location,  X-shape tips, and mid-plane mass density,  are all self-consistent in the Milky Way galaxy bulge.
\end{abstract}




\section{Introduction}

A bimodal distribution has recently been discovered in the distribution of   red clump giants in the Galactic bulge 
 \citep[]{mcwilliam10,nataf10}. 
The observed distributions can be explained with a vertical X-shaped structure in the bulge region 
\citep{saito11,li12,gerhard12,ness12}. Proper motions in the X-shaped bulge imply that the bulge 
is rotating and exhibits non-circular motion \citep{poleski13,vasquez13}.
The X-shape is primarily comprised of moderate metallicity stars, [Fe/H] $> -0.5$  \citep{ness12}.

X-shaped or boxy/peanut-shaped bulge structures are associated 
with galactic bars \citep{bureau06,gardner13,erwin13} and are due to a buckling instability
\citep{raha91,merritt94}, orbits associated with vertical resonances \citep{combes90,pfenniger91}
or resonant trapping into a vertical Lindblad resonance during bar growth \citep{quillen02}.  
Orbits supporting
the X-shape in N-body simulations are near banana-shaped periodic orbits 
\citep{combes90,pfenniger91,patsis02,martinez06}.
Three dimensional N-body simulations with galactic bars exhibit X- or boxy/peanut-shaped bulges \citep{combes90,raha91,atha05,debattista05,martinez06,li12,gardner13}
and these have been used to model observed boxy/peanut-shaped bulges in 
galaxies and their velocity distributions, as seen
from line widths  \citep{bureau06,mendezabreu08,gardner13}.
 
Because resonances are often narrow, their location is strongly
dependent on the calculated values of rotation and oscillation periods.  
For example, a resonant model has 
placed tight constraints on the Milky Way's bar pattern speed 
\citep{minchev07, gardner10,gardner10b}.
If the X-shape in our Galaxy is related to a vertical Lindblad resonance,
then its location gives a tight relation between the bar pattern speed, the angular rotation rate
and the vertical oscillation frequency.  This can give a potentially important  
constraint on the mass distribution in the inner region of our Galaxy.
Furthermore if the resonance has drifted, then it may be possible to also place constraints
on the past evolution of the Milky Way's bar and bulge by identifying its affect on
the velocity distributions.

In this paper we expand on the Hamiltonian resonant model for the dynamics near a vertical resonance
by \citet{quillen02}.  This model provides a theoretical framework for predicting the location
and inclinations of  orbits in the vicinity of the resonance.
We compare predictions made with this model
to numerical simulations of barred galaxies that exhibit peanut shaped bulges.
We test the assumption that
the X-shape bulge in the Milky Way is associated with a vertical resonance
by comparing the mid-plane mass density predicted from a vertical resonance model
to that inferred from observations.

\begin{figure*}
\begin{center}
\includegraphics[width=4.8in]{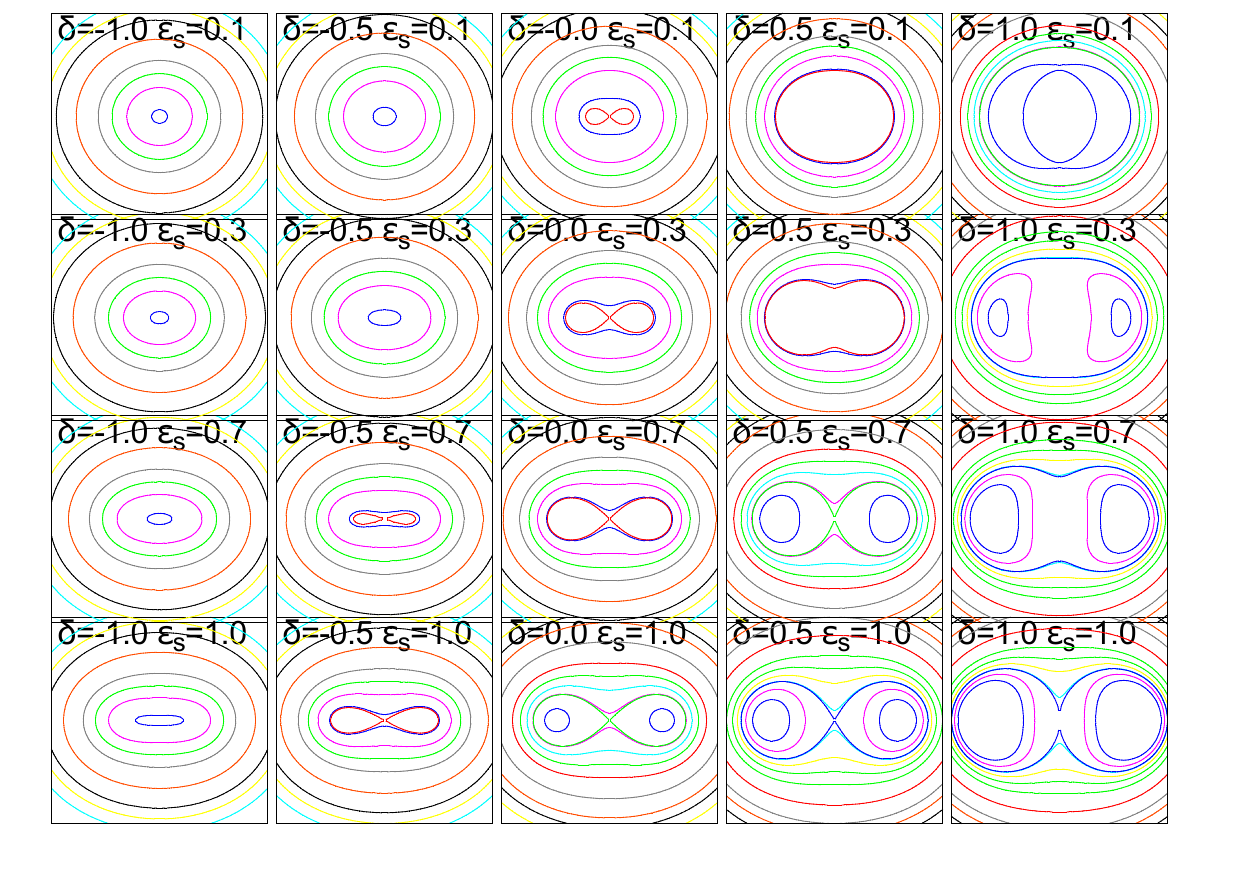} 
 \caption{Level curves are shown for the Hamiltonian in equation \ref{eqn:Ham} showing bar growth.  
 Level curves of the second order Hamiltonian in equation \ref{eqn:Ham} are shown to illustrate classes of
 periodic orbits for different values of $\delta$ or distance to resonance.
 The level curves are shown in a coordinate system with $(x,y) = \sqrt{2J_z}( \cos\phi,\sin\phi)$,
 so radius on the plot is equal to $\sqrt{2 J_z}$.   Higher radius in each panel corresponds to higher orbital inclination.
 Fixed points correspond to orbits that are periodic in $\theta_z$ and $\theta$.
 The stable periodic orbits are found at  $\phi = 0,\pi$ and these correspond to upward oriented banana-shaped
 orbits and downward oriented banana shaped orbits.  The banana-shaped orbits reach high inclination 
 along the bar major axis and lie slightly below the mid-plane along the bar minor axis.
 On resonance there are no planar orbits.  
 If the product of coefficients $a \epsilon_s$ is positive rather than negative, then each panel is rotated
 by 90$^\circ$ and the orbits are below the mid-plane at the ends of the bar and reach high inclinations 
 along the bar minor axis. 
From top to bottom each row is meant to represent a different time.
 From left to right each panel shows a different radial location in the galaxy from 
 small radius (left, low $\delta$) to larger radius (right higher $\delta$).
  Bar growth is shown with increasing perturbation strength $\epsilon_s$ from top to bottom.
  \label{fig:ham_grow}
 }
\end{center}
\end{figure*}

\section{Hamiltonian model for a vertical resonance}

Orbital dynamics in a galaxy can be modeled with an axisymmetric Hamiltonian with the addition of
perturbations from non-axisymmetric structures such as a bar (e.g., \citealt{cont75}).
The family of banana-shaped periodic orbits have two vertical oscillations 
per rotation period in the frame rotating with the bar \citep{combes90,pfenniger91,patsis02}.
At this commensurability,
\begin{equation}
\nu \approx 2(\Omega - \Omega_b), \end{equation}
where $\Omega$ is the angular rotation rate in the plane, $\nu$ the frequency of vertical oscillations (both at a 
mean or guiding orbital radius, $r_g$),  and $\Omega_b$ is the bar pattern angular rotation rate. 
The above commensurability is also known as a vertical Lindblad resonance and its role
in accounting for boxy peanut shaped bulges was suggested by \citet{combes90}.
In the bar frame there are two vertical oscillation periods per rotation period.  
The resonance is often called a 2:1 vertical Lindblad resonance.
By integrating the above expression we find a resonant angle or resonant argument
\begin{equation}
\phi \equiv \theta_z + 2(\theta - \Omega_b t) \label{eqn:phi}
\end{equation}
that is slowly moving near the resonance.  
Here $\theta_z$ is the angle associated with vertical oscillations and $\theta$ the azimuthal angle in the galactic
plane.
The periodic banana-shaped periodic orbits have constant values of resonant angle $\phi = 0$ or $\pi$.
The value of the constant value 
specifies the orientation of the periodic orbit (banana pointed up or banana pointed down).
When the angle $\phi$ is fixed, variations in $\theta$ are related to variations in $\theta_z$, however no
epicylic or radial oscillation is specified.  By considering a single dimension of freedom, the vertical one,
we have ignored the radial degree of freedom.  We refer to constant $\phi$ orbits as periodic orbits, however,
they would only be periodic in three dimensions if they were also periodic in their radial degree of
freedom.  When the orbit is periodic in three dimensions and reaches apocenter when at high inclination above
or below the mid-plane, then the orbit is banana-shaped.  

Near a resonance, only perturbative terms to the Hamiltonian that contain 
slowly varying angles need be considered.  Those with rapidly varying angles
only cause small perturbations to action variables and so can be neglected.
Following \citet{quillen02}, 
the dynamics  near a vertical Lindblad resonance can 
be modeled with a Hamiltonian
\begin{equation}
H(J_z, \phi) = a J_z^2 + \delta J_z + \epsilon_s J_z \cos(2\phi) \label{eqn:Ham}
\end{equation}
 in a galaxy that is symmetrical about the mid plane.
Here  $J_z$ is the action variable associated with vertical oscillations that is conjugate
to the angle $\theta_z$.
At low inclination,
action angle variables, $J_z, \theta_z$, are related to positions and velocities in cylindrical coordinates
\begin{eqnarray}
z &\approx& \sqrt{2J_z \over \nu} \cos \theta_z \nonumber \\
v_z &\approx& -\sqrt{2J_z \nu} \sin \theta_z \nonumber \\
\theta_z &=& \nu t + {\rm constant},  \label{eqn:coord}
\end{eqnarray}
where $z$ is the height above the Galactic plane and $v_z$ the vertical velocity component.
The coefficient $a$ depends only on the symmetric mass distribution in the galaxy.
The first two terms in the Hamiltonian (equation \ref{eqn:Ham}) arise from the unperturbed system alone
or the axisymmetric gravitational potential.
The coefficient $\epsilon_s$ depends on the bar perturbation strength and shape.
The coefficients $a, \delta$ depend on angular momentum (or mean radius) 
and are derived more rigorously in our appendices \ref{ap:action} and \ref{ap:can}, expanding on the 
calculations by \citet{quillen02}.

The coefficient $\delta$ describes distance from resonance,
\begin{eqnarray}
\delta &\approx& \nu - 2(\Omega - \Omega_b).  \label{eqn:aa}
\end{eqnarray}
The vertical oscillation frequency, $\nu$, depends on the axisymmetric (or azimuthally averaged) 
mass distribution with
\begin{equation}
\left. \nu^2 \equiv {\partial^2 V_0 \over \partial z^2} \right|_{z=0} \label{eqn:nu}
\end{equation}
evaluated in mid-plane at approximately the mean radius $r_g$ of the orbit 
(see appendix \ref{ap:action} for description in terms of angular momentum and all three action angle variables).
Here $V_0(r,z)$ is the gravitational potential derived from the axisymmetric mass distribution.
The above one dimensional Hamiltonian depends not on $J_z,\theta_z$ but on $J_z,\phi$.
Canonical transformations reduce the full three dimensional Hamiltonian to the one dimensional
version given in above equation \ref{eqn:Ham} and this is shown in appendix \ref{ap:can}.

As the orbit transverses different radii, the star crosses regions with
different values of ${\partial^2 V_0 \over \partial z^2}$.   This sensitivity is taken into account
with Hamiltonian terms that depend on both $J_z$ and $J_r$, the action
variable associated with epicyclic motion  (see appendix
\ref{ap:action}).  A term proportional to $J_r J_z$ can introduce a small shift
in the location of resonance (see appendix \ref{ap:shift}), here described by the coefficient $\delta$.

The bar gravitational potential perturbation's vertical dependence, $V_b(r, \theta,z,t)$, can be approximated to low
order in $z$ with  a sum of Fourier components
 \begin{equation}
V_m (r, \theta, z, t) = \left[ C_m(r)  + C_{mz}(r) z^2\right] \cos( m(\theta - \Omega_b t)). \label{eqn:V_m}
\end{equation}  
Each $m$ is a Fourier component of the bar's gravitational potential and  we have retained
the dependence on $z$.
Above it is assumed that the gravitational  potential is symmetric about
the mid-plane and that when the bar grows only the Fourier amplitudes, described by the coefficients,
$C_m,  C_{mz}$, increase in strength. 
Here $C_m$ has units of gravitational potential, (km/s)$^2$, and $C_{mz}$ has units
(km/s)$^2$ kpc$^{-2}$ or frequency$^2$.
For a bar that is similar shape on either side of the mid-plane (buckle-free), 
we need only consider $m=2$ and $4$ Fourier components.  The $m=4$ term contributes
to the 2:1 resonance associated with banana-shaped orbits and with resonant angle given
in equation \ref{eqn:phi}.
Inserting the expression for $z$ (equation \ref{eqn:coord}) into the above potential expression we can
write the potential term that is proportional to $z^2$ in terms of action variables as 
\begin{equation}
{C_{mz} \over 4} \left({2J_z \over  \nu} \right) \left[ \cos( 2 \theta_z - m(\theta-\Omega_b t)) 
+ \cos (2 \theta_z + m(\theta-\Omega_b t)) \right]
\end{equation}
where the first term is important near an inner m:1 vertical resonance and the second term
is appropriate for an outer vertical resonance.  Henceforth we neglect the outer resonance term.
For $m=4$ the resonant angle is equivalent to $2\phi$ as defined in equation \ref{eqn:phi}
and we find the coefficient in equation \ref{eqn:Ham},
\begin{equation}
\epsilon_s = {C_{4z}  \over 2 \nu}. \label{eqn:epss}
\end{equation}
This coefficient has units of frequency.

If the bar buckles during growth we can also consider a time dependent perturbation that includes an
asymmetric term
\begin{equation}
V_2 (r, \theta, z, t) = B_{2z}(r,t) z \cos(2( \theta - \Omega_b t)).
\end{equation}
Inserting $z$ in 
 action angle variables we find a perturbation term in the form
 \begin{equation}
 {B_{2z} \over 2 }\sqrt{2J_z \over  \nu} \left[ \cos (\theta_z - 2(\theta - \Omega_b t)) + \cos (\theta_z - 2(\theta - \Omega_b t))
 \right].
\end{equation}
In this case we would consider a Hamiltonian 
\begin{equation}
H(J_z, \phi) = a J_z^2 + \delta J_z + \epsilon_b J_z^{1/2} \cos(\phi) \label{eqn:Ham_b}
\end{equation}
with 
\begin{equation}
\epsilon_b =  {B_{2z} \over  \sqrt{2 \nu}}. \label{eqn:espb}
\end{equation}
While the Hamiltonian given in equation \ref{eqn:Ham} is equivalent to a second order mean motion resonance,
that given in equation \ref{eqn:Ham_b} is equivalent to a first order mean motion resonance
(see \citealt{M+D} chapter 8).
The order refers to the power of $J_z^{1\over 2}$ in the perturbation term,
corresponding to the power of eccentricity or inclination, 
depending upon the setting.

When the galaxy is symmetric about the mid plane it may seem mysterious that the 2:1 vertical
resonance strength depends on the $m=4$ Fourier coefficient of the gravitational potential.
We can consider the morphology of the banana-shaped orbit in a frame moving with the bar.  
It is high above the mid plane at the ends
of the bar, on the bar major axis, and below the mid plane  on the bar minor axis.  Because
the gravitational potential is symmetrical about the mid plane, the orbit reaches high points in the potential
both along the major axis and along the minor axis.  Thus a star in the orbit experiences positive extrema
in the potential at four points in the orbit.  Consequently the $m=4$ perturbation associated with the
bar excites the 2:1 vertical resonance.  

We can qualitatively consider the coupling when the galaxy is buckling.  Because of the buckle the potential
is low along the bar major axis and high along the bar minor axis. For an orbit oriented with the
buckle,  
the $m=2$ Fourier component from the bar perturbation excites the resonance.

\begin{figure*}
\begin{center}
\includegraphics[width=4.8in]{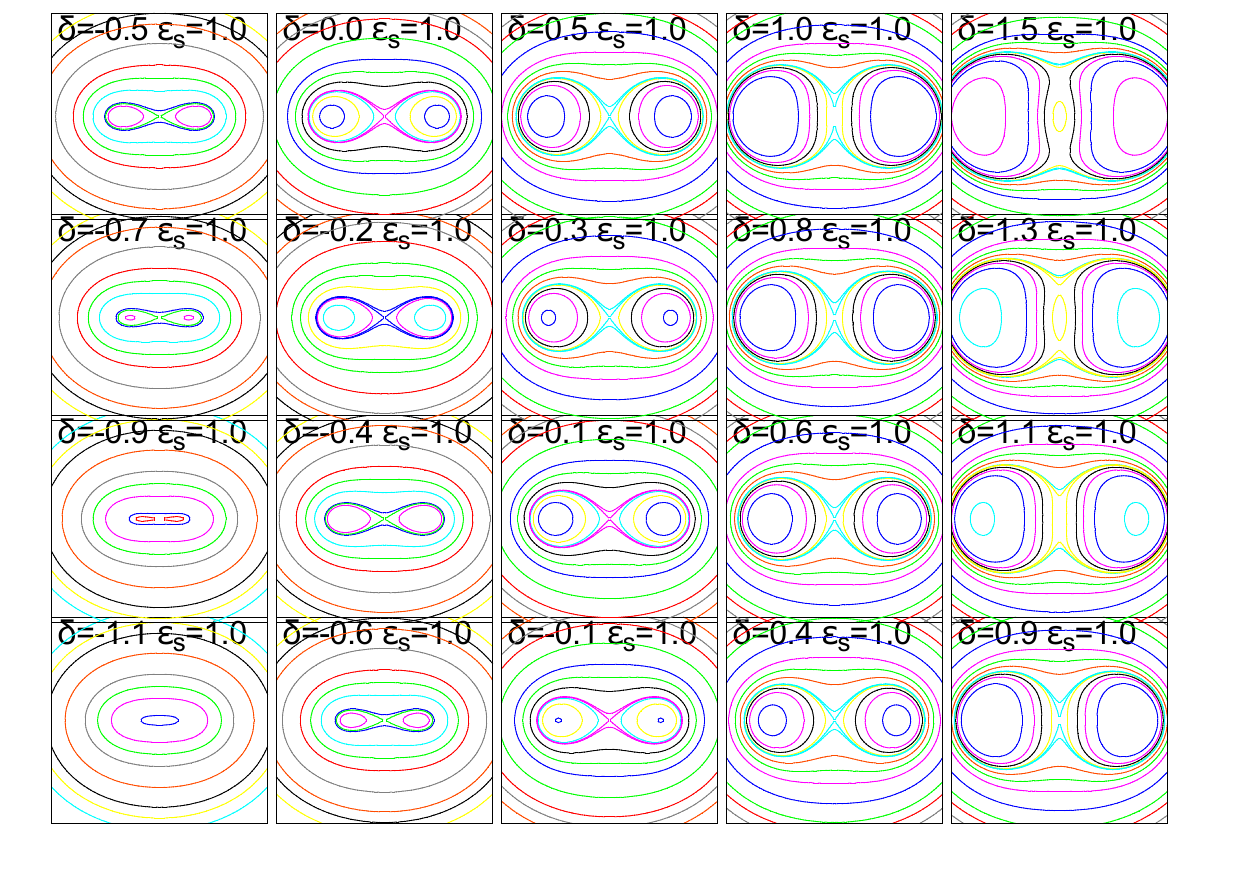} 
\caption{
Similar to Figure \ref{fig:ham_grow} except
 the bar slows down from top to bottom panels but is kept at the same strength.  
The evolution is similar if the disk thickens.  Both processes decrease $\delta$
and move the resonance outward in radius.
 On the  rightmost panel, as $\delta$ decreases the island of low inclination orbits
fades away.  
Stars originally in low inclination orbits in this island at the outer edge of the peanut
 are pushed out of the plane and into high inclination orbits just inside the separatrix.
 As the resonance continues to drift (see second column from right) the separatrix shrinks and these stars then
 must cross the separatrix.  Eventually they are left in high inclination orbits
that circulate outside resonance and so do not support the peanut shape.
Stars are heated vertically as they pass through resonance.  
While they are just inside the separatrix they librate around $\phi=0$ or $\phi=\pi$ and
support the peanut-shape.
A signature of adiabatic bar slowing or disk thickening would be a population of stars
at the outer edge of resonance, just inside the separatrix, with height $z_{sep}$ (equation \ref{eqn:zsep}).
\label{fig:ham_slow}
}
\end{center}
\end{figure*}

\subsection{Resonance size scales} \label{sec:ressize}

The coefficient $\delta$ describes the distance from resonance and so can be ignored when considering
the characteristic momentum and timescale in resonance.
These size scales 
 can be estimated from the coefficients $a,\epsilon_s$ or
from $a, \epsilon_b$ (see \citealt{quillen06} equation 7).
For a first order resonance,
the important timescale, $\tau_{lib}$ (a libration timescale), and momentum scale, $J_{res}$, 
\begin{eqnarray}
\tau_{lib} &=&  \left|{\epsilon_b^2 a }\right|^{-{1\over 3}}\nonumber  \\
  J_{res} &=& \left|{ \epsilon_b \over a}\right|^{2 \over 3}, \label{eqn:taulib}
\end{eqnarray}
and for the second order resonance
\begin{eqnarray}
\tau_{lib} &=&  \left|{\epsilon_s }\right|^{-1}\nonumber \\
  J_{res}& =& \left|{\epsilon_s \over a}\right|.  \label{eqn:Jres}
\end{eqnarray}
We refer to a libration frequency, $\omega_{lib} = \tau_{lib}^{-1}$.

As $\delta$ is a frequency, we can estimate the width of the resonance by considering
where $\delta$ is are similar in size to the libration frequency.
At low inclination,
the region where the resonance is important is where $|\delta| \lesssim \tau_{lib}^{-1}$ or
\begin{eqnarray}
|\delta | \lesssim  
\left\{
\begin{array}{lcl}
  \left|{\epsilon_b^2 a }\right|^{1\over 3} &   \quad {\rm for}   \qquad & {\rm first~order}  \\
  \left|{\epsilon_s }\right|                          &  \quad {\rm for}   \qquad &\rm{ second~order}
  \end{array}
\right.,
\end{eqnarray} 
where the first and second expressions correspond to first and second order resonances, respectively.
At high inclination, the mean inclination of the orbit contributes to the resonance width.
For $J_z \gtrsim J_{res}$ the libration timescale can be estimated from the frequency of libration in a pendulum 
model or
\begin{eqnarray}
\tau_{lib} &\approx& \sqrt{ \epsilon_s J_z \over a} \qquad {\rm for} \qquad {\rm second~order}  \\
              &\approx &  \sqrt{ \epsilon_b J_z^{1/2} \over a} \qquad {\rm for} \qquad {\rm first~order} 
\end{eqnarray}


\subsection{Hamiltonian Level Curves}

In Figures \ref{fig:ham_grow} and \ref{fig:ham_slow} we show level curves of the second order Hamiltonian 
(equation \ref{eqn:Ham})
in a canonical coordinate system with $(x,y) = \sqrt{2J_z}( \cos\phi,\sin\phi)$ so that radius on the plot scales
with orbital inclination or the maximum height of the orbit above the midplane.  
Panels from left to right show increasing values of $\delta$ or radius.
Convention is given here for $a < 0$, $\epsilon_s >0$ and $\delta $ defined as in equation \ref{eqn:aa}.
In this Figure, labels correspond to $\delta$ and $\nu$ normalized so that $a=-1$.
The frequency setting distance to resonance, $\delta$,
is expected to increase with increasing radius, and be negative inside resonance and positive outside resonance. 
As the bar slows down, reducing $\Omega_b$,  or the disk thickens, reducing the vertical oscillation
frequency $\nu$,  the distance to resonance, $\delta$, decreases, moving the resonance outward to larger radius.

On either side of resonance in Figure \ref{fig:ham_grow} (left or right panels) corresponding to large positive
or large negative $\delta$,   orbits at the origin remain at low inclination in the mid-plane.   
In the intermediate regions, $ \left| {\delta  \over  \epsilon_s} \right| < 1$, there are no planar orbits.
Only a single fixed point exists for ${\delta \over |\epsilon_s|} <  -1 $, two stable fixed points exist for 
$ \left| {\delta  \over  \epsilon_s} \right| < 1$ and three for ${\delta \over  |\epsilon_S|} >1$.

Fixed points, corresponding to periodic orbits in $z$, $\theta$ and $\theta_z$ in the bar's corotating frame,
 can be found by setting Hamilton's equations to zero and solving for their values of $J_z, \phi$.
The stable fixed points satisfy $\phi  = 0$ or $\pi$.   In the region where there are two stable fixed points
\begin{equation}
J_{z,fixed}(\delta)  = {\delta + \epsilon_s \over (-2 a)} 
 \qquad {\rm for}\qquad \left|{\delta\over \epsilon_s }\right| <1. \label{eqn:jfixed}
\end{equation}
The fixed points correspond to banana-shaped orbits where the orbit can reach a high inclination
at the ends of the bar.  
Fixed points at $\phi=0$ correspond to periodic orbits in the family denoted BAN$+$, upward
facing banana-shaped orbits, and those at
$\phi=\pi$ to BAN$-$,  a downward facing banana-shaped orbit family (e.g., \citealt{pfenniger91,martinez06}).
The two periodic orbit families are predicted by the Hamiltonian model but exist over a narrow range
in $\delta$.   At large negative $\delta$ the only fixed point is low inclination.  At large positive $\delta$
periodic orbits only exist in the mid plane or at extremely high inclinations.

The fixed point value of $J_z$ ranges from 0 where $\delta = - |\epsilon_s|$
to $J_{z,fixed} = \left|{\epsilon_s \over a}\right|$ where $\delta =  |\epsilon_s|$.  
The maximum value for the fixed point $J_{z,fixed}$ (at $\delta = \epsilon_s$), in the region where
there are no planar orbits, is equal to  
the $J_{res}$ value estimated from dimensional analysis (equation \ref{eqn:Jres}).
For $a$ negative and $\epsilon_s$ positive, fixed points have $\phi=0, \pi$.

In the region $|\delta/\epsilon_s|<1$, where there are no planar orbits, using 
the Hamiltonian (equation \ref{eqn:Ham}) we can 
write the energy as a function of $J_{z,fixed}$
\begin{eqnarray}
H &=& a J_{z,fixed}^2 + J_{z,fixed}(\delta + \epsilon_s) = 3 a J_{z,fixed}^2 \nonumber \\
&=& {3\over 4} a z_{f}^4 \nu^2  \label{eqn:Hfixed}
\end{eqnarray}
For the last step we have used the relation between $z$ and $J_z$ (equation \ref{eqn:coord})
and denoted $z_f$ as the maximum height of the periodic orbit with $J_{z,fixed}$.
This predicts a quadratic relation between the heights of the BAN+ and BAN- families
as a function of the Jacobi integral of motion.  
With the addition of terms depending upon the radial
degree of freedom,
our Hamiltonian is the Jacobi constant (see the appendices \ref{ap:action} and 
\ref{ap:can}). 
While \citet{pfenniger91,martinez06} have plotted 
the height of the periodic orbits of the BAN+ and BAN- families as a function of Jacobi
constant, up to now a functional relation has been lacking.  In future the Hamiltonian model
could be tested with fits to the bifurcation diagram.  The bifurcation diagram
could also be predicted 
for different simulation snapshots taking into account variations in the coefficient $a$,
the location of the resonance and the resonance strength.

If the product $a \epsilon_s$ is positive then the level curves are rotated by 90$^\circ$, 
and  fixed points have $\phi =  \pm \pi/2$, corresponding 
to  periodic orbits in the ABAN (anti-banana) family \citep{pfenniger91}.
The orbits are near the mid-plane at the ends of the bar
and are at high inclination along the bar minor axis. 
These are figure eight or infinity-shaped when projected on the $xz$ plane (with the $x$ axis is oriented on
the bar major axis, see Figure 8 by \citealt{martinez06}) and are usually unstable \citep{skokos02}.
The product $a \epsilon_s$ is dependent on radius.  It is possible that periodic orbits in the ABAN
family could appear in a region where $a \epsilon_s$ is negative.  Fixed points in our low order model lie on a line
(so you either have BAN$+$,BAN$-$ or ABAN but not both) but a higher order model (in $J_z$)
could give simultaneously both classes of periodic orbit families.

The Hamiltonian level curves can also be used to explore when an orbit supports or contributes
to a peanut shape in the galaxy.
Orbits that have $\phi$ librating around $0$ would spend more time above the plane at the bar ends
and less time below the plane there.   Those librating about $\pi$ would spend more time below
the plane at the bar ends.  These orbits would support a peanut shape as
they are coherent with the bar.
 However orbits that circulate about the origin could not provide good support for a peanut shape.
 For $|\delta/\epsilon_s|<1$
 the division between the two classes of orbits occurs on an orbit that contains
 the unstable fixed point at the origin and is called the separatrix.
 The separatrix contains two unstable orbits and separates orbits librating around $\phi=0$ and $\pi$
 from those circulating about the origin nearer the mid plane.
 The level curves in Figure \ref{fig:ham_grow} illustrate that for $\delta < -\epsilon_s$ orbits do not
support a peanut shape.  However, for $\delta > \epsilon_s$ high inclination islands exist that would support
a peanut shape.  In between, all  orbits within the separatrix and those near the separatrix support the peanut, 
but those at 
high inclinations do not.

Also of interest is the separatrix height.  
Where $\delta = |\epsilon_s|$ the separatrix consists of two circles that touch at the origin (see right lower
panel in Figure \ref{fig:ham_grow}).
The circles have a maximum distance
from the origin corresponding to 
$J_{z,sep} = \left|{2\epsilon_s / a} \right|$
giving a maximum height in these orbits of
\begin{equation}
z_{sep} = 2 \left| {\epsilon_s \over a\nu}\right|^{1/2} \label{eqn:zsep}
\end{equation} 
we we have used equation \ref{eqn:coord} relating $z$ and $J_z$.

\subsection{Bar growth, bar slowing and disk thickening}

Figure \ref{fig:ham_grow} illustrates how phase space varies during bar growth,
whereas Figure \ref{fig:ham_slow} illustrates how phase varies as the bar slows down or the disk thickens.
In these figures each column is meant to represent a different radius and each row
a different time.
From top to bottom in Figure \ref{fig:ham_grow} we show variations in the Hamiltonian (equation \ref{eqn:Ham}) 
level curves
as  the coefficient of bar perturbation strength $\epsilon_s$ increases. 
From left to right the coefficient $\delta$, controlling distance to resonance, increases.
  In  figure \ref{fig:ham_slow} from top to bottom $\delta$ is slowly decreased to illustrate
the effect of bar slowing or disk thickening, with increasing time going from top to bottom.

Recall that volume in phase space
is conserved if evolution is adiabatic.  Stars originally in the mid-plane would be located at the origin 
in Figure \ref{fig:ham_grow}.  If bar growth is adiabatic then those stars must remain in a small area or volume.
This means that they would remain near the fixed points (corresponding to banana-shaped periodic orbits)
 in these panels.    
When the vertical oscillation frequency, $\nu$, the coefficient $a$, and the bar pattern speed, $\Omega_b$,
do not vary then $\delta(r)$ is constant.  If the bar grows in strength, then $\epsilon_s$ increases in time.
In this case phase space varies as shown going downward in Figure \ref{fig:ham_grow}.
If the bar strength increases adiabatically, within the region $|\delta| < |\epsilon_s|$,
stars originally in the mid-plane  remain near periodic orbits (fixed points) and are lifted into high inclination orbits
and remain near a periodic orbit family
\citep{quillen02}. 
For $|\delta| > |\epsilon_s|$ orbits exist in the mid-plane at all times and so stars originally  
at low inclination can remain there.  
The vertical height of the bulge as a function of radius would depend on the height
of the periodic orbits (equation \ref{eqn:jfixed}).  The peanut height would be
zero  where $\delta = - |\epsilon_s|$ and reach a maximum of
\begin{equation}
z_{max} = \sqrt{ \left|{2 \epsilon_s \over a \nu}\right|}  =  \sqrt{C_{4z}  \over a \nu^2 }
\label{eqn:zmax} 
\end{equation}
where $\delta = |\epsilon_s|$. 
Above we have used equation \ref{eqn:epss} for $\epsilon_s$.
The height, $z_{max}$, is the maximum fixed point height within the region where there are no in-plane 
or low inclination orbits.
There are higher fixed points at larger $\delta$ but these exist at radii where there are also planar orbits
(see panel on upper right in Figure \ref{fig:ham_slow}).

\citet{quillen02} proposed that a bow-tie or X-shape would be seen because of 
the linear dependence of the fixed point action variable $J_z$ on the distance to resonance, $\delta$. 
Using expressions for $\delta, a, \nu, C_{4z}$ as a function of radius, 
the height of a peanut could approximately be  predicted using 
\begin{equation}
z(\delta) = z_{max} {1 \over 2} \left(1 + {\delta \over \epsilon_s} \right)
  \qquad \qquad {\rm for} \qquad \left|{\delta \over \epsilon_s }\right| <1,  \label{eqn:zdelta}
\end{equation}
if only $\epsilon_s$ varies during bar growth and bar growth is adiabatic.
This prediction is appropriate if  the resonance is wide.  In other words, the region spanned by
$\left|\delta/\epsilon_s \right| <1$ must be wide enough to match the extent of observed peanut shapes.
As we will show below, the resonance is not  wide enough to confirm this expectation.

If the bar growth is not adiabatic  then stars originally near fixed points would not be restricted to remain
  near fixed points or period orbits.
   If the resonance is thin and weak (corresponding to small $\epsilon_s$), it would have a slow libration frequency,
  and even slow evolution in the galaxy may not be adiabatic.
  In this case the distribution of stars is heated by the resonance even during bar growth.   
  In the region $|\delta/\epsilon_s| <1$
 there are no orbits in the mid-plane,  so stars should have a height distribution at least as large
 as $z(\delta)$ following bar growth.    \citet{quillen02} primarily discussed resonance trapping due 
 to adiabatic bar growth but  showed that even when a bar grew moderately quickly, an X-shape 
 was seen in simulated stellar height distributions.  The adiabatic limit can be estimated from the square
 of the libration timescale \citep{quillen06}.  Hence variations in the coefficients $\epsilon_s, \delta$
 can be considered adiabatic if for example $|\dot \epsilon_s |\lesssim \epsilon_s^2$ or
 $|\dot \delta| \lesssim  \epsilon_s^2$.
 
 \begin{figure*}
\begin{center}
\includegraphics[width=4.8in]{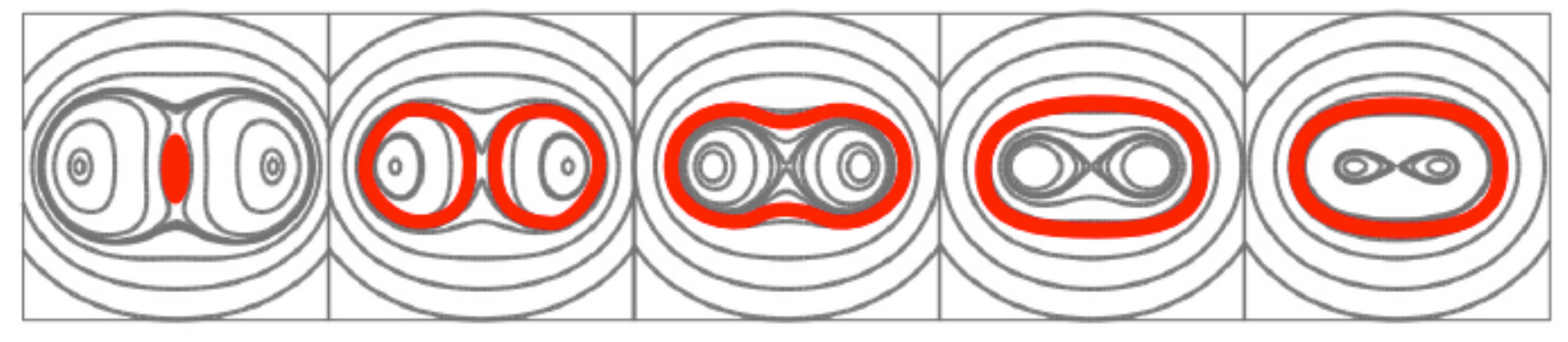} 
\caption{
Hamiltonian level curves showing the evolution in phase space of stars originally located exterior to the resonance
in the mid plane as the resonance drifts outward.
In the leftmost panel stars near the origin, in the mid plane, and just outside resonance are in the red area.   As the resonance moves outward, (moving from left to right panels) stars are first forced out of the mid plane and into orbits just interior to the separatrix (second panel from left). 
Inside the separatrix (panel second from left) stars librate about $\phi=0$ (right red oval) or $\pi$ (left red oval) 
and so support an X-shape. Then as the
separatrix shrinks, the stars leave the separatrix and are left at high inclination but no longer supper the peanut or X-shape (rightmost two panels).  
Outside the separatrix (rightmost three panels) the angle $\phi$ oscillates about the origin.
We associate stars in the vicinity of the resonance separatrix, shown in the red area in the second panel on left, 
and in the middle panel as stars in the X-shape.  
Stars interior to the separatrix would be banana shaped, either oriented
up or down depending upon the value of the nearest periodic orbit.   
Stars just exterior to the separatrix would have morphology similar to the sum of upward and downward banana-shaped orbits.  Because they spend little time near $\phi = \pm \pi/2$ they would support the peanut shape
even though $\phi$ oscillates instead of librating.
\label{fig:ham_red}
}
\end{center}
\end{figure*}

When the bar slows down, or the disk thickens, the parameter $\delta$ increases.  
This setting is illustrated
in Figure \ref{fig:ham_slow}  showing level curves with fixed perturbation strength $\epsilon_s$ but 
$\delta$ decreasing from top to bottom panels. 
As the resonance moves outward,
stars originally in the mid plane just outside resonance are forced out of the mid plane
and into high inclination orbits.
If the evolution is adiabatic, stars originally near fixed points are pushed back into the mid plane.
The separatrix shrinks in size (moving downward on the figure).
Stars librating about fixed points eventually cross the separatrix. Afterwards their orbits take them around
the origin so the angle $\phi$ oscillates and they no longer support the peanut or X-shape. 
A star that originally is outside of resonance, and in the mid plane,
is first heated to high inclination when it enters resonance, when
it must enter an orbit just within the separatrix.  
Then as the separatrix shrinks the star must leave the resonance,
but it will leave at a high inclination.  If the resonance continues to drift outward then it will leave
high inclination stars behind with angle $\phi$ circulating rather than librating.
The resonance heats the stars to a height approximately equal to  the separatrix height
(equation \ref{eqn:zsep}). 
The fate of stars originally in the mid plane just exterior to resonance
is illustrated in Figure \ref{fig:ham_red} for adiabatic drift.
If the drift is not adiabatic then one can think of an instantaneous mapping between panels
in Figure \ref{fig:ham_slow} rather than continuous evolution of area preserving volumes as shown
in Figure \ref{fig:ham_red}     

The resonance size scale and timescale can be used to estimate
 how long it takes an orbit to be heated by the resonance.
For example, a star placed at the unstable fixed point near the origin would exponentially move
away reaching an height of order $z_{max}$ in a timescale of order $\tau_{lib}$.

If the bar speeds up or if gas is accreted adding mass to the mid plane and raising the vertical oscillation frequency,
then the resonance moves inward instead of outward.  In this case we consider Figure \ref{fig:ham_slow}
but evolution is upward on this figure instead of downward.  Particles trapped near fixed points would be lifted to 
even higher inclination.  In this case an X-shape would also be seen but it would extend to $\delta > \epsilon_s$.
This situation is similar to the capture of Pluto in resonance with Neptune.  If the bar speed increase is adiabatic
stars would be confined to high inclination orbits, near fixed points, at a radius where mid plane orbits are allowed.   
   
 The role of the disk thickness in influencing the vertical oscillation frequency was
a key property of a time dependent resonant trapping model for disk thickening and bulge growth in the absence
of a bar by  \citet{sridhar96a,sridhar96b}.
Here also the resonance itself can cause the disk to thicken, reducing the vertical oscillation frequency $\nu$
and the  perturbation strength $\epsilon_s$ and possibly decreasing $a$.
A time dependent Hamiltonian could in future be used to construct an instability model for the growth
of the peanut.

\begin{figure*}
\begin{center}
\includegraphics[width=4.8in]{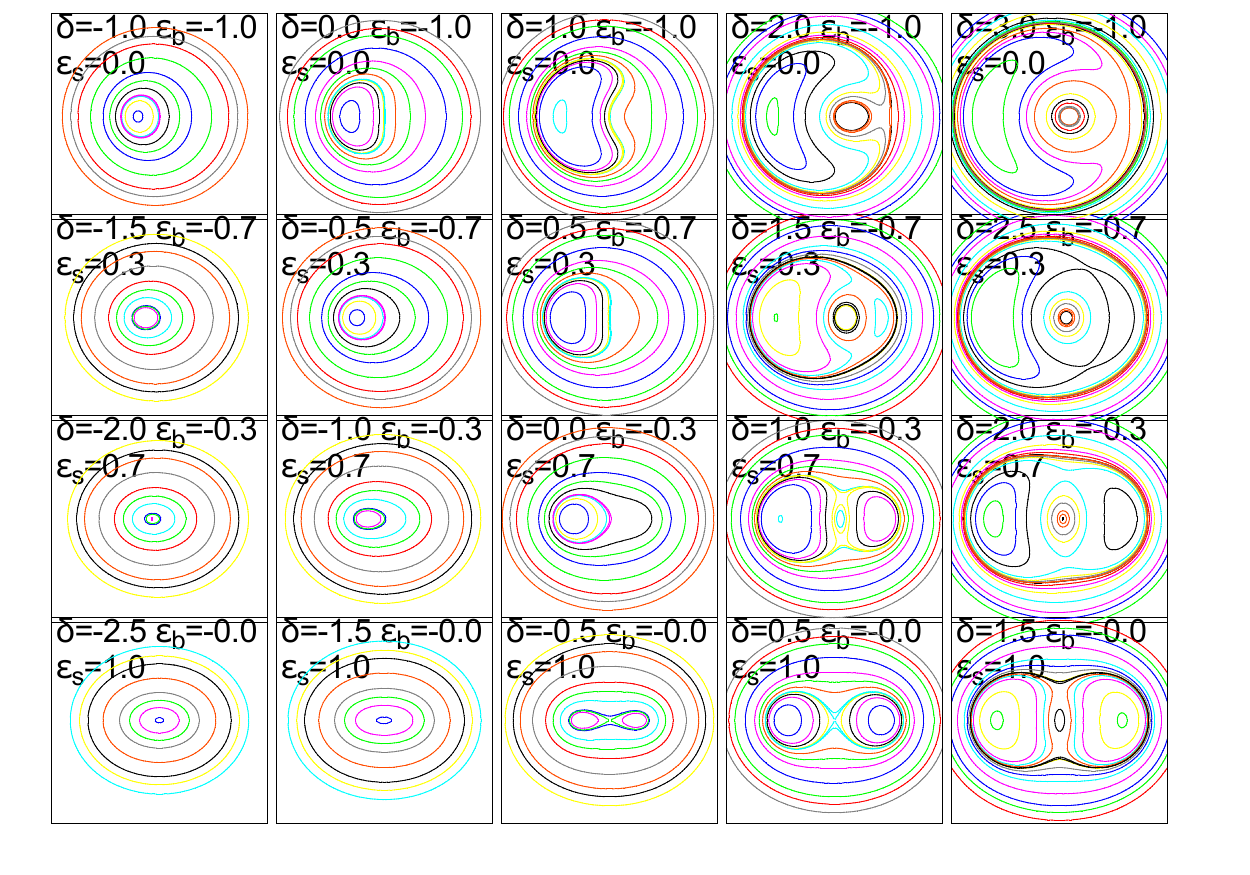} 
\caption{
Level curves of the Hamiltonian in equation \ref{eqn:Ham_1} are shown showing the transition between
first and second order that would take place following bar buckling.
From left to right the coefficient $\delta$, setting distance to resonance increases.
From top to bottom, $|\epsilon_b|$ decreases while $\epsilon_s$ increases.
On the bottom periodic orbits giving both upward and downward banana shaped  are present 
($\phi=0$ and $\phi=\pi$).  However on the top, orbits supporting the buckling dominate on the outer side of resonance and those pointed opposite are found on the inner side.  If the transition is adiabatic
then orbits trapped in one family (orbits supporting the bar buckle) could remain at high inclination.
Afterward more stars could be found in one family of banana shaped periodic orbits than
the other.   It is possible that an asymmetry could persist in the phase space distribution after the bar
buckled.
  \label{fig:Ham_1}
}
\end{center}
\end{figure*}

\subsection{Transition from bar buckling to a symmetric potential}

When a bar buckles the Hamiltonian resembles the first order one, equation \ref{eqn:Ham_b}.  However
as the buckling dies away, $\epsilon_b$ drops.  Weaker second order terms become
more important and the Hamiltonian resembles the second order one, equation \ref{eqn:Ham}.  
We can consider a Hamiltonian that contains both first and second order terms or
\begin{equation}
H(J_z , \phi) = aJ_z^2 + \delta J_z + \left[ \epsilon_s J_z + \epsilon_b J^{1/2} \right] \cos 2 \phi \label{eqn:Ham_1}
\end{equation}
We expect $a,\epsilon_b <0$ and $\epsilon_s >0$ (see Figure \ref{fig:gS0_vertshape} below showing
potential shapes).  
Level curves for this Hamiltonian are shown in Figure \ref{fig:Ham_1} showing the effect of 
 decreasing $|\epsilon_b|$ and increasing $\epsilon_s$. 
 This mimics the transition between
  bar buckling and a symmetric state. Both first order and second order Hamiltonians have fixed points
 at $\phi=0,\pi$, corresponding to upwards and downwards facing banana shaped periodic orbits.
 However, one class has much higher inclination than
 the other for first order Hamiltonian (top panels in Figure \ref{fig:Ham_1}) and $\delta >0$.
 As  symmetry about the mid plane is restored, both families of banana shaped 
 periodic orbits are present and they have similar inclination.   If the transition is adiabatic, then
 one family of banana shaped orbits could be more populated.  An asymmetry may persist in the
 number of stars in each periodic orbit family after bar buckling.
  

\begin{figure*}
\begin{center}
\includegraphics[trim={0.0cm 0.3cm 0.0cm 0.0cm},clip,width=6.2in]{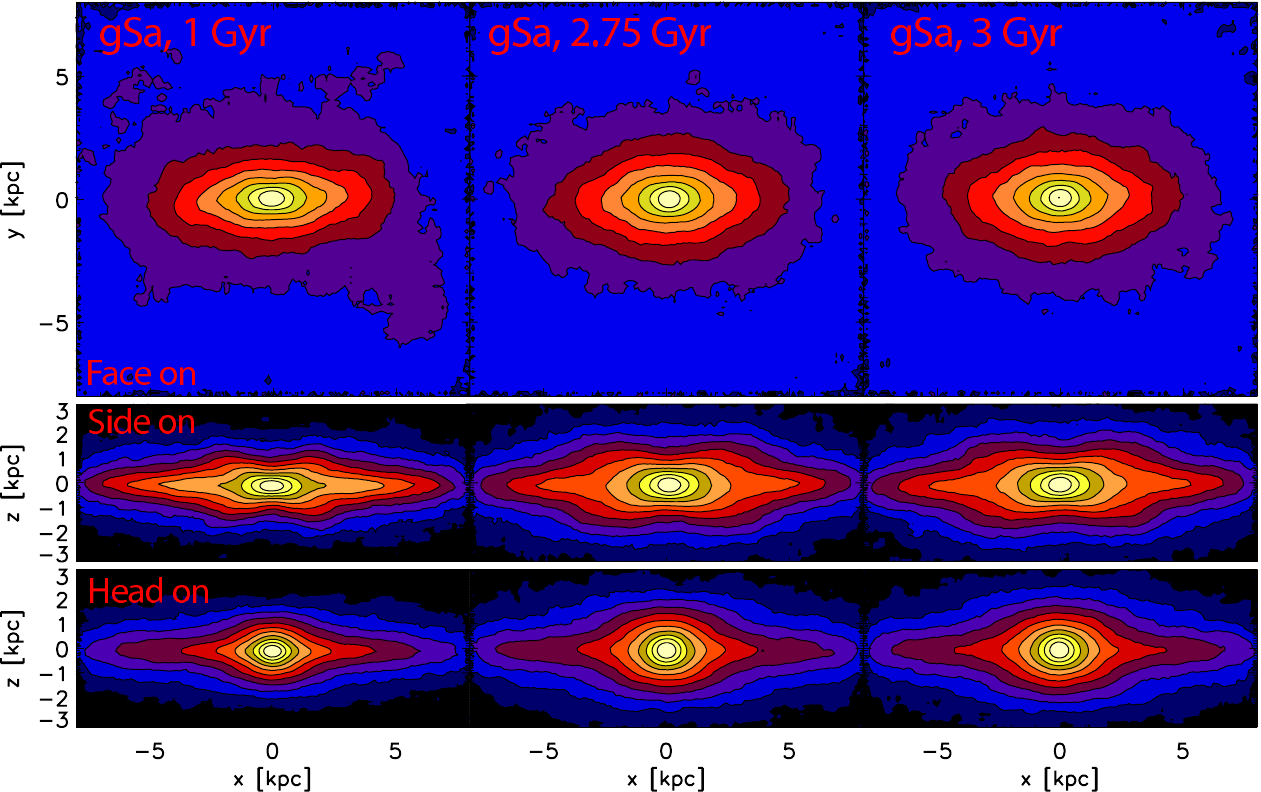} 
 \caption{Simulation snapshots are shown for three different times (1.0, 2.75, 3.0 Gyr) 
 during the gSa GalMer simulation.
 Top panel shows projected disk density on the $xy$ plane with $x$ aligned with the bar.
 The middle and bottom panels show projected density on the $xz$ and $yz$ planes.
 Axes are in kpc.  Only disk stars are shown.
The bar is not at a fixed pattern speed, but continues to slow down.
The peanut shape continues to grow.  
  \label{fig:snap_gSa}
 }
\end{center}
\end{figure*}

\begin{figure*}
\begin{center}
\includegraphics[trim={0.0cm 0.3cm 0.0cm 0.0cm},clip,width=6.2in]{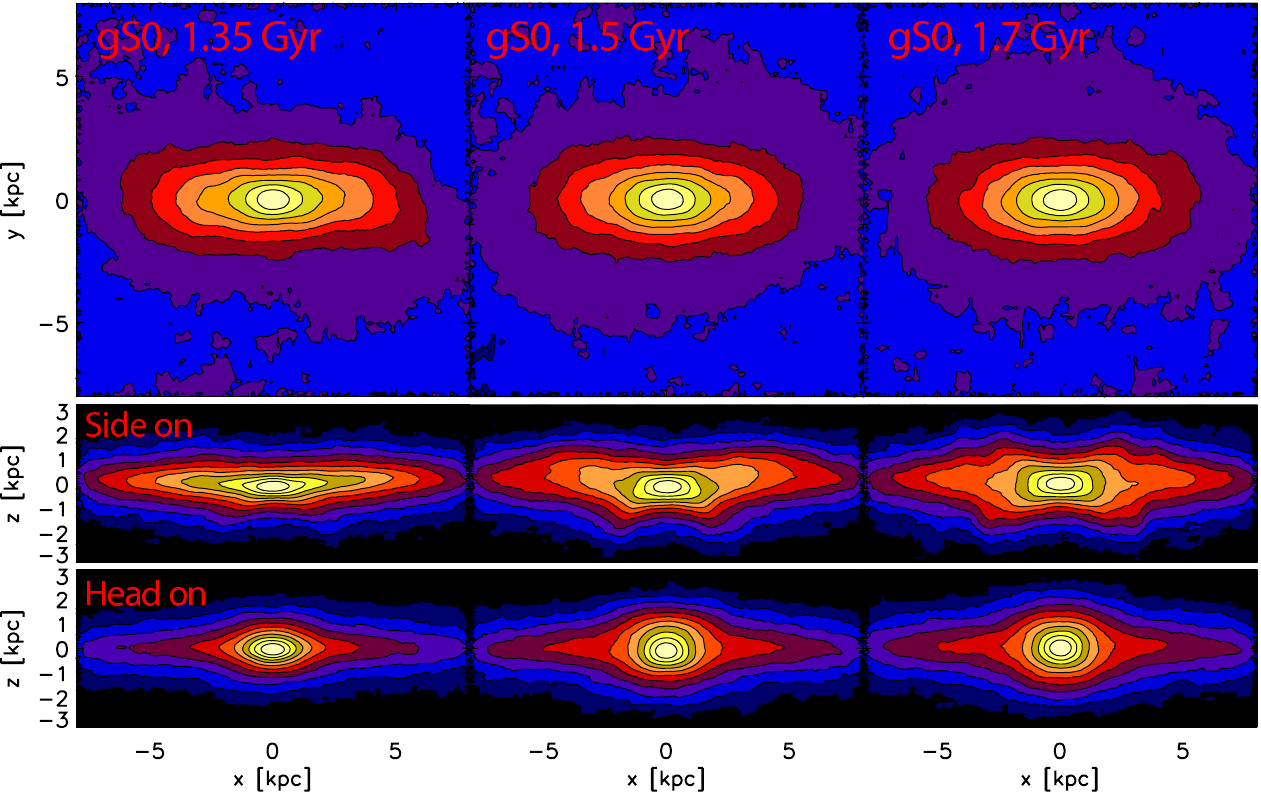} 
 \caption{Simulation snapshots are shown for the gS0  simulation during bar buckling and
 at 1.35, 1.5 and 1.7 Gyr.  Similar to Figure \ref{fig:snap_gSa}.  
 During bar buckling the bar exhibits a W shape with two outer peaks
 separated by a large radius above the plane and two inner because less separated below the plane.
 The peaks of the W are closer together at later times than at earlier times.
 \label{fig:snap_buckle}
 }
\end{center}
\end{figure*}

\begin{figure*}
\begin{center}
\includegraphics[trim={0.0cm 0.3cm 0.0cm 0.0cm},clip,width=4.2in]{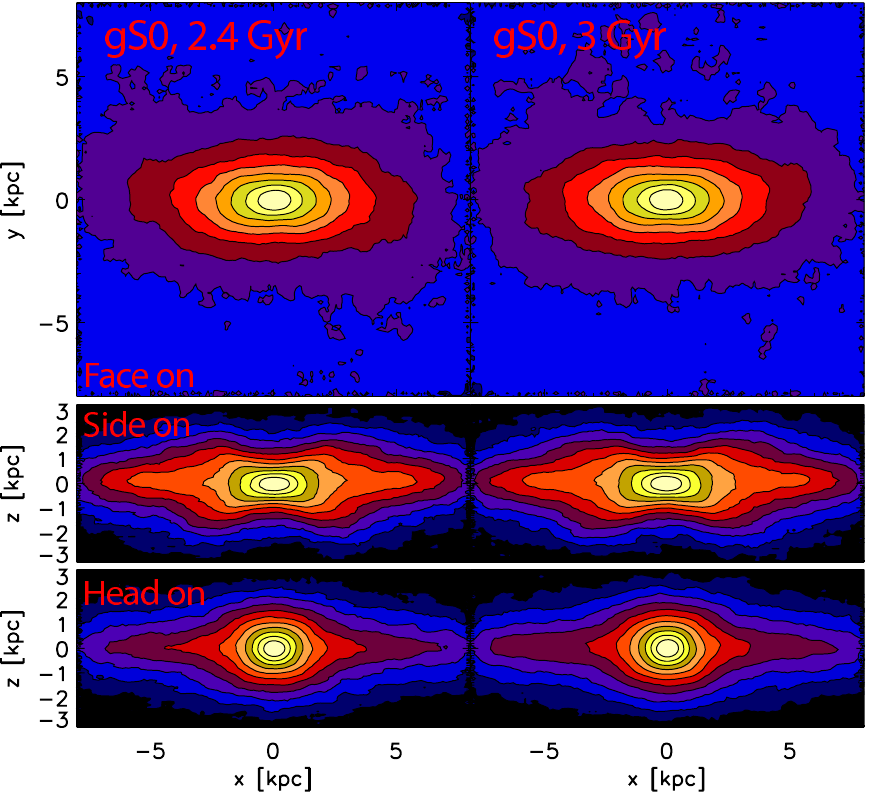} 
\caption{Simulation snapshots are shown for the gS0  simulation at 2.4 and 3.0 Gyr long after bar buckling.
 Similar to Figure \ref{fig:snap_gSa}.
The peanut length is longer than during bar buckling.
 The bar is not at a fixed pattern speed, but continues to slow down.
  \label{fig:snap_gS0}
 }
\end{center}
\end{figure*}

\begin{figure*}
\begin{center}
$\begin{array}{cc}
\includegraphics[height=3.0in]{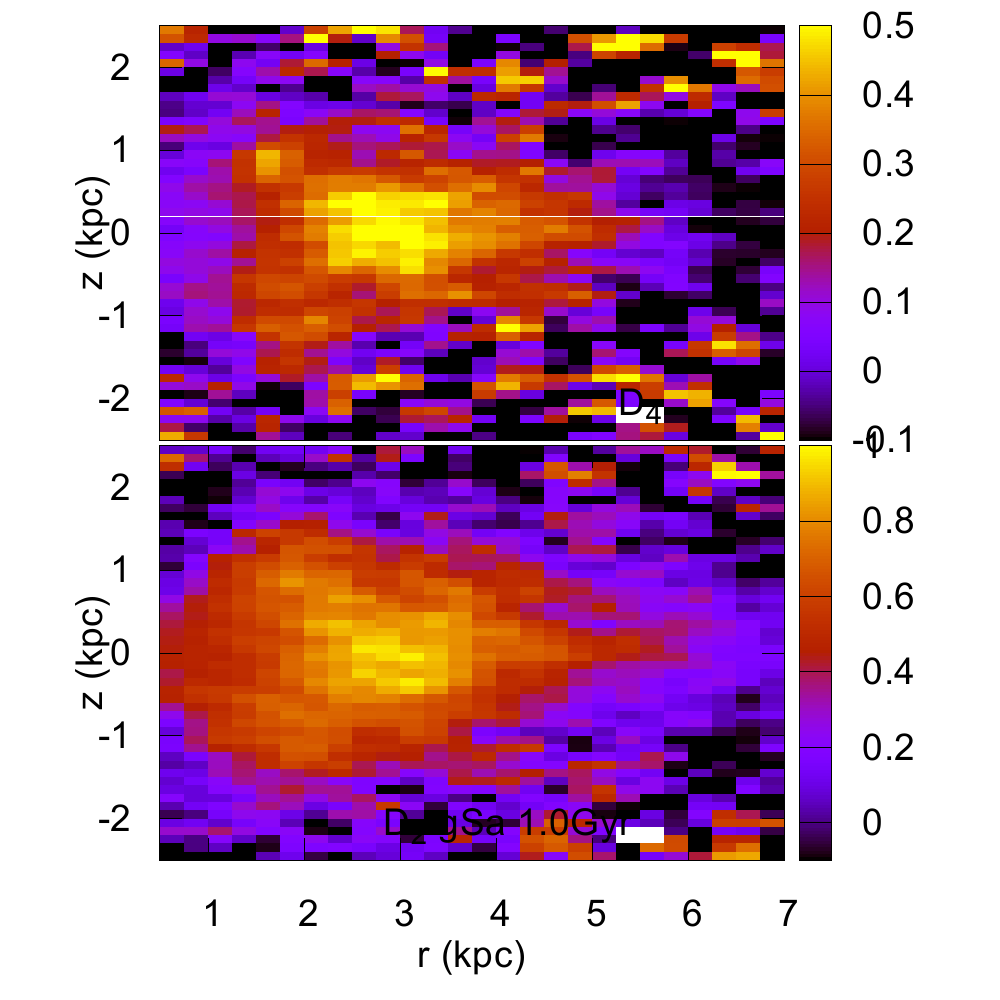} &
\includegraphics[height=3.0in]{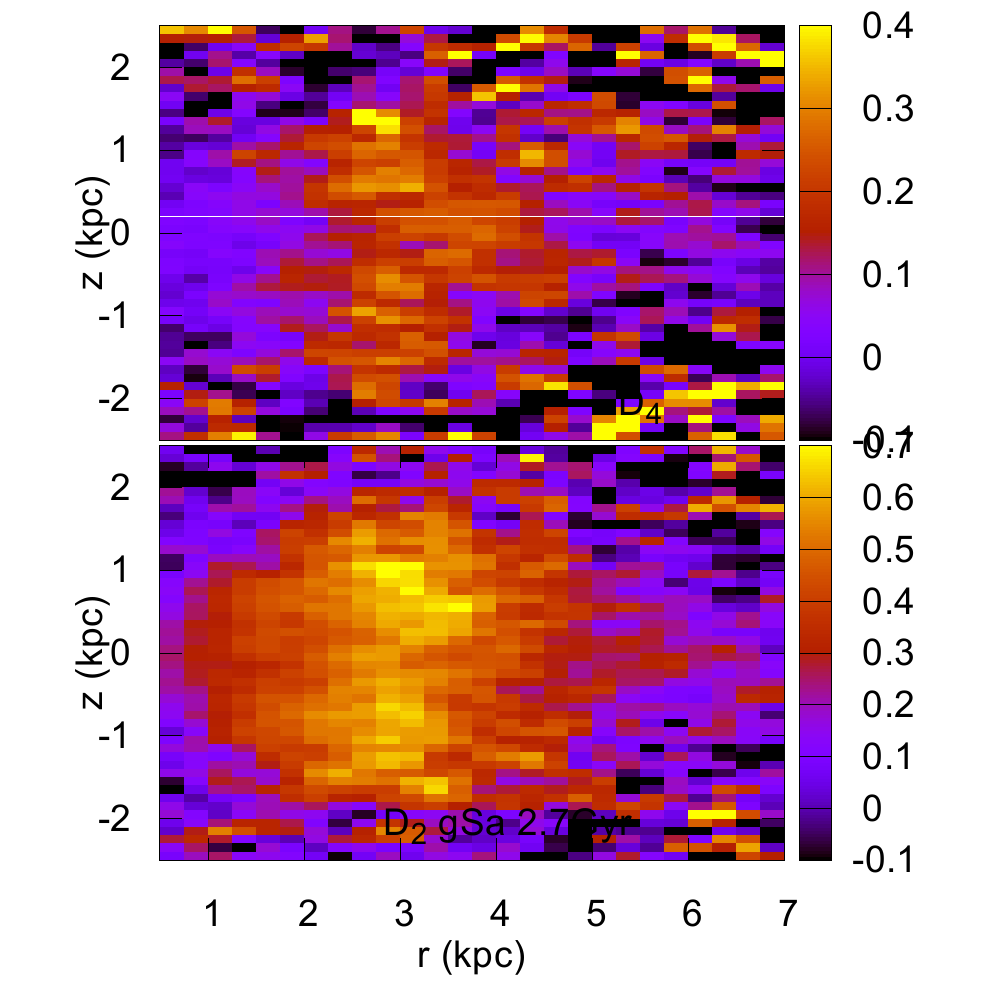} 
\end{array}$
\caption{
Fourier components of the disk density as a function of radius and height.
The  $m=2$ and $m=4$ Fourier coefficients of the density distribution are shown, divided by
the axisymmetric value,  for the gSa simulation
at 1.0 (left) and 2.7 Gyr (right).
The top of each panel shows the $m=4$ Fourier coefficient and the bottom of each panel
the $m=2$ coefficient as a function of radius and of height.  The value at each position is shown divided
by the axisymmetric value at that radius and height. 
   The $x$ axes are radii and the y axes are height $z$ above or below the galactic plane,  both in kpc.
The X-shape is visible at later times in the shape of the $m=2$ Fourier coefficient (see panel on lower right).
\label{fig:gSa_denshape}
}
\end{center}
\end{figure*}

\begin{figure*}
\begin{center}
$\begin{array}{ccc}
\includegraphics[trim={0.2cm 0 0.6cm 0},clip,height=2.9in]{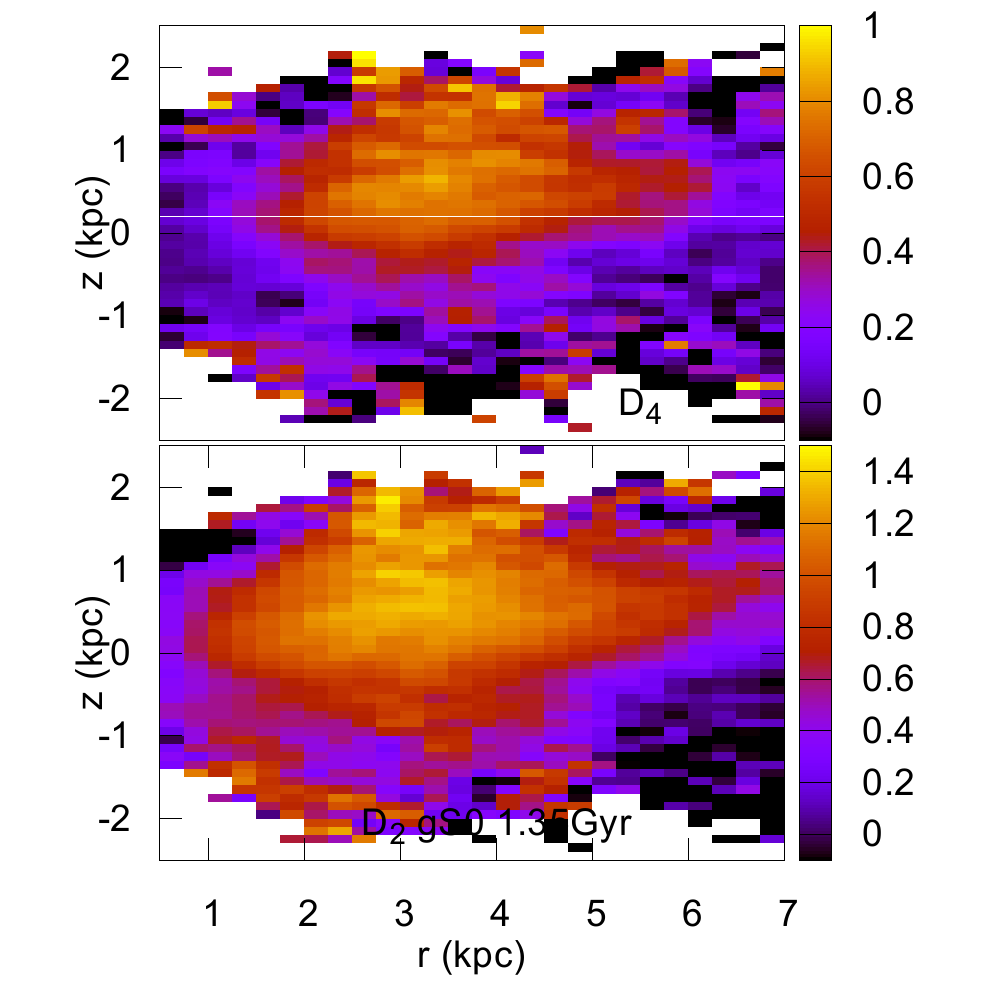} &
\includegraphics[trim={1.6cm 0 0.6cm 0},clip,height=2.9in]{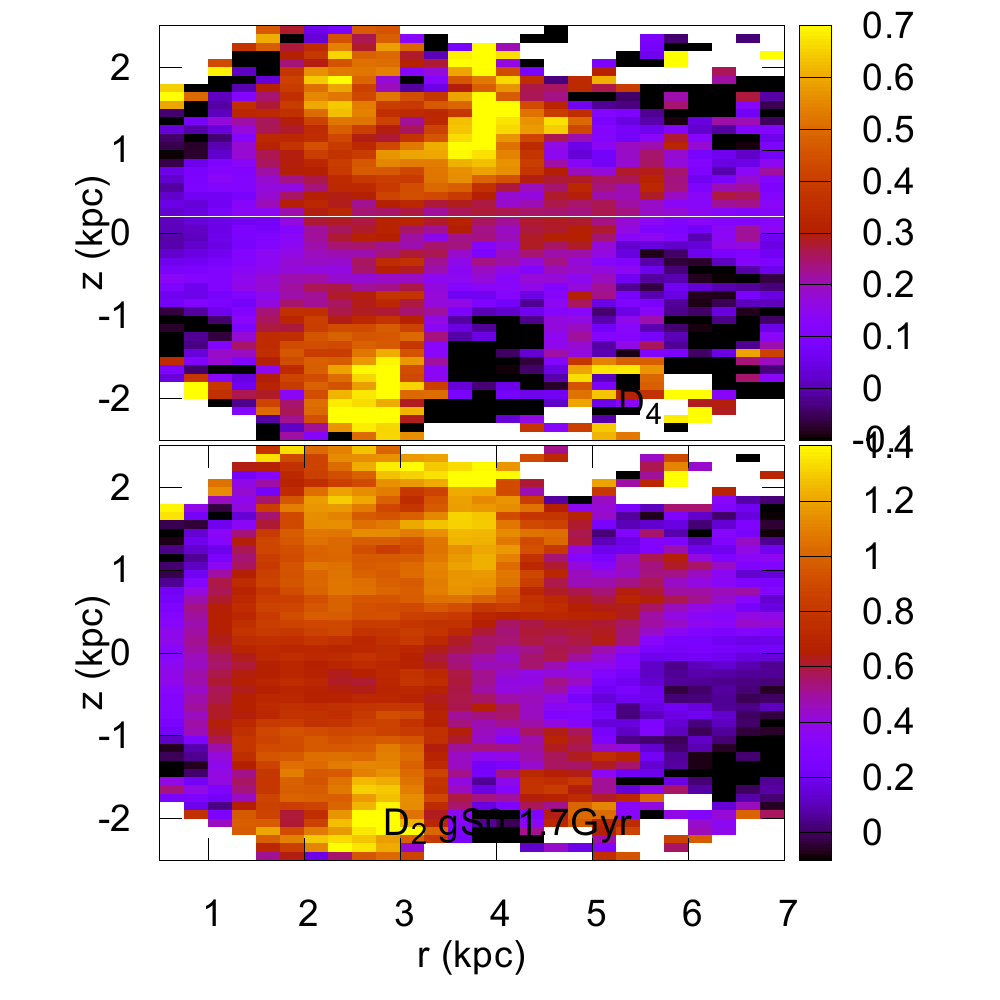}  &
\includegraphics[trim={1.6cm 0 0         0},clip,height=2.9in]{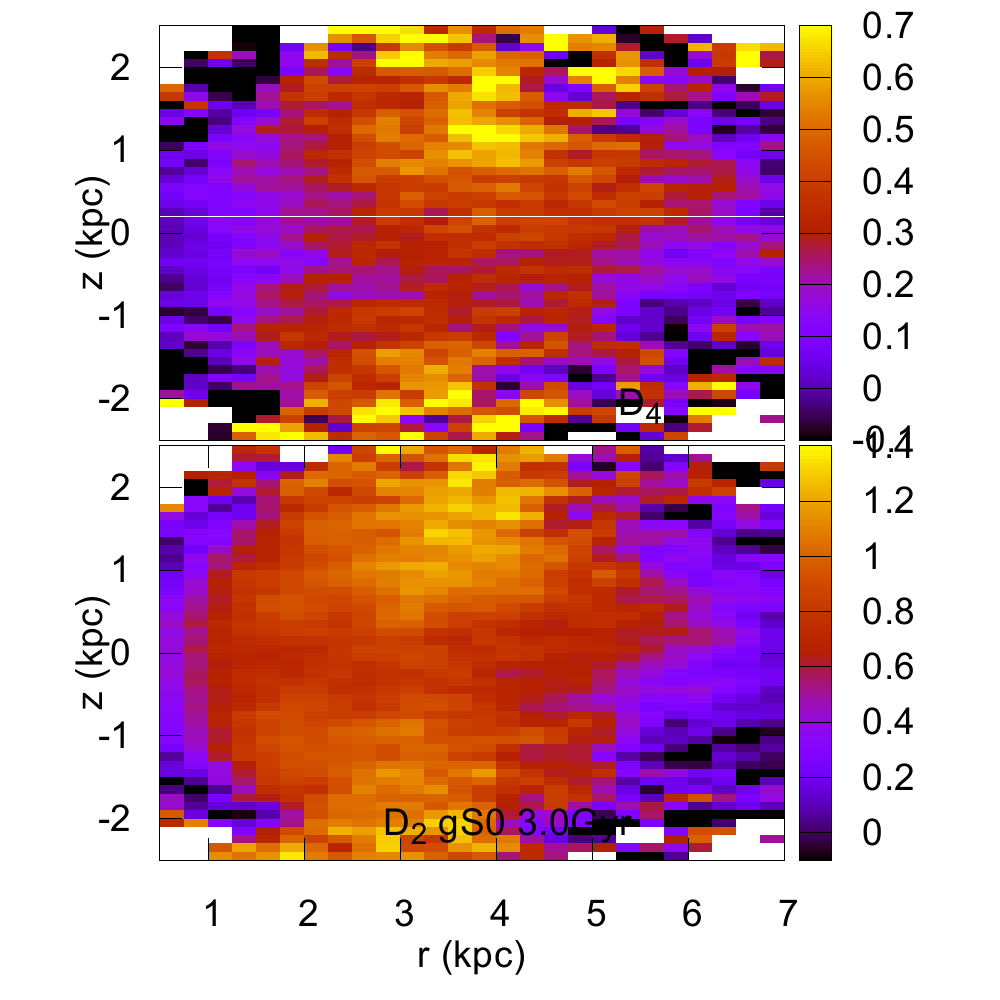} 
\end{array}$
\caption{
Normalized Fourier $m=2$ and $m=4$ components of the density as a function of radius and height.
Similar to Figure \ref{fig:gSa_denshape} except for the gS0 simulation and at 1.35, 1.7 and 3.0 Gyr. 
\label{fig:gS0_denshape}
}
\end{center}
\end{figure*}

\section{Peanut shaped bulge formation in numerical simulations}

To test the resonant model for a peanut-shaped bulge  we examine
peanut-shaped distributions present in N-body simulations.
%
We study isolated giant disk galaxy simulations denoted `gSa' and `gS0' from the GalMer database \citep{dimatteo07,chil10}.
The simulations include 320,000 (gS0) or 240,000 (gSa) disk particles, and 160,000 dark matter halo particles.
The initial halo and bulge are modeled as Plummer spheres.
The initial gaseous and stellar discs follow Miyamoto-Nagai density profiles.
The Toomre parameter of both stars and gas is taken to be $Q = 1.2$ as the initial condition of the Tree-SPH simulations.
GalMer simulations outputs have velocities in 100 km/s, positions in kpc, potential in units of (100 km/s)$^2$
and timesteps in units of Gyr.
For a discussion of the properties of the isolated disk GalMer simulations see \citet{minchev12}.
These simulations were chosen because they have many disk particles and displayed strong bars.
In the gS0 simulation the bar buckles, whereas the bar did not buckle in the gSa simulation.
A comparison of the two simulations is particularly interesting because of this difference in bar evolution. 
Both simulations exhibit strong peanut-shaped bulges at later times.

In Figure \ref{fig:snap_gSa}
two snapshots of the stars alone from the gSa simulation are shown, the first one at 1~Gyr, just after bar growth
and the second one at 2.74~Gyr, at the end of the simulation.
As can be seen from the vertical disk density profiles, the galaxy is strongly peanut-shaped and
the peanut shape continues to grow after bar formation.
There is no evidence of bar buckling, but 
bar buckling is not required for boxy/peanut-shape formation. 
\citet{friedli90} reported that N-body simulations that forced z-symmetry did grow boxy/peanut-shaped bulges but more slowly than those that did not force the symmetry.

In Figure \ref{fig:snap_buckle} and \ref{fig:snap_gS0} we show snapshots of the gS0 simulation.
Figure \ref{fig:snap_buckle} shows the gS0 simulation just after bar growth and during the bar buckling phase.
Figure \ref{fig:snap_gS0} shows the gS0 simulation at later times.  
During the bar bucking phase, the peanut shape is quite strong but smaller than at later times.
The peanut shape resembles a W-shape with two upper widely separated peaks at large radius above the plane
and two less widely
separated peaks below the plane.  
As was true in the gSa simulation, the peanut shape in the gS0 simulation becomes longer at later times.
In both simulations discussed here, the bar pattern speed does not remain fixed, 
but slows down after bar growth. 
Pattern speeds as a function of time are shown in Figure 2 by \citet{minchev12}).

We compute the Fourier coefficients of the stellar disk density  from the simulation snapshots 
in a grid of radial and vertical positions;
$S_m(r,z) \equiv {1 \over \pi} \int \rho(r,z,\phi) \cos m\phi d\phi $, where $\rho(r,z,\phi)$ is the stellar density.
In Figures \ref{fig:gSa_denshape} and \ref{fig:gS0_denshape} we show these Fourier coefficients, as a function of $r,z$, 
divided by the axisymmetric average of the density at each radius and height, 
$S_0(r,z) \equiv {1 \over 2\pi} \int \rho(r,z,\phi) \cos m\phi d\phi $.
The peanut-shape is particularly evident at later times in these images.
The peanut-shape in these normalized Fourier components is wider (in height) than the peanut shape
in the actual density distribution. 
This implies that
above the disk, the fraction of stars aligned with the bar is larger than the fraction of stars aligned with
the bar in the mid plane.   

During bar-buckling, the first order Hamiltonian model should model the resonance (equation \ref{eqn:Ham_b}).
Inside resonance (negative $\delta$ and at smaller radius), periodic orbits
are found at $\phi=\pi$, whereas outside resonance (positive $\delta$), periodic orbits have $\phi=0$
(see Figure \ref{fig:Ham_1}).
Outside resonance banana shaped periodic orbits oriented with the buckling support the buckling, 
having high $z$ at the ends of the bar,
whereas inside resonance, periodic orbits have low $z$ at the ends of the bar.
The W-shape in Figure \ref{fig:snap_buckle} 
may be consistent with this Hamiltonian model.    The two inner radius apexes of the W may correspond to
the banana shaped orbits inside resonance with the banana ends pointing down, 
whereas the two outer upper ends of the W could be from 
 banana-shaped orbits with the two ends oriented upward. 
 The width of the resonance is set by the libration frequency ($\tau_{lib}^{-1}$)
and depends on the resonance strength (equation \ref{eqn:taulib}). 
As the resonance weakens, the libration frequency
decreases and the resonance is strong over a smaller range of radius.
Consequently, when the asymmetric perturbation is strong at early times, 
the two peaks in the W are far apart, but they approach each other
as the buckle weakens at later times.   The progression is as seen in Figure \ref{fig:snap_buckle}.
This progression of the classes of orbits has also been illustrated by \citet{martinez06} by searching for
periodic orbits in N-body snapshots.
When the galaxy becomes more symmetrical the first oder resonant term weakens and the second order
one dominates.  The
resonant model becomes second order (equation 
\ref{eqn:Ham}), 
and both sides of resonance contain both classes of banana shaped orbits, as shown in Figure \ref{fig:ham_grow}.
The transition is illustrated in a varying Hamiltonian model in 
Figure \ref{fig:Ham_1} from top to bottom showing the growth
of the downward facing banana shaped periodic orbits appearing as the buckling dies away.

Depending upon the speed of evolution of the perturbations, the distribution
of upwards oriented banana-shaped orbits may  differ from the distribution of 
downward oriented banana-shaped orbits after the buckling has dissipated.   

\subsection{Estimating oscillation frequencies  in the simulations }

We describe how we measure the angular rotation
rate, $\Omega$, vertical oscillation frequency $\nu$, and epicyclic frequency, $\kappa$, as a function
of radius in individual simulation snapshots.

Because the halo is live, the center of the galaxy does not remain fixed.  
To measure quantities as a function distance from the galaxy center,
the bulge center at each timestep must be taken into account.
At every timestep we computed centroids in the central region using stars with coordinate
radius inside $r<2$ kpc.
Velocity and position centroids computed from all stars in this central region were subtracted before making
additional measurements.
In the annulus $4<r<5$~kpc 
we computed the $m=2$ Fourier components of the mass distribution projected
onto the mid-plane.
This gives a measurement of the bar orientation angle.
The stars were then rotated so that the bar was oriented horizontally for all subsequent computations. 
Bar pattern speeds
 were measured from the advance of the bar angle as a function of time and were consistent with 
 those plotted in Figure 2 by \citet{minchev12}.

Many GalMer simulations are archived with an array giving the gravitational potential at the position of every star.
We used this to estimate the azimuthally averaged potential in the mid plane as a function of radius and height.
To reduce the graininess caused by the halo particles,   
we also computed the Fourier components by direct summation
as a function of height using the positions and masses of the stars alone.

\begin{figure}
\begin{center}
\begin{tabular}{c}
\includegraphics[width=3.4in]{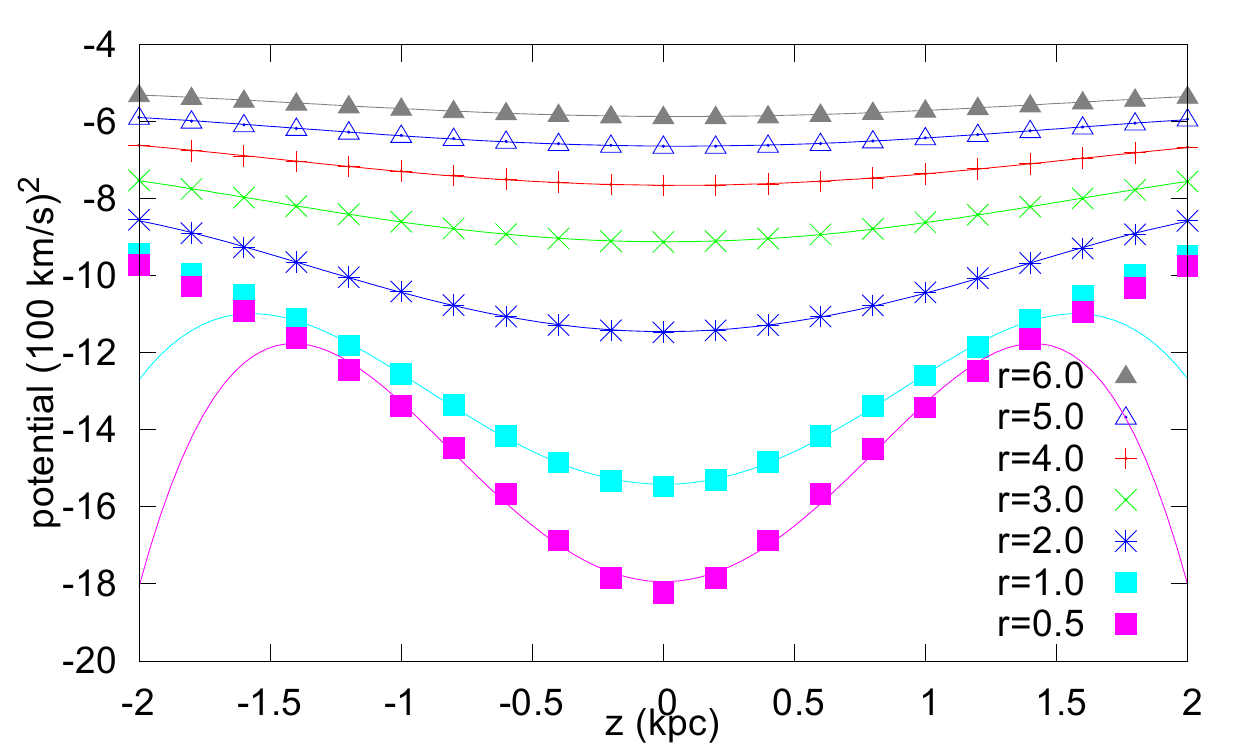} \\
\end{tabular}
 \caption{Shape of the azimuthally averaged gravitational potential as a function of $z$ 
 at different galactocentric radii in the mid-plane.  
Fits to the potential are shown as lines.   These fits are used to estimate the  vertical oscillation
frequency, $\nu$, and the parameter $\lambda$.
 This figure shows fits for the gS0 simulation at 1.7 Gyr.
  \label{fig:vp} 
 }
\end{center}
\end{figure}

Vertical slices of the azimuthally averaged gravitational potential at different radii
are shown in Figure \ref{fig:vp} for a single snapshot.   
The profiles are centered on zero implying that the centroiding 
was effective.
At each radius,  we fit a quadratic function as a function of $z$.
Both the profiles (measured from a simulation snapshot) and the fits are shown in Figure \ref{fig:vp}
at two different times.
These functional fits are used to compute the  vertical oscillation frequency, $\nu$, and parameter 
\begin{equation}
\lambda = \left.{\partial^4 V_0 \over \partial z^4}\right|_{z=0}
\end{equation}
at different radii.  The parameter $\lambda$ is needed for computation of the $a$ coefficient in the Hamiltonian
(see appendices \ref{ap:action} and \ref{ap:can}).
We checked that the measurements for $\lambda$ and $\nu$ were not strongly dependent
upon the form of the function fit to the potential or the method used to compute
the azimuthal average.

\begin{figure}
\begin{center}
\begin{tabular}{c}
\includegraphics[width=3.4in]{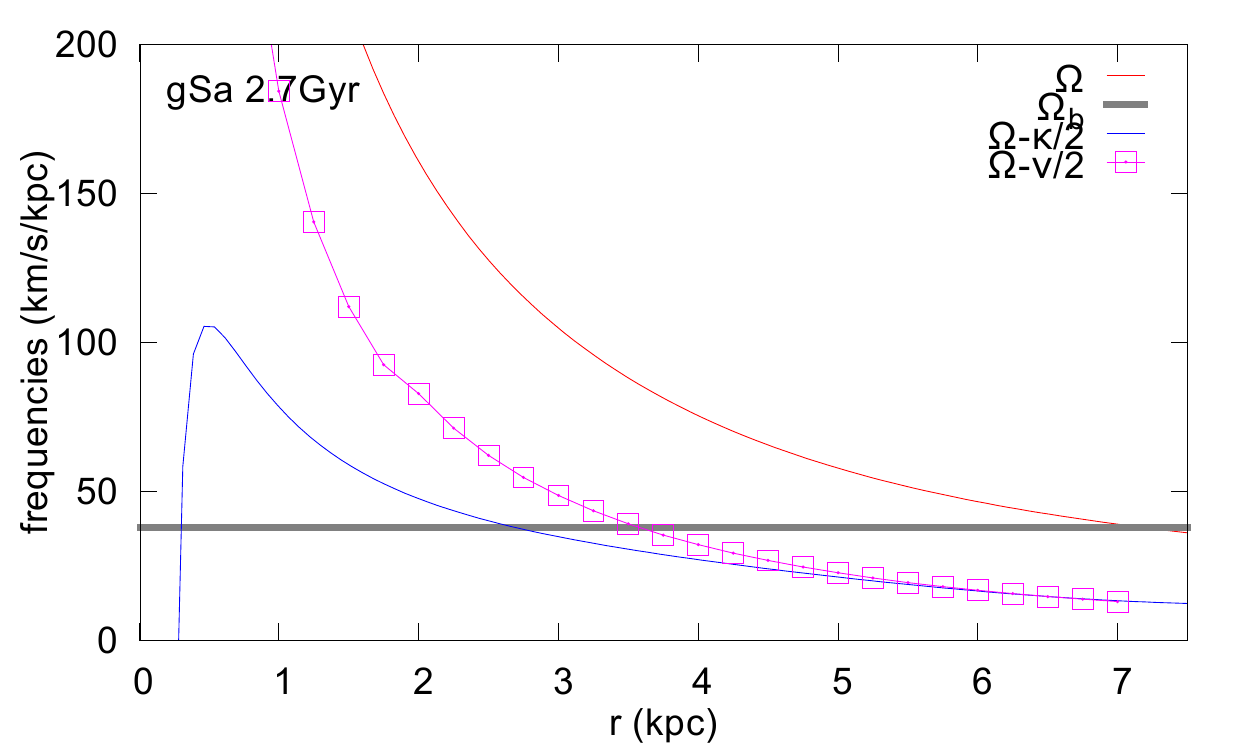} \\
\includegraphics[width=3.4in]{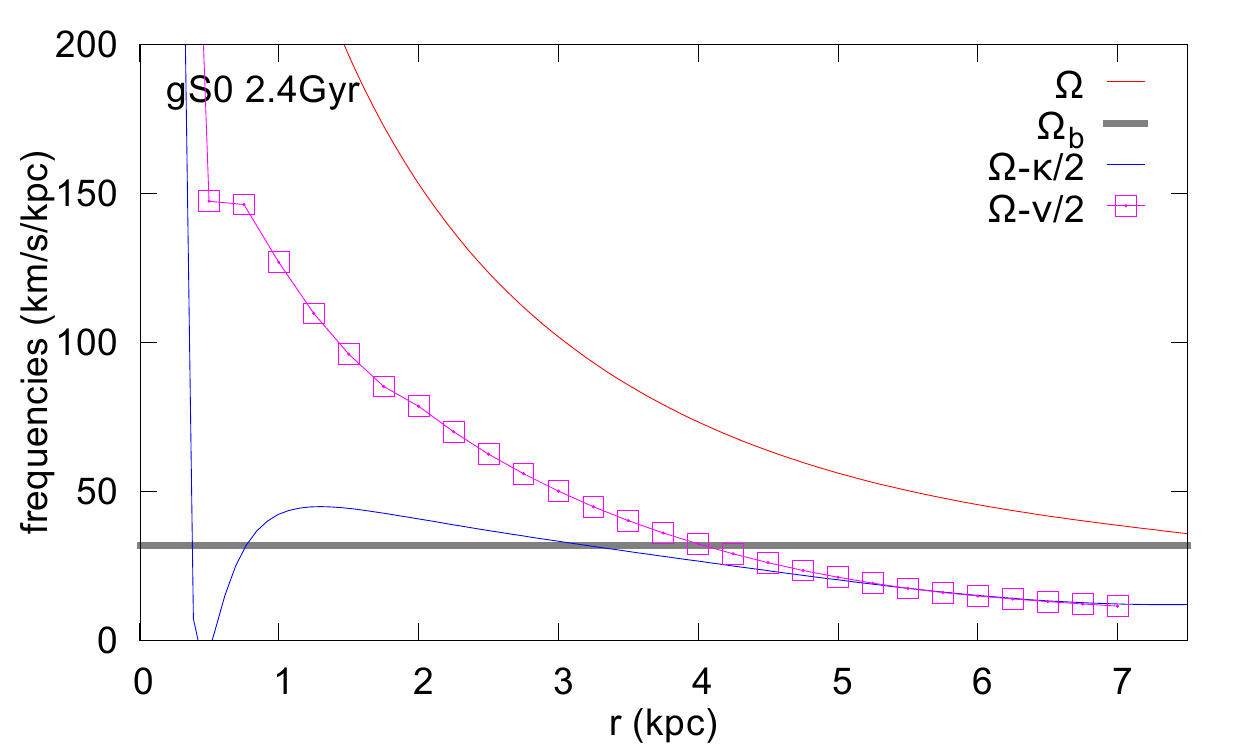}
\end{tabular}
 \caption{Location of vertical and Lindblad resonances
 as estimated from the potential fitting. 
 A polynomial function that included inverse and logarithmic terms was fit to the mid-plane potential.
This function was then differentiated to estimate the angular rotation rate 
and epicylic frequency, $\Omega$ and $\kappa$.
Derivatives of the functions fit to the vertical shape of the axisymmetric components
(with an example shown in Figure \ref{fig:vp}) are used to estimate the vertical oscillation frequency, $\nu$.
The approximate location of the vertical resonance can be estimated from where 
the bar pattern speed line, $\Omega_b$ (shown as a horizontal line),  crosses $ \Omega - \nu/2$ (plotted as points).
If the bar slows down, or the disk thickens, the vertical resonance can be at a larger radius
at later times.
The vertical resonance (associated with banana orbits) 
lies outside the Lindblad resonance (where $\Omega_b$
intersects $\Omega - \kappa/2$; plotted with a blue line). Consequently,  
orbits in the vicinity of the vertical resonance should be aligned with the bar.
a) For the buckling-lacking gSa simulation at $t=2.7$ Gyr.
b) For the bar-buckling  gS0 simulation at $t=2.4$ Gyr.
For both simulations the length of the peanut-shape is approximately consistent with  the radius
of the vertical resonance.
   \label{fig:res} 
 }
\end{center}
\end{figure}

 A polynomial function, $ f(r) = a_0 + a_1 r + a_2r^2 + a_3 r^3+ b_0 \log(r) + b_1/r + b_2/r^2$,
was fit to the azimuthally averaged potential, $V_0(r)$ in the mid-plane.  
Inverse and logarithm terms were used so as to allow the potential to diverge at the origin.
The coefficients $a_0, a_1, a_2, a_3$ have units of (km/s)$^2$, (km/s)$^2$ kpc$^{-1}$, (km/s)$^2$ kpc$^{-2}$
and (km/s)$^2$ kpc$^{-3}$ respectively.  
The coefficients $b_0, b_1, b_2$  have units (km/s)$^2$, (km/s)$^2$ kpc$^{1}$, and (km/s)$^2$ kpc$^{2}$.
This function
was then differentiated to estimate the angular rotation rate and epicyclic frequency
\begin{equation}
\Omega(r) \equiv \sqrt{{1 \over r} {\partial V_0 \over \partial r}} \qquad \qquad
\kappa(r) = \sqrt{3 \Omega^2 + {\partial^2 V_0 \over \partial r^2}},
\end{equation}
both evaluated at $z=0$.
These frequencies are shown along with the bar pattern speed at
early and late times in the simulations in Figure \ref{fig:res}.
The bar pattern speed is shown as a horizontal line in this figure.
The location where the bar pattern speed crosses $\Omega$ is approximately consistent
with a corotation radius estimated from 1.2 times the bar length.
The approximate location of the vertical resonance can be estimated from where 
the bar pattern speed line, $\Omega_b$,  crosses $ \Omega - \nu/2$.
Because the bars slow down and the disk thickens, the vertical resonance is at a larger radius
at later times in both simulations.

\citet{combes90} found that $\nu \sim \kappa$
near the 2:1 vertical resonance and so the 2:1 vertical and Lindblad resonances were essentially on top of
one another.
However, Figure \ref{fig:res} shows that the vertical resonance lies outside the Lindblad resonance 
(where $\Omega_b$ intersects $\Omega - \kappa/2$).  
 Orbits are primarily aligned with the bar outside the Lindblad resonance and
perpendicular to the bar inside the Lindblad resonance.  
Consequently,  orbits in the vicinity of the vertical resonance should be aligned with the bar.
Periodic orbits (in three dimensions) would be banana shaped.  They would be elongated along the bar
and reach high inclinations at the ends of the bar.  If the vertical resonance were
inside the Lindblad resonance then the periodic orbits could be at high inclinations at the ends of the bar
but the orbits would be aligned perpendicular to the bar.    

It is convenient that the Lindblad resonance (where $\kappa \sim 2(\Omega -\Omega_b)$) is
not in the same place as the vertical 2:1 Lindblad resonance (where $\nu \sim 2(\Omega -\Omega_b)$.
Near the vertical resonance, the frequency $\Delta = \kappa - 2(\Omega -\Omega_b)$, setting distance
to the 2:1 Lindbald resonance,
is not slow and so bar potential perturbations with form $\cos (\theta_r - 2 (\theta-\Omega_b t))$
can be taken into account with an approximation to first order in epicyclic amplitude 
(see appendix \ref{ap:radial} where the
eccentricity of a periodic orbit is estimated).
BAN+ and BAN- periodic orbits are those that have both fixed $J_r$ and $J_z$. 
The angular momentum sets the distance $\delta$ to the vertical resonance.
For different values of $L$, we expect that the Jacobi integral of BAN+ and BAN- orbits is more strongly dependent 
on the $J_z$ value of the periodic orbit 
(see equations \ref{eqn:jfixed}, \ref{eqn:Hfixed} ) in the vicinity of the vertical resonance than 
on the $J_r$ of the periodic orbit.

\citet{pfenniger91} discussed orbits that have a single vertical oscillation per rotation period in 
the bar frame.  Periodic orbits in this family were called `anomalous' \citep{heisler82}.
This resonance is associated with the resonant angle $\theta_z - (\theta - \Omega_b t)$
and so occurs where $\nu \sim \Omega - \Omega_b$.
We find that  $\nu > \Omega$ even at small radius in these simulations and so 
we did not find a region where $\nu \sim \Omega-\Omega_b$ corresponding 
to a 1:1 vertical resonance associated with `anomalous orbits'.   These orbits could exist in 
simulations of other galaxies.

Because the vertical resonance is not on top of the Lindblad resonances, orbital eccentricity
within the bar
can be estimated to first order in the epicyclic amplitude (see appendix \ref{ap:radial}).
We have estimated  the resonance location using a mean radius, but the orbits 
should have a distribution with eccentricity approximately equal to that of the oval 
periodic orbits  in the mid plane.   
 If the orbital eccentricity is $e$ then the distance along the bar axis
 is approximately $(1+e)r_c$ where $r_c$ is the average radius. 
As the bars in both are simulations are relatively weak, we ignore the eccentricity of
the orbit when comparing the location of the peanut (as seen along the bar major axis)
to the location of the resonance (estimated from the mean radius). 

For both simulations the location of the vertical resonance (where $\delta \sim 0$) is approximately
consistent with the location of the peanut shape.  
 In the gS0 simulation during bar buckling  the disk thickness
increased at a radius of about 2 kpc.   
However the peanut shape at later times grew to larger radii of 3-4 kpc.
The location of the peanut-shape seen in the mass distribution is consistent
with the location of the vertical resonance at both early and later times, confirming the results of the
periodic orbit study by \citet{combes90,martinez06}.
This  suggests that the location of the  vertical resonance
is more important than a previous history of bar buckling in determining the final location and height
of the peanut-shape.   The gSa simulation supports this interpretation as the peanut length is also consistent
with the location of the resonance, even though the bar did not buckle in this simulation.

\subsection{Bar perturbations}

\begin{figure*}
\begin{center}
$\begin{array}{cc}
\includegraphics[height=3.0in]{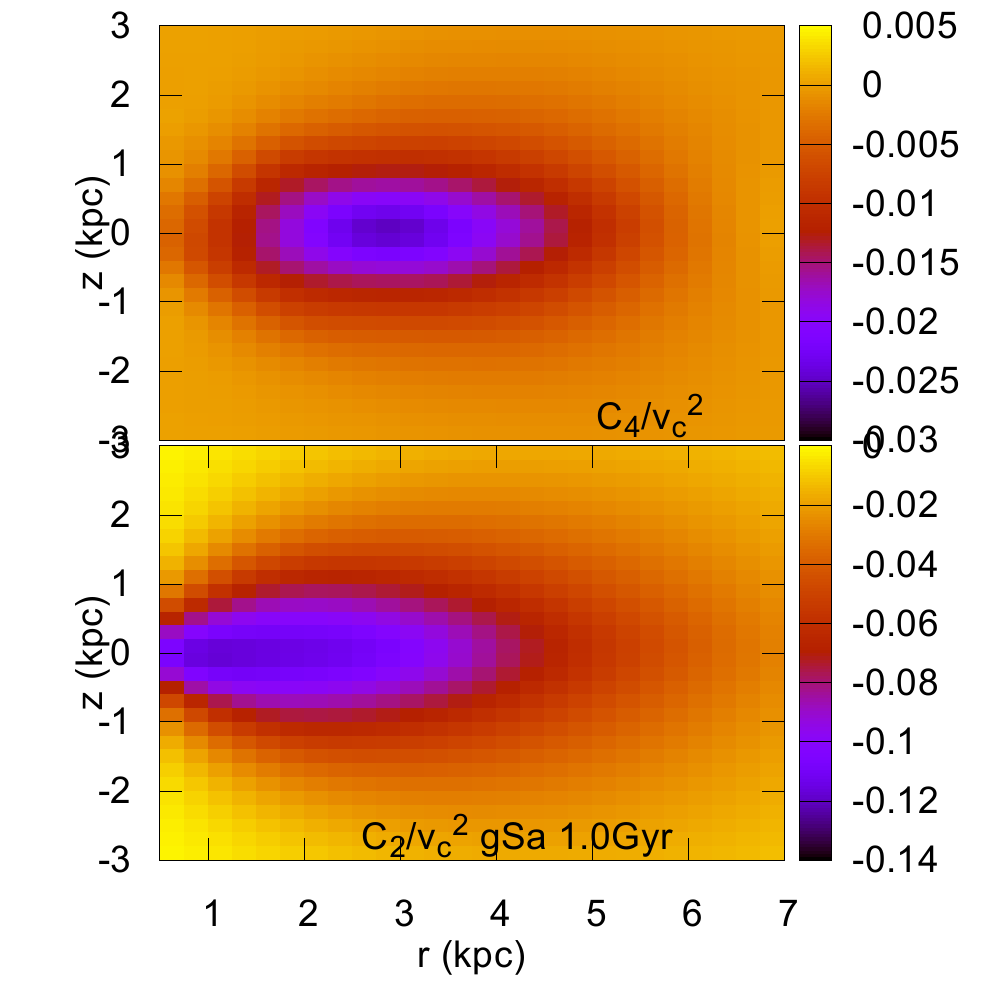} &
\includegraphics[height=3.0in]{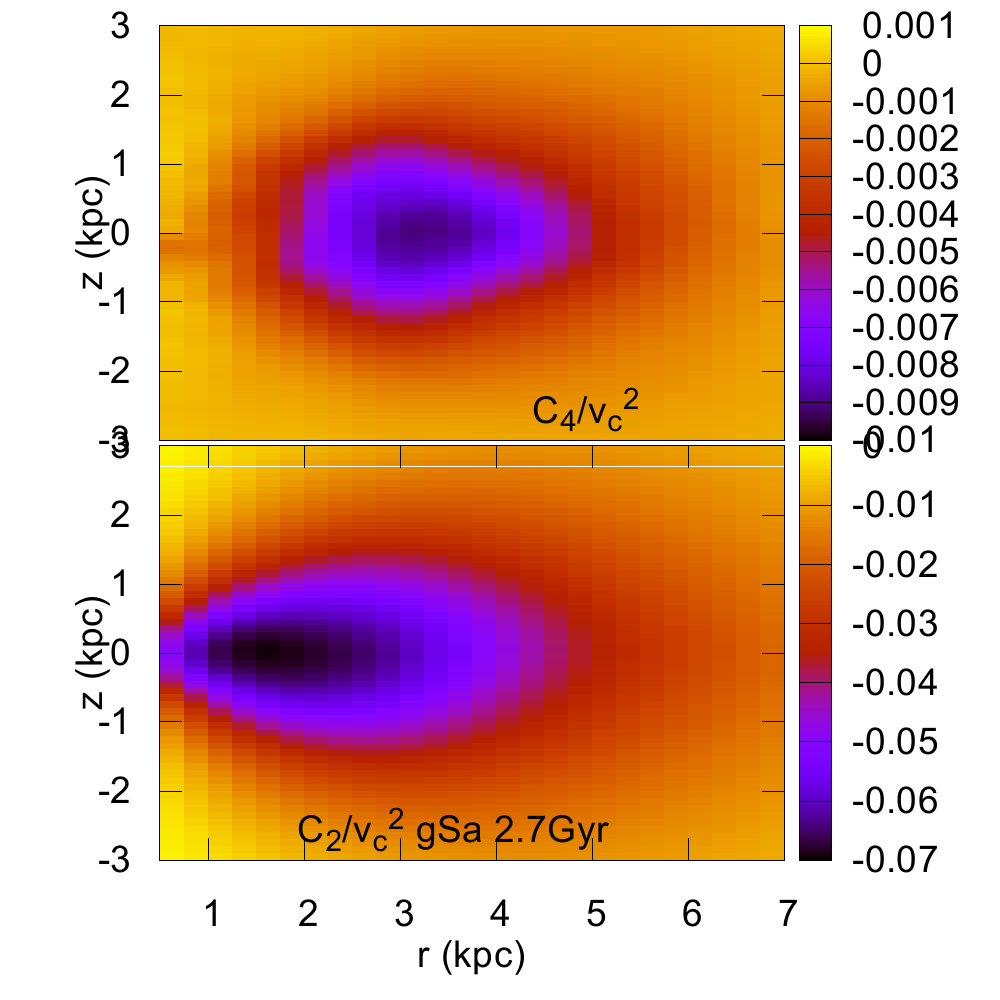} 
\end{array}$
\caption{
The bar perturbation $m=2$ and $m=4$ Fourier coefficients  of the gravitational potential normalized by the square of the circular velocity.
The coefficients are shown
 as a function of radius and $z$ and
computed from the gSa simulation at time 1.0 and 2.7 Gyr, left and right, respectively.  
The top of each panel shows the $m=2$ Fourier coefficient and the bottom of each panel
the $m=4$ coefficient.    The $x$ axes are radii and the y axes are $z$, both in kpc.
The potential Fourier components are shown in units of (100 km/s)$^2$. 
\label{fig:gSa_vertshape}
}
\end{center}
\end{figure*}

\begin{figure*}
\begin{center}
$\begin{array}{ccc}
\includegraphics[trim={0.4cm 0 0.4cm 0},clip,height=2.8in]{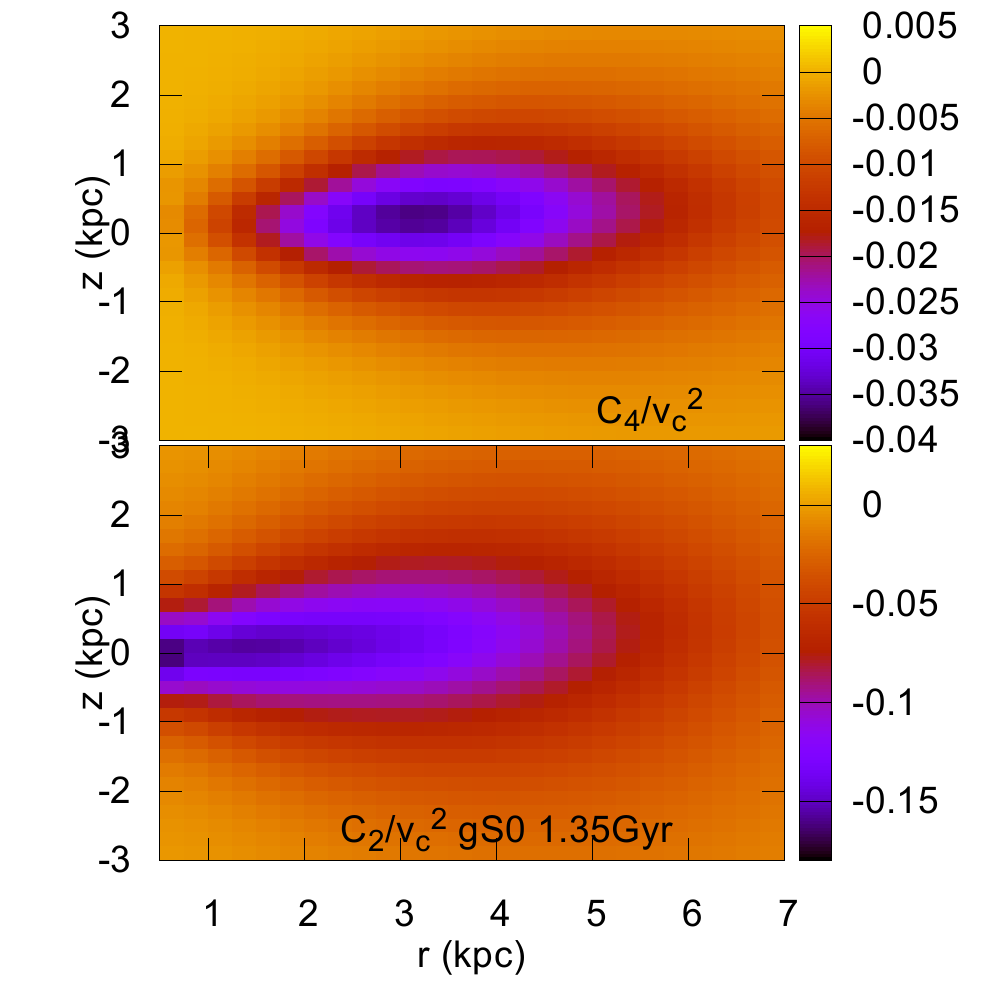} &
\includegraphics[trim={1.6cm 0 0.4cm 0},clip,height=2.8in]{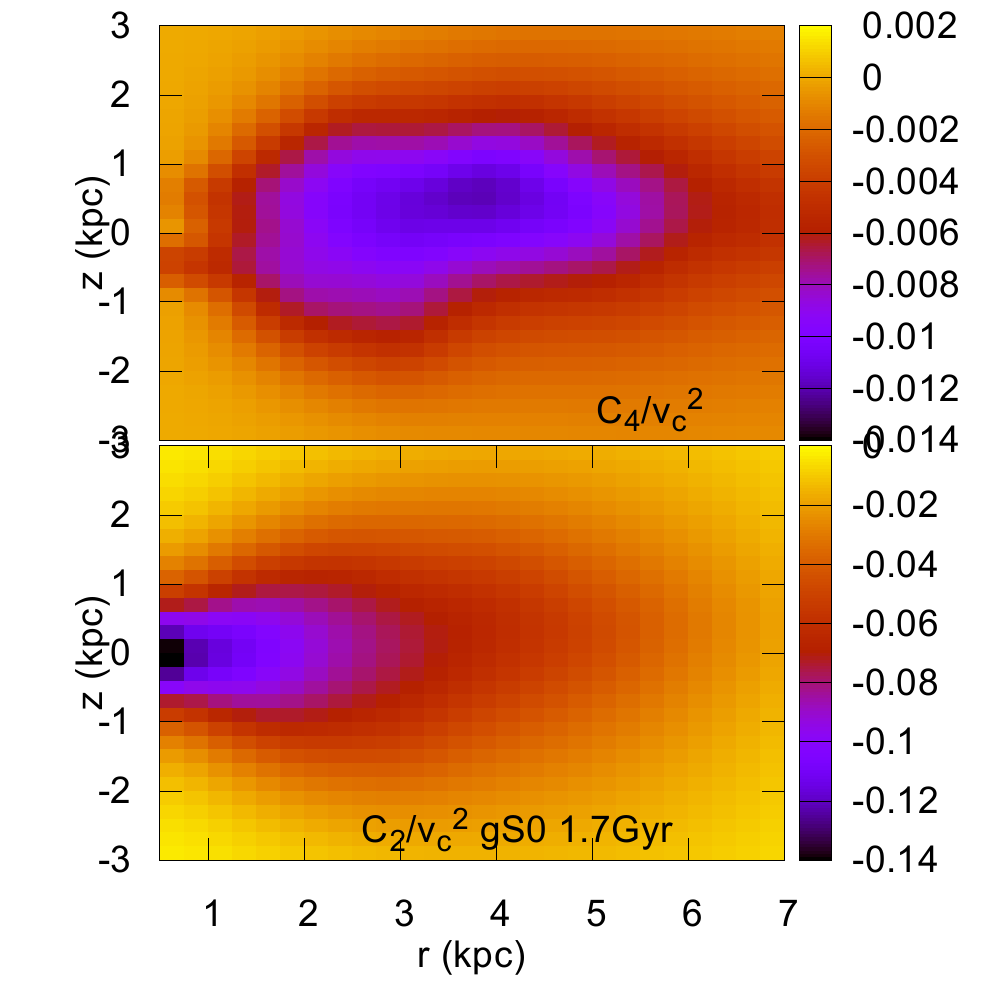}  &
\includegraphics[trim={1.6cm 0 0         0},clip,height=2.8in]{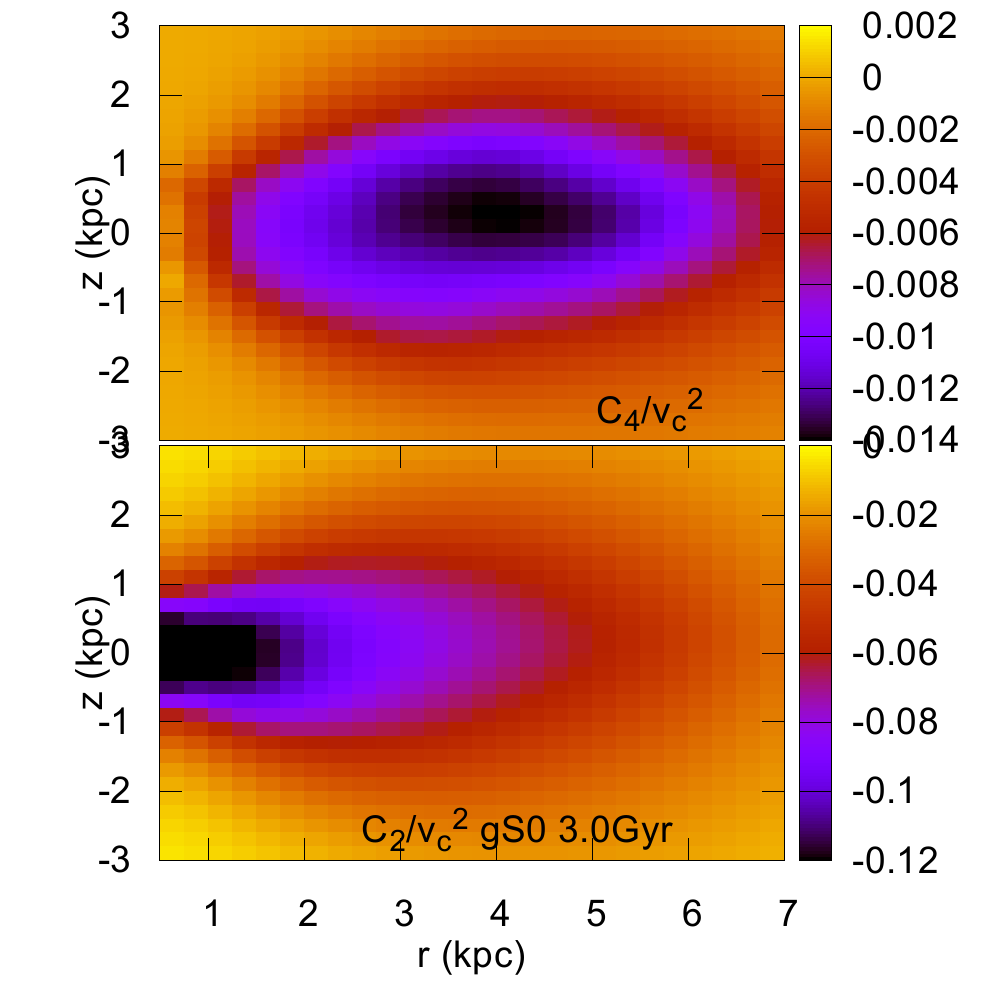} 
\end{array}$
\caption{
The bar perturbation $m=2$ and $m=4$ Fourier coefficients  of the gravitational potential normalized by the square of the circular velocity.
Similar to Figure \ref{fig:gSa_vertshape} except for the gS0 simulation and at 1.35, 1.7 and 3.0 Gyr. 
\label{fig:gS0_vertshape}
}
\end{center}
\end{figure*}

\begin{figure}
\begin{center}
\begin{tabular}{c}
\includegraphics[width=3.4in]{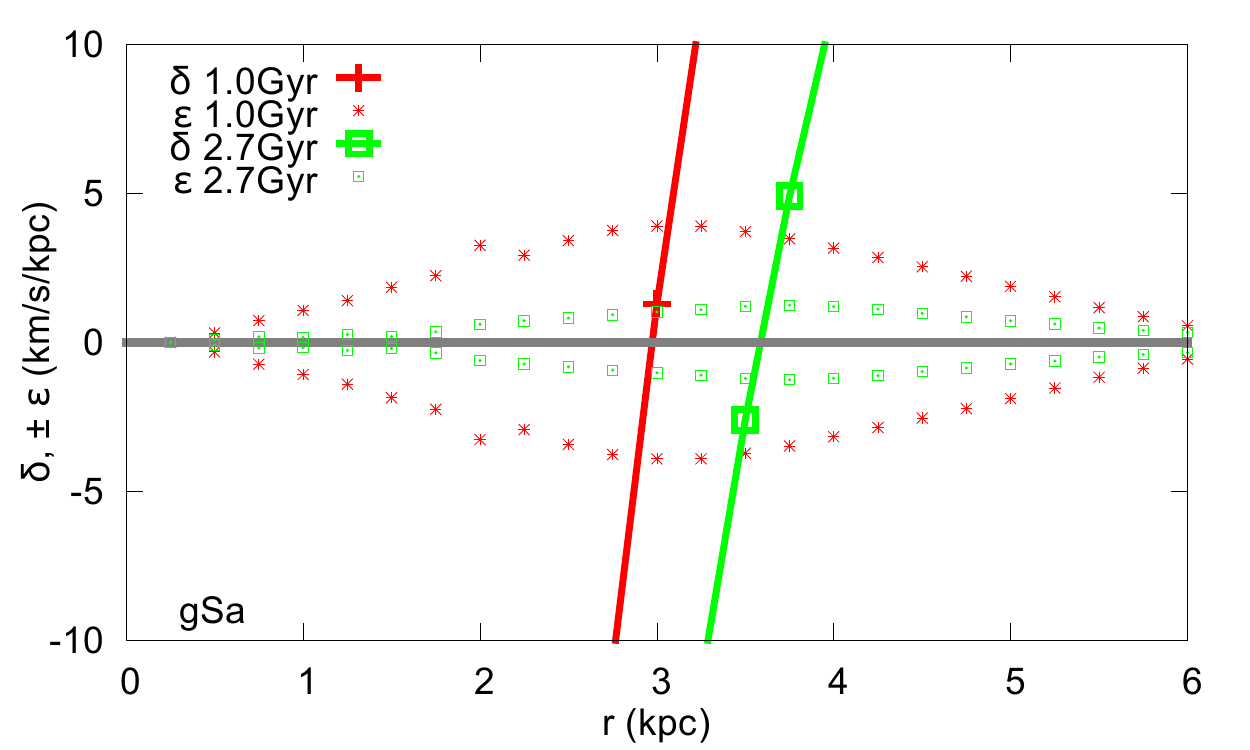} \\
\includegraphics[width=3.4in]{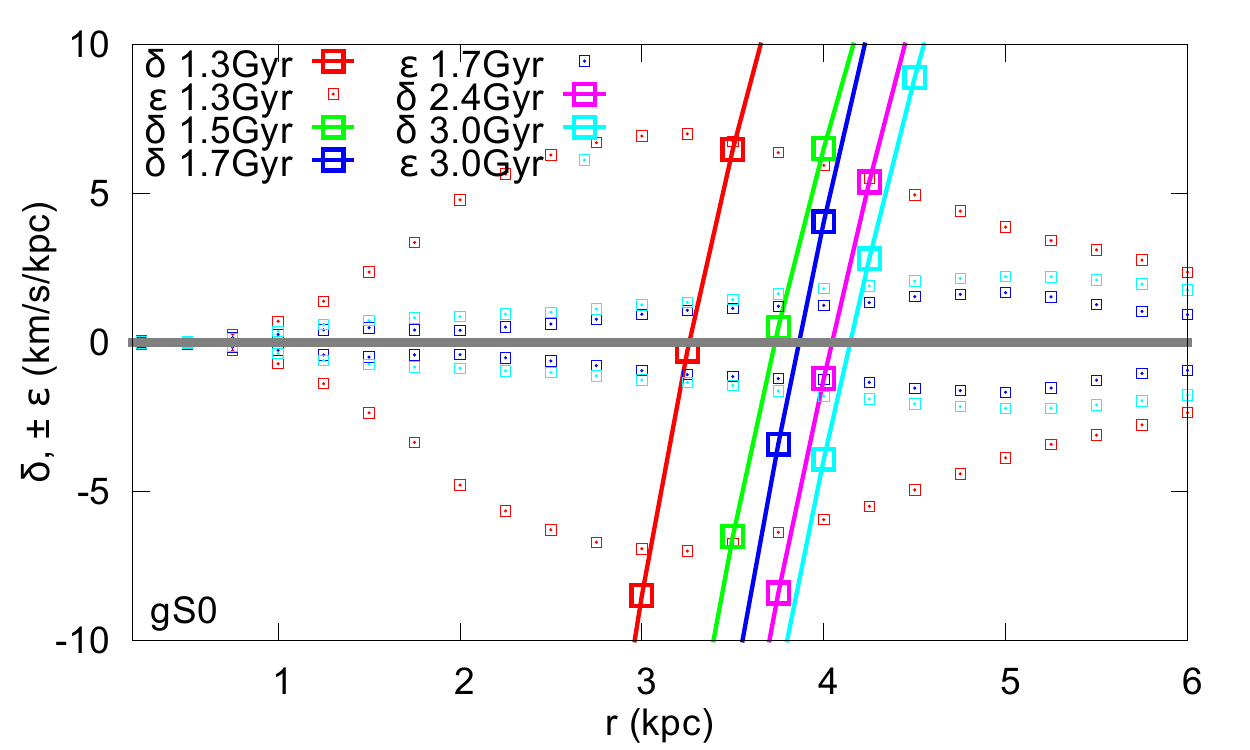}
\end{tabular}
\caption{
How the vertical resonance location varies at different times.
Here distance to resonance $\delta = \nu - 2( \Omega - \Omega_b )$ is plotted as a function of radius
but at different times in the simulation.
Also plotted are perturbation strength $\epsilon_s$ values (approximately the resonant libration frequency) 
estimated from the $m=4$ Fourier components of the potential.
Values for $\delta$ and $\pm \epsilon_s$ for each snapshot are given the same color.
The width of the resonance and extent of the peanut-shape or bow-tie shape can be estimated by taking
the radius where $\delta$ for each color crosses the lower set of similar color points (where $\delta = -\epsilon_s$)
to where $\delta$ crosses the higher set of similar color points (where $\delta = \epsilon_s$). 
a) In the buckling-lacking gSa simulation.  
b) In the bar-bucking gS0 simulation.  At times $t = 1.35$ -- 1.7 the bar buckles and the disk thickens.
This deceases the vertical oscillation frequency, $\nu$, moving the resonance outward.  
In both simulations the bar slows down at later times, moving the resonance outward.
  \label{fig:res_all} 
 }
\end{center}
\end{figure}

The $m=2$ and $m=4$ 
Fourier coefficients as a function of $z$ and $r$ of the gravitational potential, 
divided by the mid plane circular velocity $v_c^2(r)$, 
are shown in Figure \ref{fig:gSa_vertshape} and \ref{fig:gS0_vertshape}.
Because we work in a frame aligned with the bar, only cosine components are nonzero.
The bar also gives strong vertical perturbations in the $m=4$ Fourier component as seen from the
top panels in these figures.  This is relevant because when the galaxy is symmetrical about the midplane,
the vertical resonance strength is dependent upon the vertical structure of the $m=4$ coefficient.

\begin{figure}
\begin{center}
\begin{tabular}{c}
\includegraphics[width=3.4in]{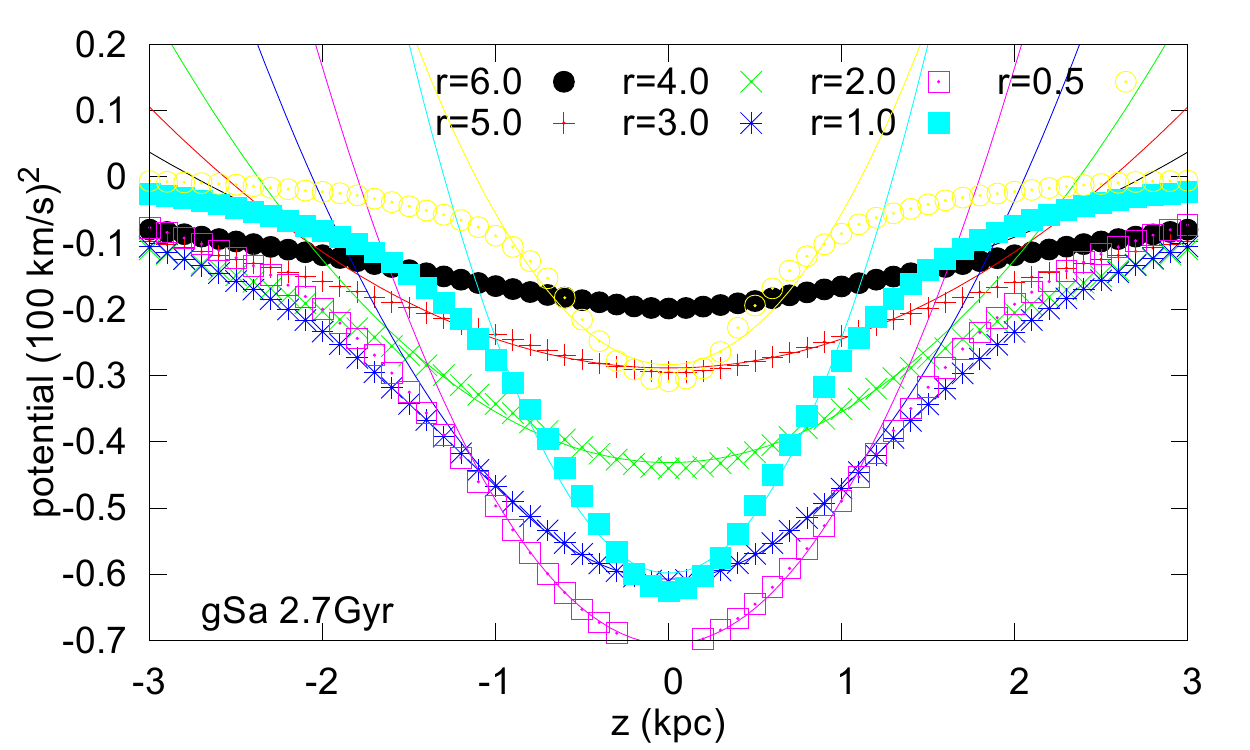} \\
\includegraphics[width=3.4in]{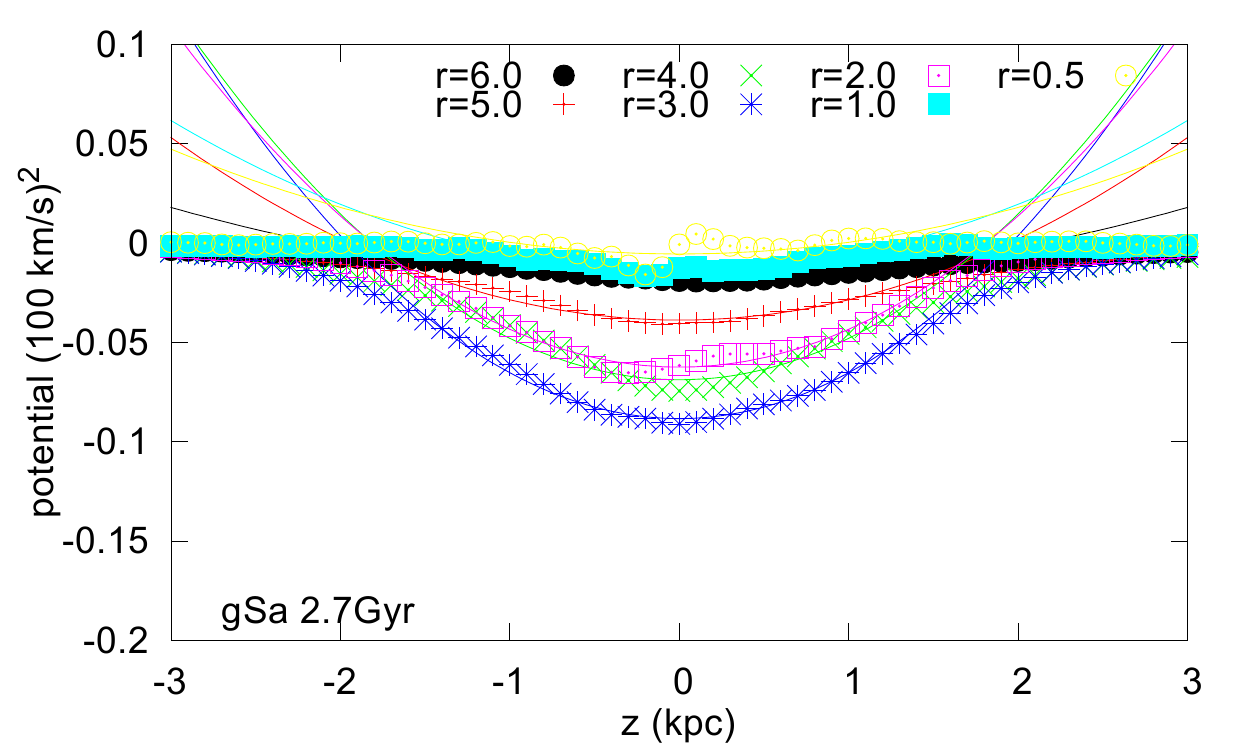} \\
\end{tabular}
 \caption{Fits to the vertical structure of  Fourier components of the bar potential.
 A quadratic function was fit to the potential profile extracted at different radii.
 The potential profile at each radius is shown
 with a different color point and the accompanying fit with the same colored line.
 Shown here are the points and fits to the gSa simulation at $t=2.7$ Gyr. 
 a) For $m=2$ giving  parameters $C_2$ and $C_{2z}(r)$.
  b) For $m=4$ giving  parameters $C_4$ and $C_{4z}(r)$.
 \label{fig:vp4} 
 }
\end{center}
\end{figure}

\begin{figure}
\begin{center}
\includegraphics[width=3.4in]{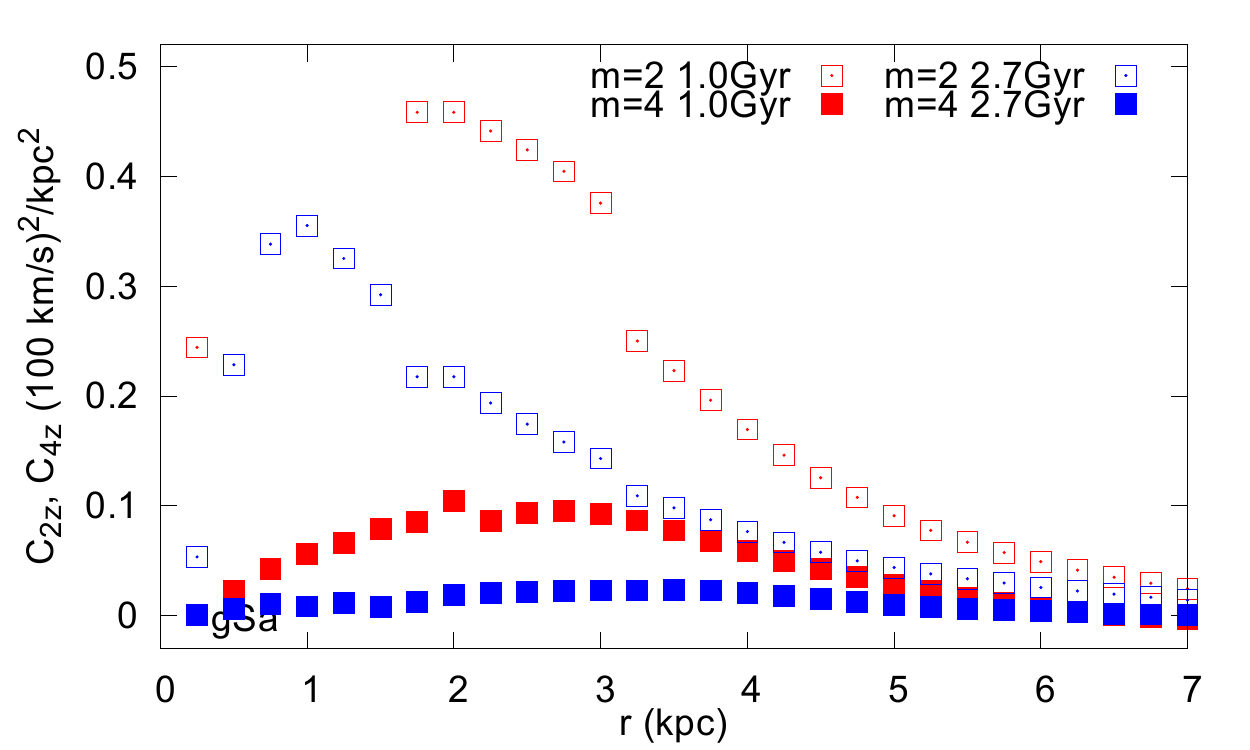} 
\includegraphics[width=3.4in]{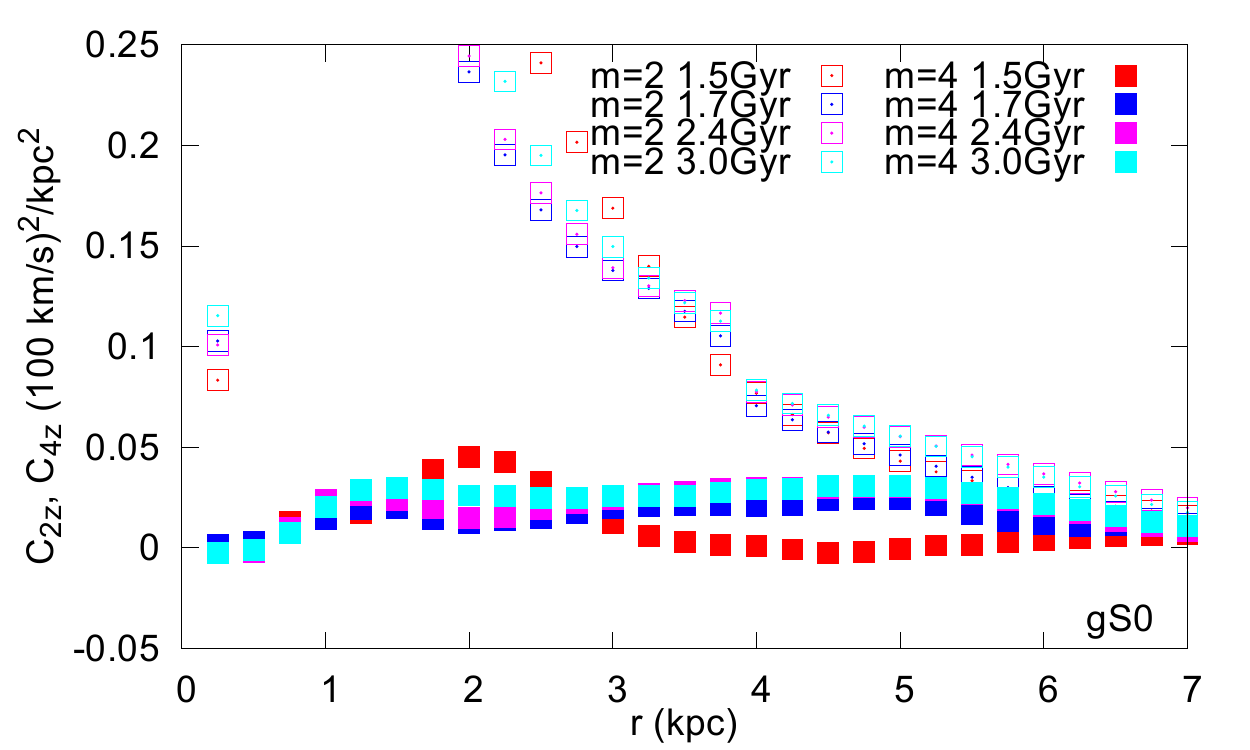} 
 \caption{Quadratic coefficients $C_{2z}, C_{4z}$ for the vertical
 bar perturbations at different radii and at different times. 
 a) For the gSa simulation.
 b) For the gS0 simulation.
 These coefficients arise from the fits to the Fourier components of the gravitational potential
 as a function of $z$.  
 The vertical perturbations from the bar are stronger  at earlier times in the gSa simulation.
  \label{fig:mom} 
 }
\end{center}
\end{figure}

Figure \ref{fig:vp4} shows examples of the Fourier components as measured from
a single snapshot at different radii and as a function of $z$.
We fit a quadratic function to each of these sets of points,  approximating the bar perturbation as
\begin{eqnarray}
V_b(\theta,r,t) &=& \left[C_2(r) + B_{2z}(r)z + C_{2z}(r) z^2\right] \cos(2(\theta-\Omega_b t)) +  \nonumber \\
			    &&  \left[C_4(r) + B_{4z}(r)z + C_{4z}(r) z^2\right] \cos(4(\theta-\Omega_b t))  \nonumber \\
			    &&  \label{eqn:Vb_all}
\end{eqnarray}
with the result of our fitting giving us values for $C_2,  B_{2z}, C_{2z}, C_4, B_{4z}, C_{4z}$ as a function
of radius.  
%
The measurements for the coefficients $C_{2z}, C_{4z}$  as a function of radius
for different snapshots in the two simulations is shown in Figure \ref{fig:mom} and illustrate
that the quadratic function is an adequate fit.

The coefficient $C_{4z}$ determines the $\epsilon_s$ coefficient for our second order Hamiltonian model.
In the vicinity of the vertical resonance with resonant angle $\theta_z - (\theta - \Omega_b t)$
(where there is a single vertical oscillation per orbit in the bar's frame)
the coefficient $\epsilon_s$ would depend on $C_{2z}$.
During bar buckling $B_{2z}$ contributes to $\epsilon_b$ for the first order Hamiltonian model.
The $C_{2z}, C_{4z}$ coefficients are positive as expected.  As long as the coefficient $a$ remains
negative, periodic orbits would be banana shaped and so support a peanut shape.

Previous work has used as a measure of bar strength the parameter
$Q_T$ \citep{combes81}. At a given radius
this is the ratio of the maximum tangential force to the azimuthally averaged
radial force in the mid-plane.
Using the $m=2$ component and equation \ref{eqn:Vb_all},
we find
$Q_T \approx {2C_2(r) \over v_c^2}$.   
We use
figures \ref{fig:gSa_vertshape} and \ref{fig:gS0_vertshape} (giving $C_2$ at $z=0$) to
 estimate $Q_T \sim 0.2$ for both simulations.
These would be classified as weak bars in comparison to the sample studied by \citet{laurikainen04}. 

\subsection{The vertical resonance location and its width}

Using the coefficients for the potential perturbation strength we can estimate the libration frequency
(equivalent to $\epsilon_s$; see equation \ref{eqn:Jres}) in resonance.  
Using equation \ref{eqn:epss} we compute the resonance libration frequency 
from the $m=4$ bar perturbation strength $C_{4z}$ (shown in Figure \ref{fig:mom}) and 
the vertical oscillation frequency, $\nu$.
Libration frequencies are also plotted in Figure \ref{fig:res_all}.  
Recall from the discussion above (section \ref{sec:ressize}) that the resonance width can be estimated
from the condition $|\delta | \lesssim |\epsilon_s|$. 
In Figure \ref{fig:res_all} we have plotted $\delta$ with estimates for $\pm \epsilon_s$ so these two
quantities can be directly compared.
 We find that  the resonance libration frequency, estimated from the bar perturbation strength,
  is approximately 1-2 km/s/kpc.  The range of radius
  that satisfies $|\delta/\epsilon_s|<1$ is small, $\approx 0.1-0.3$ kpc.  This range
  in radius corresponds to $dr/r \sim 0.05$ about resonance commensurability 
  or in angular momentum 
  r $dL/L \sim 0.05$. 
 The resonance can only affect orbits in a small region of phase space. 
 
The resonant width is significantly narrower than  the extent
of the peanut seen in the Fourier components (for the density Figures \ref{fig:gSa_denshape} and \ref{fig:gS0_denshape}, for the potential Figures \ref{fig:gSa_vertshape} and \ref{fig:gS0_vertshape}).
\citet{quillen02} suggested that an X-shape or bow-tie shape was formed due to the linear dependence of
the fixed point height on $\delta$ between $-\epsilon_s < \delta < \epsilon_s$.
Here we find that this radial region is far too small to account for the extent of the observed peanuts
or X-shape in the Galactic bulge.  Hence the resonance capture model fails to predict the X-shape.

The association of the resonance with the feature is robust.   The Hamiltonian model implies that banana shaped
orbits are not found distant from resonance.   We see that the resonance moves 
outwards during the simulation, consequently we must consider the situation illustrated in Figure \ref{fig:ham_slow}.
In this setting stars originally in the mid plane outside of resonance are lifted to high inclination
as they encounter the resonance.   They must go into orbits just inside the separatrix until the separatrix
shrinks leaving them in high inclination orbits but with circulating $\phi$ and so no longer
supporting the peanut-shape.  While the stars are inside the resonance, their height would
be similar to the separatrix height  (equation \ref{eqn:zsep}).
In this heating model, the only stars that support the peanut shape are those that
have just entered the resonance or those that have just left and remain near the separatrix.  In this scenario,
the peanut shape is  due to a small number of stars in or near resonance.   
This implies that they must have very similar angular
momentum values.   This follows as orbits that support the peanut only exist
in the vicinity of resonance.
The X- or peanut shapes seen in galaxies must trace the narrow volume 
of the possible orbits that can support the peanut.

Because the $C_{2z}$ and $C_{4z}$ coefficients depend on the second derivative of the gravitational potential
with respect to $z$,
 $\partial^2 V_b/dz^2$,
we expect that
\begin{eqnarray}
 |C_{2z} | &\sim& \left|{C_2  \nu^2 \over v_c^2}\right| \nonumber \\
|C_{4z} |&\sim& \left|{C_4  \nu^2 \over v_c^2}\right|,  \label{eqn:C4z_approx}
\end{eqnarray}
where $C_2/ v_c^2$ and $C_4/ v_c^2$
are the amplitudes of the $m=2$ and $m=4$ Fourier
coefficients of the potential in the mid plane due to the bar perturbation. 
These can be estimated using the images shown in Figures \ref{fig:gSa_vertshape} and \ref{fig:gS0_vertshape}.
This implies that the resonant width or frequency $\epsilon_s \sim \nu {C_4 \over v_c^2}$
or ${\epsilon_s \over \nu }\sim {C_4 \over v_c^2}$.
The mid plane $m=4$ Fourier component of the potential are at most 0.01 (in units of $v_c^2$)
and so we estimate that ${\epsilon_s \over \nu} \sim 0.01$. This is approximately consistent
with the 1-2 km/s/kpc frequency we measured for $\epsilon_s$ directly.

For a fixed bar shape, setting $C_4$,   the quantity $C_4/v_c^2$ should not significantly vary.
However if the disk thickens then $\nu$ decreases.  This would have the effect of reducing $\epsilon_s$
and so the strength  of the resonance.    The frequency $\epsilon_s$ is weaker at later times than earlier ones
for the gSa simulation (as seen from the distance between red points and between green points showing
$\pm \epsilon_s$ in Figure \ref{fig:res_all}a.  The reduction may be in part due to disk thickening rather
than the bar becoming rounder.  

\subsection{Vertical excitation in resonance} 

We now discuss the vertical excitation caused by the resonance. 
In the resonant heating models $z_{sep}$ sets the height of the peanut.
Recall that $z_{max}$ is the height of the fixed points (or periodic orbits) where $\delta=\epsilon_s$,
at the outer edge of the resonance and $z_{sep}$ is the maximum height of an orbit
in the separatrix at the same $\delta$ (equation \ref{eqn:zsep}).
Equation \ref{eqn:zsep} depends on the $a$ coefficient that depends on fourth order derivatives
of the potential.
The coefficient $a$ is derived in appendix \ref{ap:can} (where $a=a_{cz}$; equation \ref{eqn:acz}) 
using a fourth order approximation in 
epicyclic and vertical oscillation amplitudes.  The coefficient given in 
equation \ref{eqn:acz} contains three terms.  The second two both have larger magnitude
than the first that we denote $a_z$ that only depends on $z$ derivatives of the
potential. 
When we compute them from our fits to the gravitational
potential,  we find that the two larger terms approximately cancel leaving a coefficient with 
$a_{cz} \sim a_z \sim - 0.05$~kpc$^{-2}$ in the vicinity of the peanut in all simulations. 
An example of the $a_z$ and $a_{cz}$ coefficients computed for
one of the simulation snapshots (computed using equations \ref{eqn:coeffs} and \ref{eqn:acz}) 
is shown in Figure \ref{fig:aacor}.  We took into account the corrections derived in appendices \ref{ap:action}
and \ref{ap:can} (see equation \ref{eqn:acz})
 and were surprised
to find that $a_z$ was a reasonable approximation to $a_{cz}$.

\begin{figure}
\begin{center}
\includegraphics[width=3.4in]{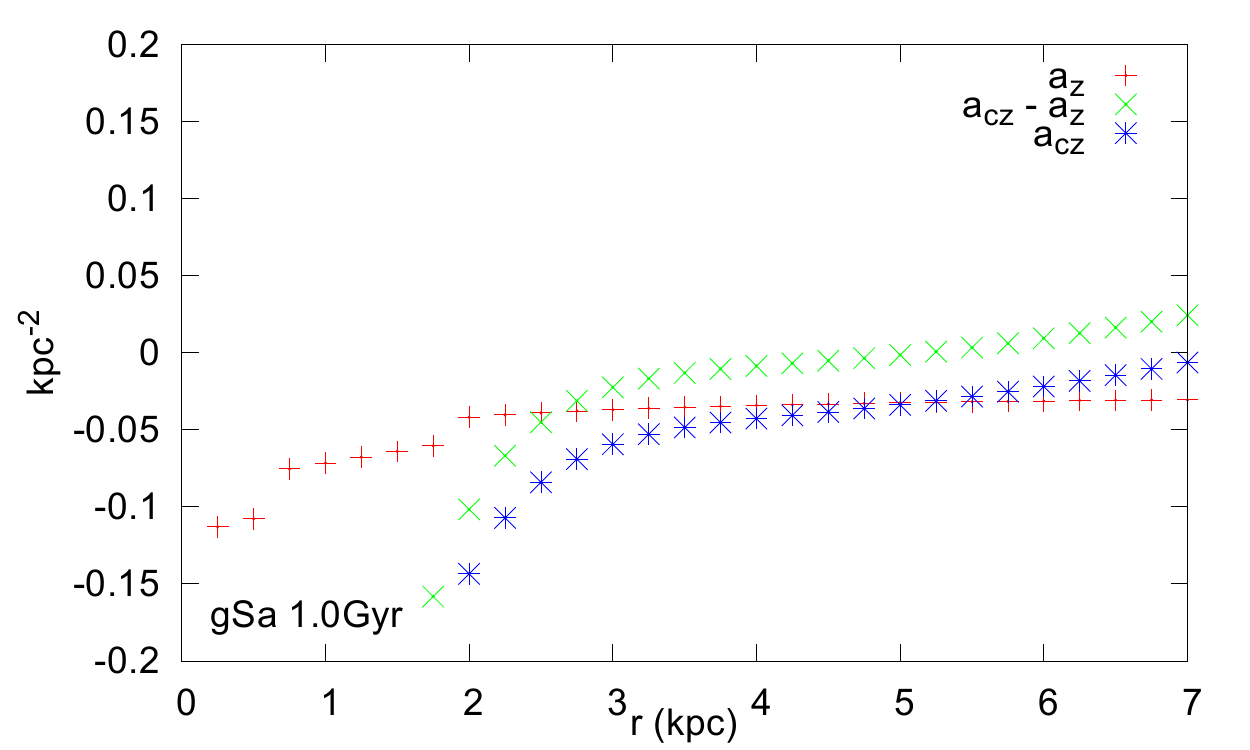} 
 \caption{Functions $a_z$ and $a_{cz}$, (computed using equations \ref{eqn:coeffs} and \ref{eqn:acz}) 
  and their difference are plotted
 here using fits to the potential of the gSa simulation at 1Gyr. 
 The coefficient $a_{cz}$ is needed to compute the heights of stars in resonance.
 \label{fig:aacor} 
 }
\end{center}
\end{figure}

\begin{figure}
\begin{center}
\includegraphics[width=3.5in]{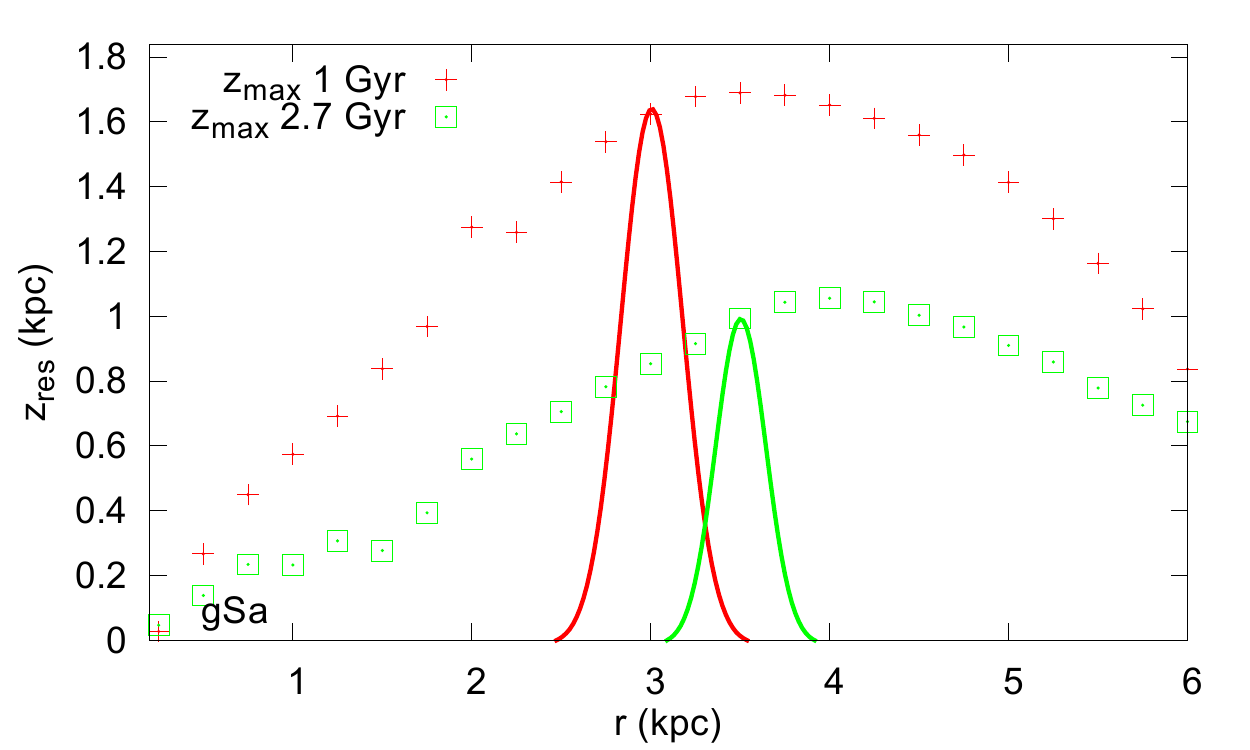} 
\includegraphics[width=3.5in]{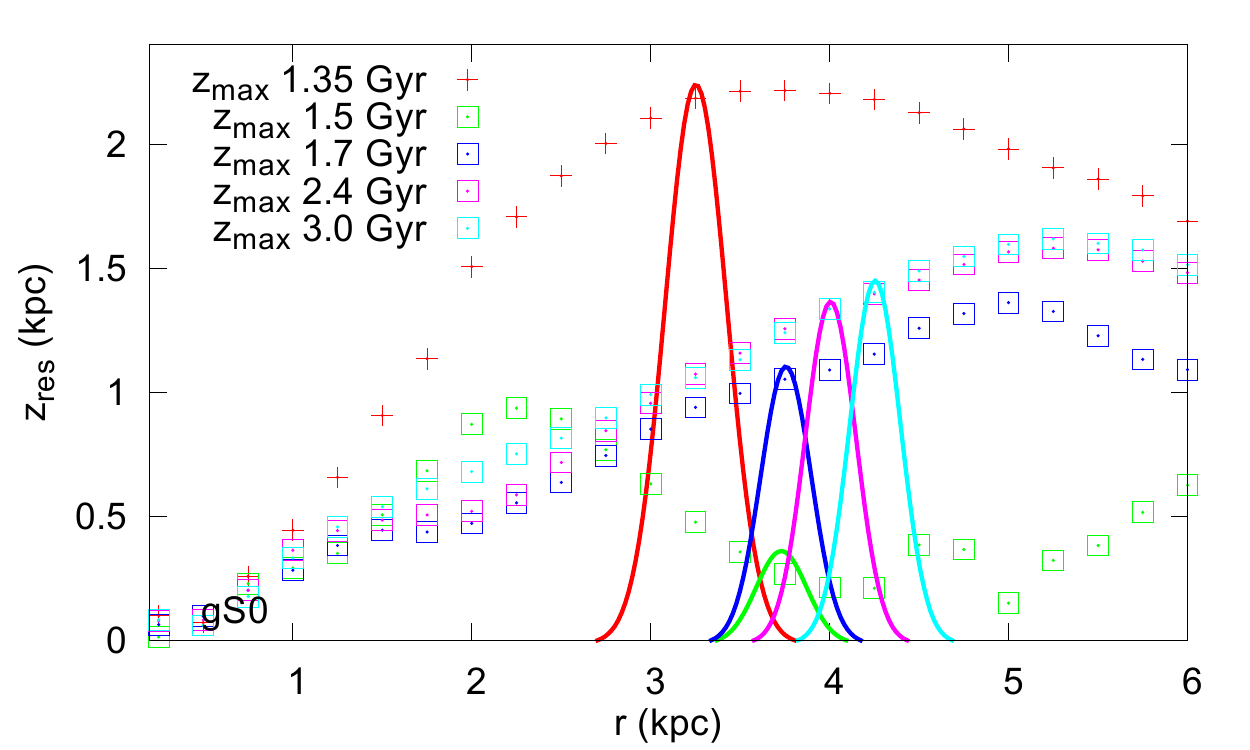}
 \caption{
 Vertical excitation height caused by the resonance.
 Small points show the
  maximum height of the separatrix, $z_{sep}$ (equation \ref{eqn:zsep}),  
  in kpc  as
  a function of radius.  The location of the resonance at different times is shown as
  solid Gaussians.
 a) For the gSa simulation and computed using potential measurements from different snapshots.
b) Similarly for the gS0 simulation.
In both simulations, the resonance  moves outward.  
As stars in the mid plane outside resonance enter the resonance, they are forced
out of the mid plane to a height approximately given by $z_{sep}$.
While in resonance, they lie just inside the separatrix.  As the separatrix shrinks they leave
resonance but remain at high inclination.
   \label{fig:zmax} 
 }
\end{center}
\end{figure}

Figure \ref{fig:zmax} shows as small points
the  separatrix height, $z_{sep}$, 
 as a function of radius calculated using equation  \ref{eqn:zsep} and $a = -0.05$~kpc$^{-2}$
and using our measurements for $C_{4z}$ and oscillation frequencies.
The height is computed for both simulations and for different snapshots.
As solid curves, the locations of the resonance are shown as narrow Gaussians.
In the resonance heating model stars are lifted to $z \sim z_{sep}$ as they enter the
resonance.  They then leave the resonance, remaining at high inclination but are no longer
oriented with and supporting the peanut shape.  As the resonance passes through
the disk, the disk is heated and should remain at approximately a height of $z_{sep}$.
Stars supporting the peanut shape and in resonance should have a height approximately
of $z_{sep}$.
From Figure \ref{fig:zmax} we see that the predicted peanut height is approximately 1~kpc,
is  consistent with the height of the observed peanut shapes.
Even though the resonance is thin and weak, because $a$ is small, the resonance
height is not insignificant.  The coefficient $a$ acts somewhat like a mass in a harmonic oscillator system. 
A weak spring
can cause a higher amplitude oscillation on a low mass than a high mass.

\subsection{Interpretation of the X-shape}

Because the resonance is thin and has moved outward,
the resonance capture model does not account for the X-shape
of peanut shaped bulges. 
Here we ask again, why is an X-shape observed?
As we have discussed above, stars only are aligned with
and support the peanut shape if they are in resonance (and have librating $\phi$ angle
near 0 or $\pi$) or are near resonance and are 
spending more time near $0, \pi$ than near $\pm \pi/2$ (see Figure \ref{fig:ham_red}).
The only population of stars likely to be in resonance (have librating $\phi$ angle)
are those that have been recently captured into resonance or those that are in the vicinity
of the resonance separatrix.  These are disk stars that 
were originally in the mid plane just exterior to resonance.  
The X-shape then must be due to the orbital distribution of these stars or the morphology of
their orbits.  Here we have associated them with the vicinity of the resonance separatrix 
at $\delta = \epsilon_s$ and have predicted their maximum height using the maximum height of
this orbit.  As can be seen from the Hamiltonian level curves (lower right panel of Figure \ref{fig:ham_grow}),
orbits within the separatrix are not periodic, instead they librate about a periodic orbit.
They would support the X-shape but would fill in the region between the top of the X and the mid plane
as $\phi$ varies between $\phi = \pm \pi/4$.  

Stars just exterior to the resonance separatrix (see Figure \ref{fig:ham_red}) spend more time
with $\phi$ near $0,\pi$ than near $\pm \pi/2$ and so would also supper the X- or peanut shape,
even though the resonant angle is circulating.
Stars that have left the resonance (after being lifted by it) and are distant from the resonance, 
are likely to be at high inclination but
no longer supporting the peanut shape.  Thus they would appear to be part of the bulge
even though they originated as disk stars.

Previous studies have found that banana shaped orbits exist over a range of Jacobi constant, $E_J$, 
\citep{pfenniger91,martinez06}.  
The Jacobi constant is the energy in the rotating frame; see the end of appendix \ref{ap:action}.
Because the resonance is narrow we find that it is only important in a very small range of angular
momentum.  However the Jacobi energy also depends on the epicyclic action 
variable $J_r$.      A periodic orbit in three dimensional space, such as in the BAN+ and BAN- families,
 is periodic in both radius and vertical height.   They only exist in the vicinity of the vertical
 resonance and this essentially sets the angular momentum value because the resonance
 is thin.  The requirement that the orbit is also periodic in radius sets the $J_r$ value and so
 the Jacobi energy.  We expect that the banana shaped periodic orbits only
 exist over a narrow range of energy.

However, within the context of the resonance heating model, stars are not captured into orbits
near the periodic orbit families but rather spend time near the vertical resonance  separatrix.
Furthermore, the vertical resonance need not affect the radial degree of freedom.
Stars of different values of eccentricity (or $J_r$) and so Jacobi energy
could be lifted in to orbits near the vertical resonance separatrix.
The resonance is most narrowly identified by its angular momentum rather than energy.
We expect that the stars supporting the X-shape would have a similar $J_r$ distribution
to the stars just exterior to resonance and so a similar distribution in Jacobi energy
as exterior to resonance.

\subsection{Velocity distributions measured along the bar}

The heating scenario discussed above makes some predictions for the velocity distributions.
Because the process is predominately one of heating, we do not expect to see a cold population
of stars near periodic orbits.  We expect that low inclination 
orbits should be depleted within resonance, due to the previous passage of the resonance.  
Outside of resonance, the disk is undisturbed and stars can be in low inclination orbits.

\begin{figure*}
\begin{center}
$\begin{array}{cccc}
\includegraphics[trim={0         0 0 0},clip,height=4in]{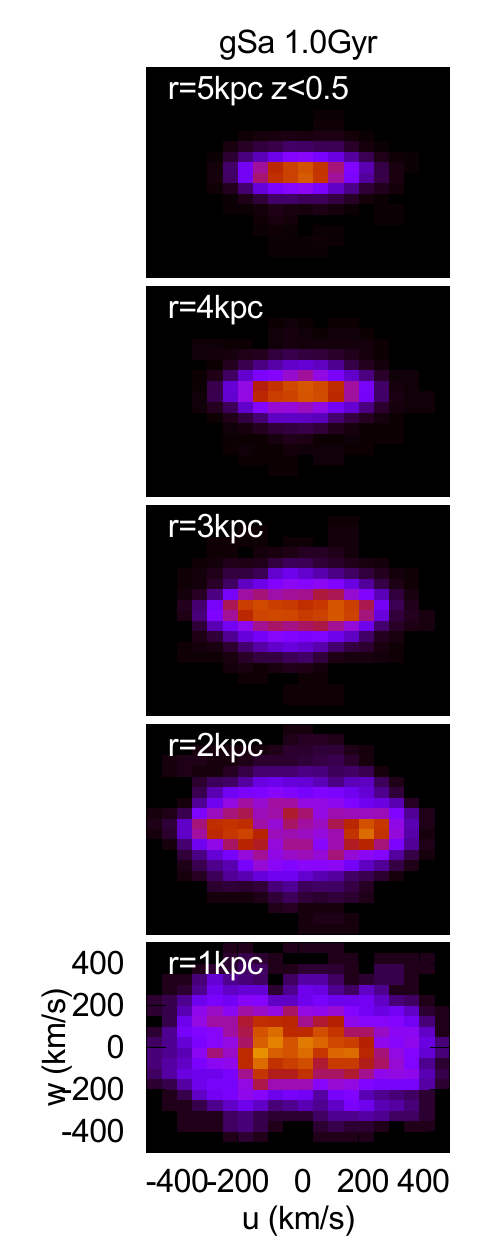} &
\includegraphics[trim={1.5cm 0 0 0},clip,height=4in]{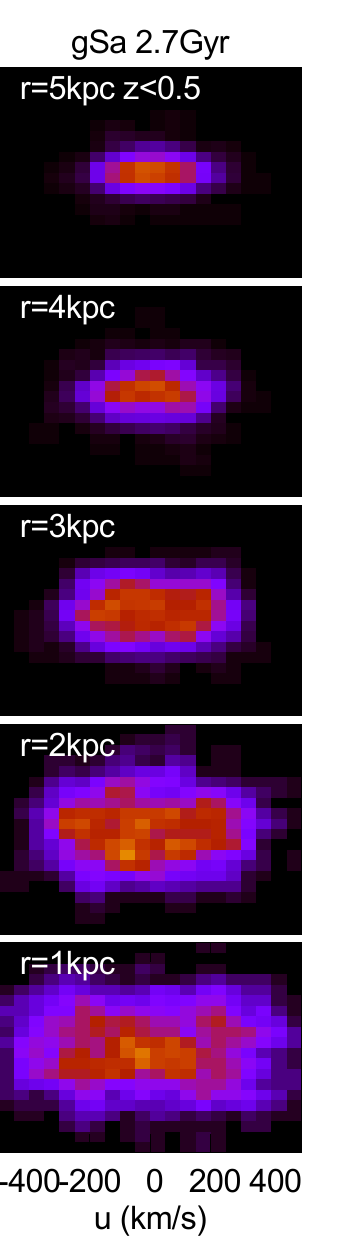} &
\includegraphics[trim={0         0 0 0},clip,height=4in]{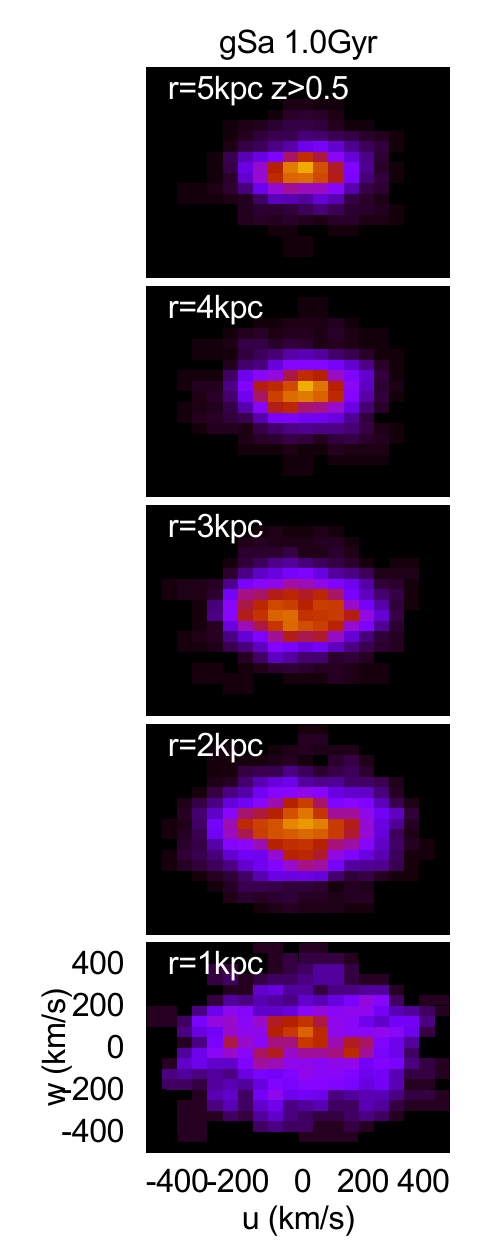} &
\includegraphics[trim={1.5cm 0 0 0},clip,height=4in]{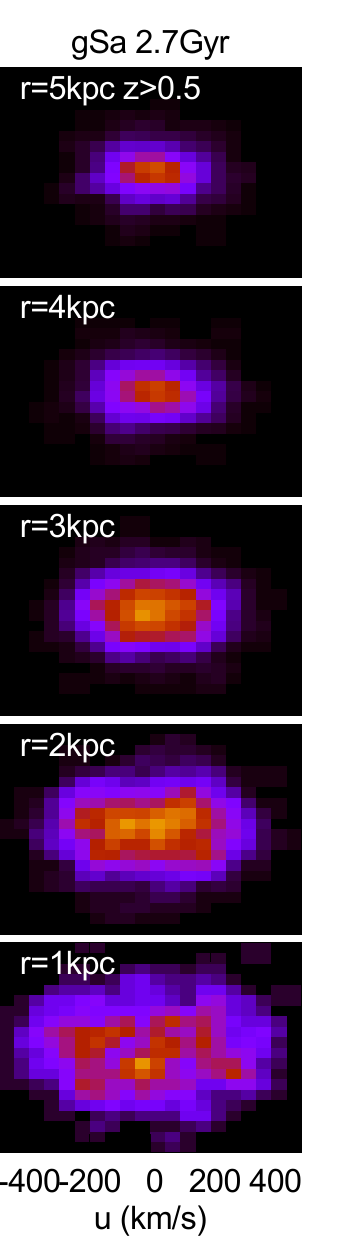}
\end{array}$
 \caption{Radial ($u$) vs vertical ($w$) velocity distributions of disk stars
 generated from snapshots of the gSa simulation.
 Each distribution is drawn from a neighborhood centered on the bar major axis at galactocentric radius
1,2,3,4,5 kpc respectively, from bottom to top.
 For the left two columns, each distribution is drawn from the mid-plane (or stars with
$|z|<0.5$ kpc) and the right two columns drawn from stars above and below the mid-plane, or stars
with $|z|>0.5$ kpc.   The first and third column (from the left) are from the snapshot at 1Gyr and the second and fourth column are from that at 2.7 Gyr.
  \label{fig:whist_gSa} 
 }
\end{center}
\end{figure*}

\begin{figure*}
\begin{center}
$\begin{array}{cccccc}
\includegraphics[trim={0.5cm 0 0.5cm 0},clip,height=4in]{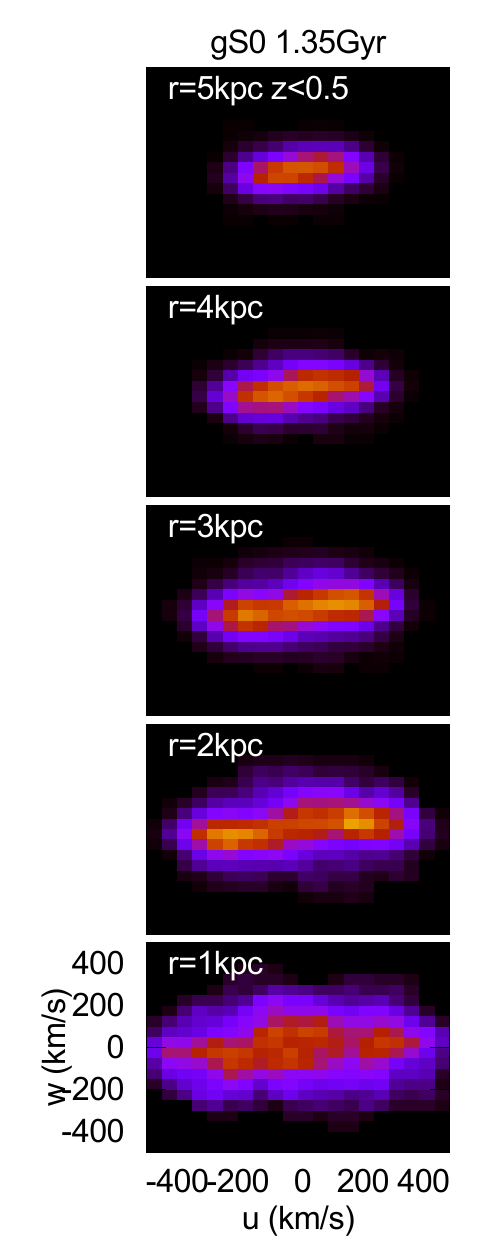} &
\includegraphics[trim={1.5cm 0 0.5cm 0},clip,height=4in]{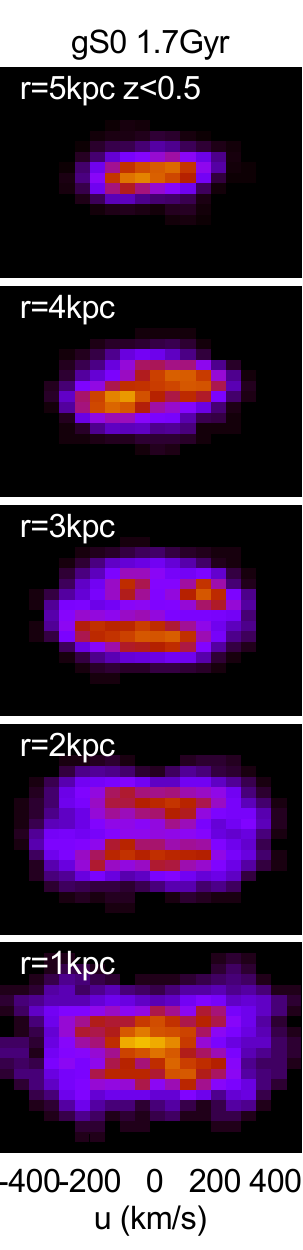}  &
\includegraphics[trim={1.5cm 0 0.5cm 0},clip,height=4in]{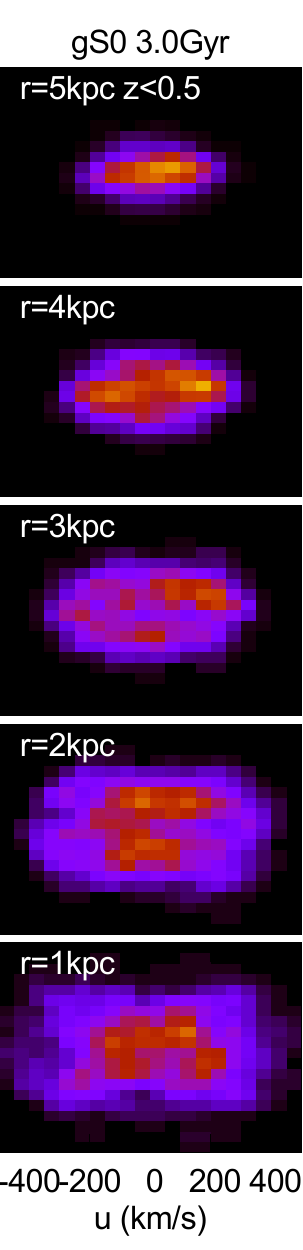}  &
\includegraphics[trim={0.5cm 0 0.5cm 0},clip,height=4in]{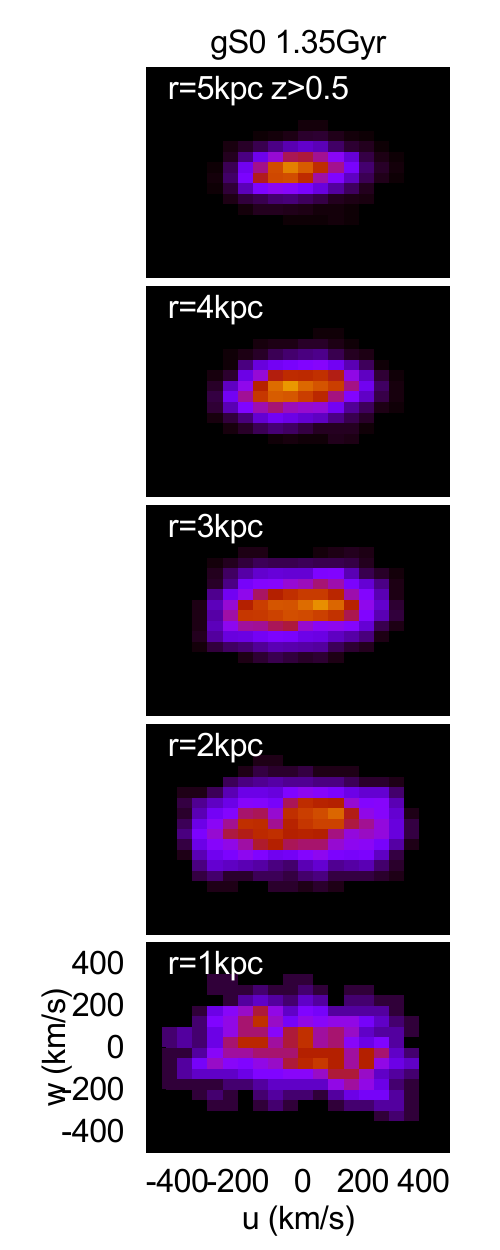} &
\includegraphics[trim={1.5cm 0 0.5cm 0},clip,height=4in]{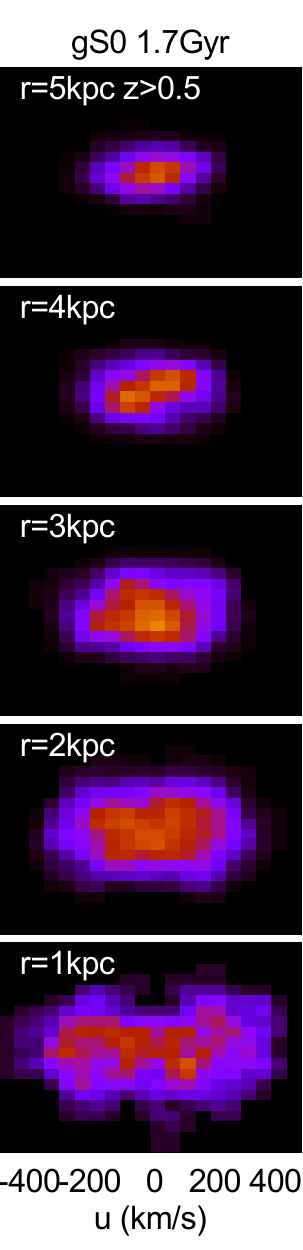}  &
\includegraphics[trim={1.5cm 0 0.5cm 0},clip,height=4in]{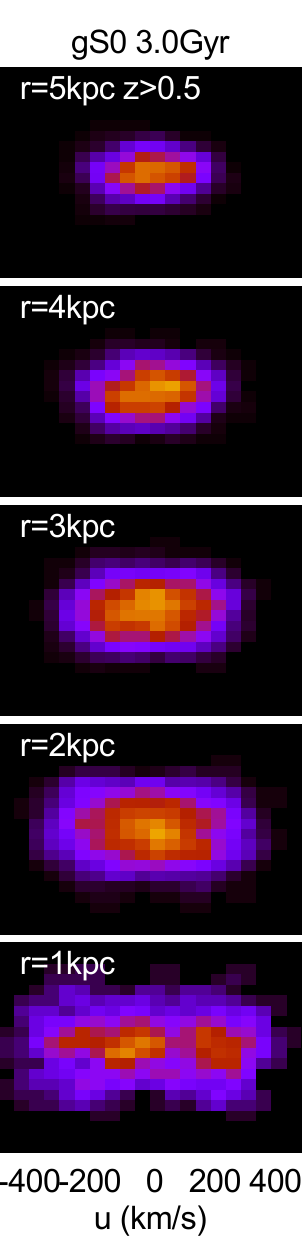}  
\end{array}$
 \caption{Radial ($u$) vs vertical ($w$) velocity distributions generated from snapshots of the gS0 simulation.
 Similar to Figure \ref{fig:whist_gSa} except for the gS0 simulation and at times 1.35, 1.7 and 3.0 Gyr.
 Within the vertical resonance, there are no orbits that remain in the mid-plane.
This gives a donut or double bar shape in the mid-plane velocity distributions for radii within the peanut-shape
at later times. 
  \label{fig:whist_gS0} 
 }
\end{center}
\end{figure*}

\begin{figure*}
\begin{center}
$\begin{array}{cccc}
\includegraphics[trim={0.5cm 0 0 0},clip,height=4in]{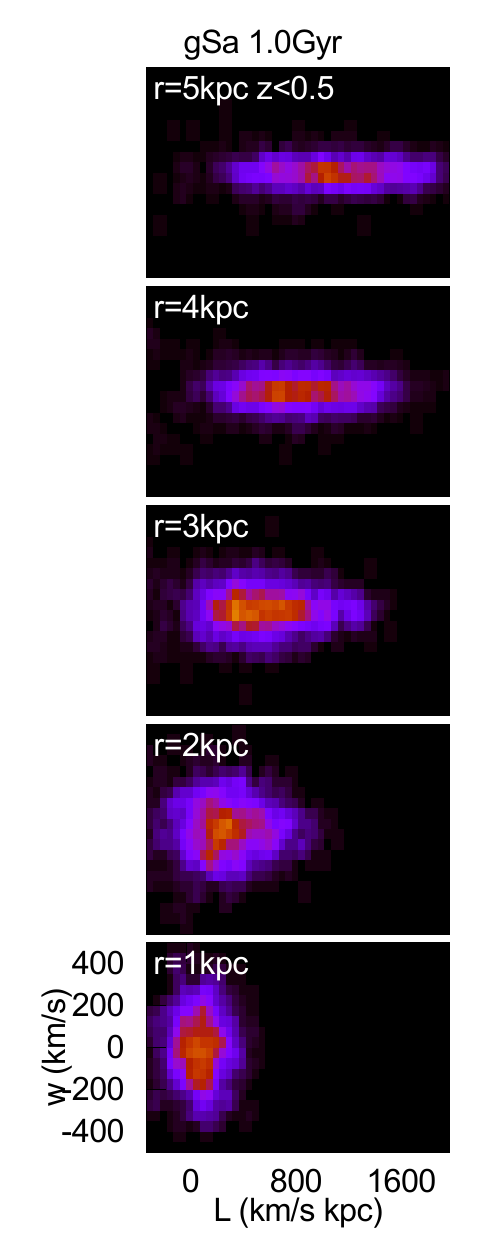} &
\includegraphics[trim={1.5cm 0 0 0},clip,height=4in]{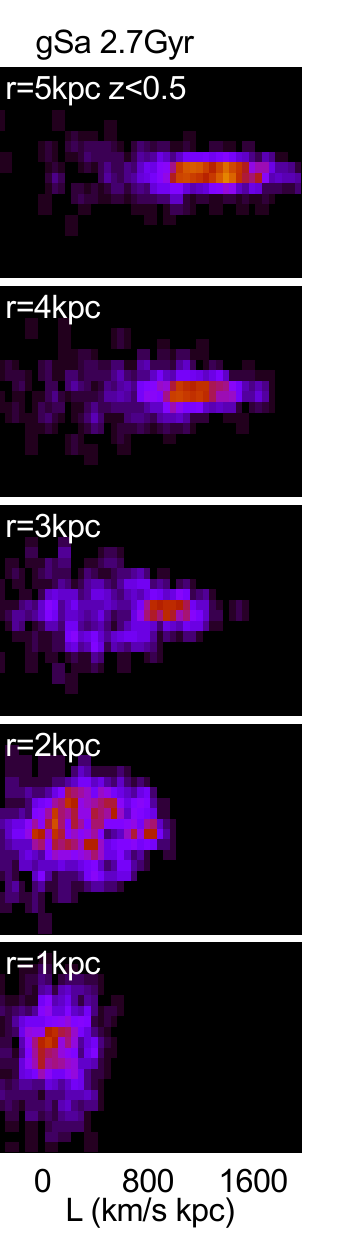} &
\includegraphics[trim={0.5cm 0 0 0},clip,height=4in]{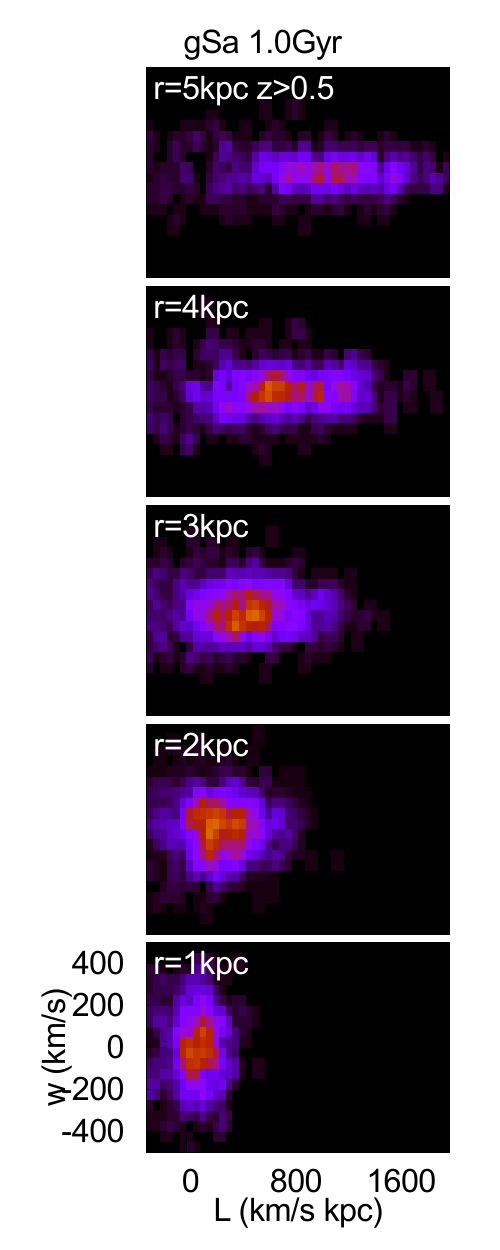} &
\includegraphics[trim={1.5cm 0 0 0},clip,height=4in]{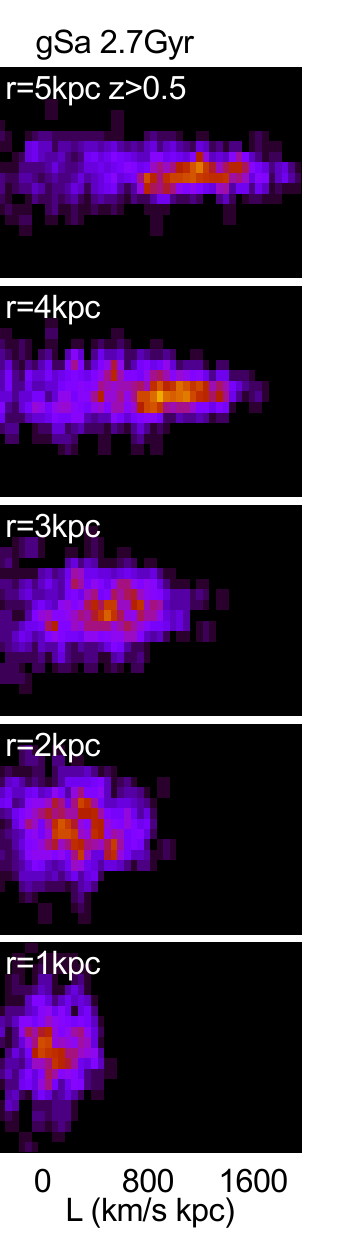}
\end{array}$
 \caption{Angular momentum vs vertical ($w$) velocity distributions of disk stars
 from snapshots of the gSa simulation.
 Similar to Figure \ref{fig:whist_gSa} except the $x$-axis corresponds to angular momentum.
 Within an angular momentum value associated with the vertical resonance, 1000 km/s kpc,
 the vertical velocity dispersion is wide.  Outside this angular momentum value, the vertical
 velocity distribution is narrow.
  \label{fig:lhist_gSa} 
 }
\end{center}
\end{figure*}

\begin{figure*}
\begin{center}
$\begin{array}{cccccc}
\includegraphics[trim={0.5cm  0 0.5cm 0},clip,height=4in]{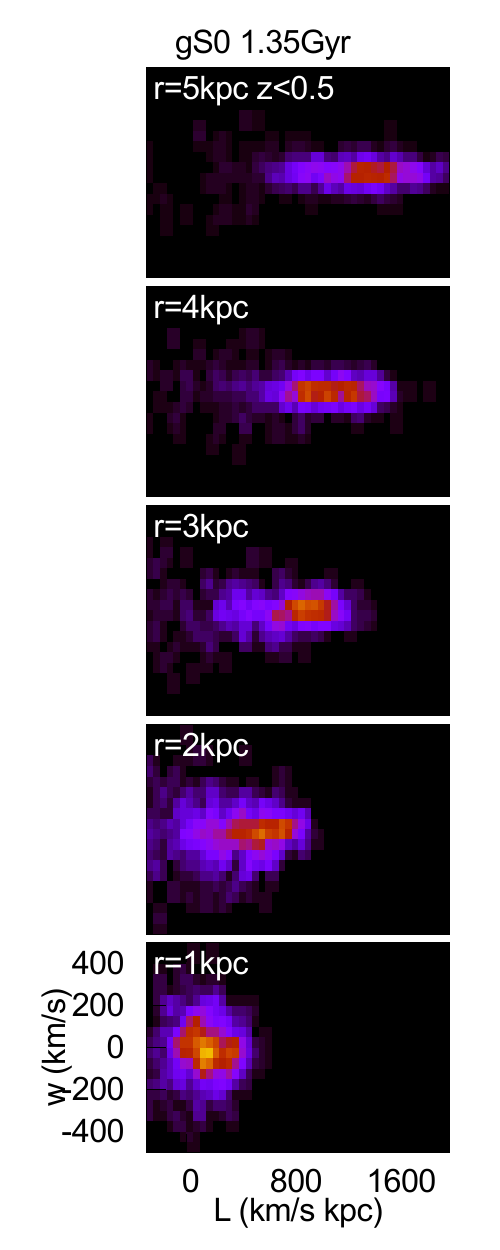} &
\includegraphics[trim={1.5cm 0 0.5cm 0},clip,height=4in]{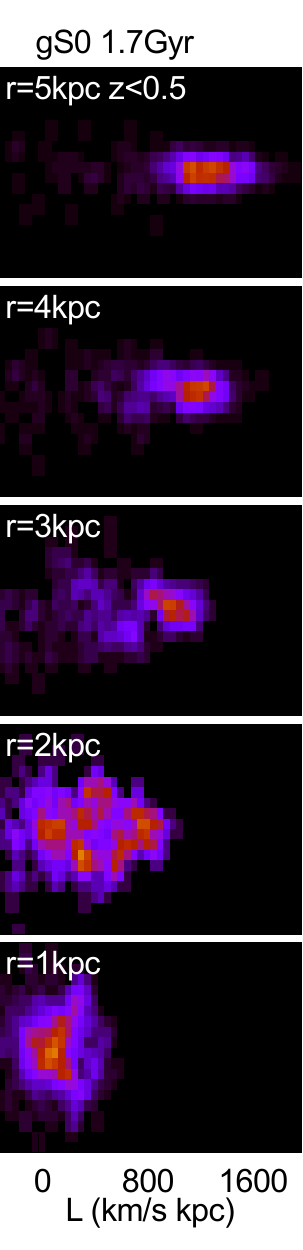} &
\includegraphics[trim={1.5cm 0 0.5cm 0},clip,height=4in]{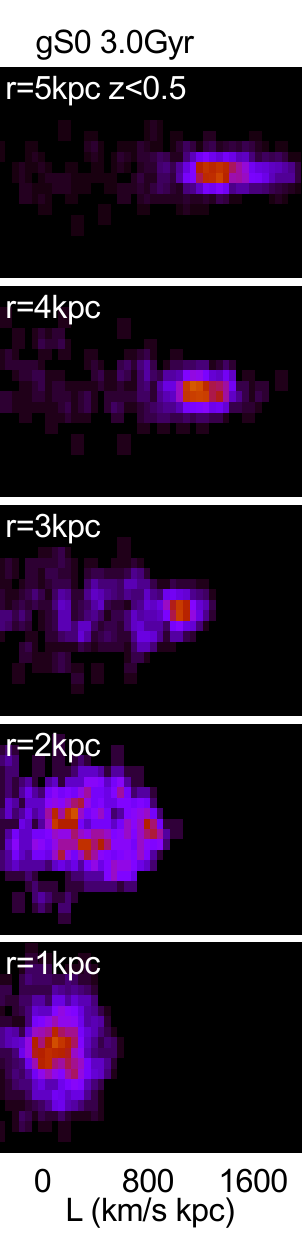} &
\includegraphics[trim={0.5cm  0 0.5cm 0},clip,height=4in]{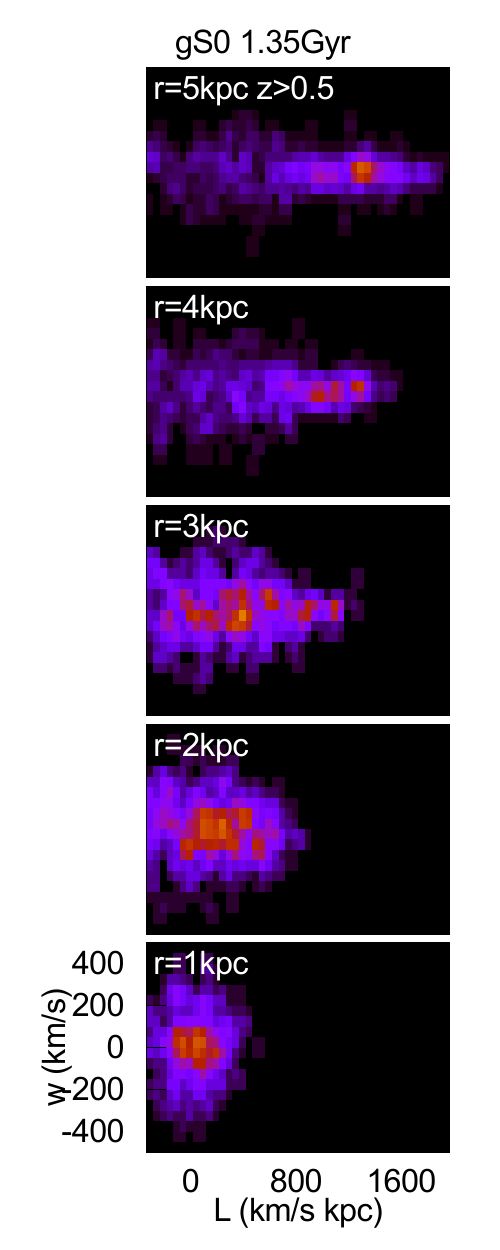} &
\includegraphics[trim={1.5cm 0 0.5cm 0},clip,height=4in]{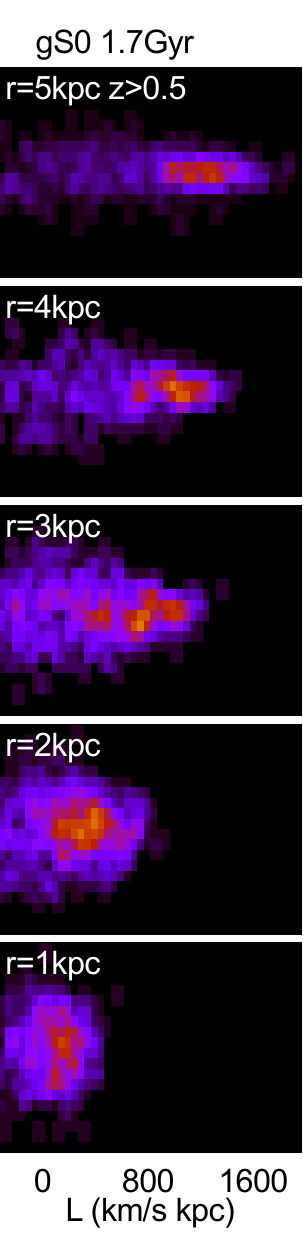} &
\includegraphics[trim={1.5cm 0 0.5cm 0},clip,height=4in]{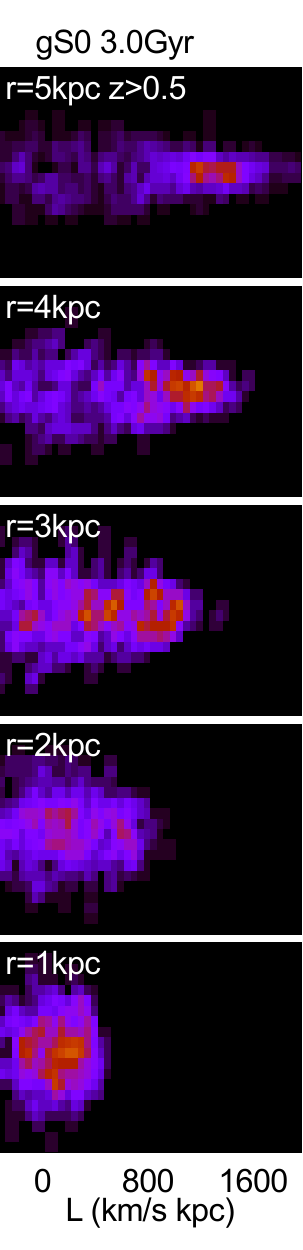} 
\end{array}$
 \caption{Angular momentum vs vertical velocity distributions from snapshots of the gS0 simulation.
Similar to Figure \ref{fig:lhist_gSa} except for the gS0 simulation.
  \label{fig:lhist_gS0} 
 }
\end{center}
\end{figure*}

We  compare the velocity distributions at different locations along the bar major axis.
At different local neighborhoods centered in
the mid-plane and on the bar major axis, we extracted
distributions of disk stars in radial velocity and vertical velocity ($w$ or $v_z$)
and in angular momentum, $L$, and vertical velocity, $w$.
In cylindrical coordinates each neighborhood has a radius (in the mid-plane) of $0.2 r_G$
where $r_G$ is the galactocentric radius of the neighborhood.
Within this cylinder, we extract velocity distributions for $|z|<0.5$ kpc and for $|z|>0.5$ kpc.
All stars within these regions were used to create the distributions.
The distributions were computed for the same times considered in previous
figures and they are shown in Figures \ref{fig:whist_gSa} -  \ref{fig:lhist_gS0}.

While the velocity distributions peak at low $w$ (planar orbits) at earlier times, they
are wider at later times.  Only at large radius is there a population of stars in planar orbits
at later times.
Within the vertical resonance, there are no orbits that remain in the mid-plane.
This gives a donut or double bar shape in the mid-plane velocity distributions in the peanut, that is particularly
noticeable at later times in Figure \ref{fig:whist_gS0} and that is not evident in the velocity distribution
above and below the plane. 
The lack of planar orbits within the resonance is consistent with the resonant
 sweeping scenario we discussed above.
In the resonant trapping model, we would expect a group of stars near periodic orbits.
However, the velocity distributions are quite wide, suggesting that the resonance
has primarily heated the stellar distribution.

The angular momentum, ($L$), vs $w$ distributions, shown in Figures \ref{fig:lhist_gSa} and \ref{fig:lhist_gS0} 
show structure, particularly
at the later times.   The lack of stars in planar orbits is visible at later times for $L<1000$ km/s kpc
in both simulations in the planar distributions (Figures \ref{fig:lhist_gSa}, \ref{fig:lhist_gS0}).
At each time and radius
there is a particular angular momentum value, below which the vertical mid-plane
velocity distribution is wide.  
We can associate this angular momentum value
with that of the vertical resonance.  Stars heated by the resonance are then
visible in the distributions seen above and below the mid-plane.
As angular momentum approximately sets the orbital period, stars in the resonance have
a particular angular momentum value.  As the vertical resonance is swept through the disk
the vertical dispersions of stars are increased.   We can identify the resonance
location from this angular momentum value of about 1000 km/s kpc, corresponding 
to approximate $r = 3$, kpc and consistent with the radius estimated for the resonance at later times
in both simulations.  Figures \ref{fig:lhist_gSa} and \ref{fig:lhist_gS0} mid-plane distributions
show that as the peanut grow, stars within the resonance increase in vertical distribution and the 
stars with high vertical dispersions are seen at increasing larger radii.

\

\subsection{Is evolution adiabatic?  }

Resonance capture is only possible when the bar growth and drift rate are effectively adiabatic.
The resonance is only effective at heating the stellar distribution if stars remain in its vicinity longer
than a resonance libration timescale.   Thus the adiabatic limit is also relevant to estimate the
effectiveness of resonant heating.
The resonant libration frequency can be used to determine whether  
variations are adiabatic.  
The libration frequency,  approximately equal to $\epsilon_s$ and plotted as
  small points shown in Figure \ref{fig:res_all},
 is approximately 1-2 km/s/kpc  $=$ Gyr$^{-1}$. 
  We can compare the square of this to the rate of bar pattern speed change.
 In the gSa simulation the bar pattern speed dropped from 40 to 30 km/s/kpc in 2 Gyr (from 
 Figure 2 by \citealt{minchev12})
 giving a rate $\dot \Omega_b \sim -5$ Gyr$^{-2}$.  This is only somewhat faster in magnitude
 than the square of the libration frequency, 
 $ \epsilon_s^2 \approx 1 -4$ Gyr$^{-2}$ (recall that the square of the libration frequency
 approximately defines the adiabatic limit \citealt{quillen06}).
 This implies that the drift rate is nearly adiabatic, particularly at later times when the
 bars don't slow does as fast.   Because the evolution is nearly adiabatic stars that encounter the
 resonance have enough time in resonance to be lifted  to $z_{max}$.  If the drift
 were significantly faster then the stars would not feel the resonance passage.
   
\section{Constraints on the Galactic X-shape Bulge}

We assume that the X-shape distribution in red clump giants in the Milky Way bulge is 
at the location of the vertical resonance.  This gives us a relation between the vertical 
oscillation frequency, $\nu$,  the angular rotation rate and the bar pattern speed at 
the location of the resonance $r_{LR}$.
Poisson's equation relates derivatives of the potential to the mass density.  Hence, the vertical resonance gives
us a constraint on the mass density in the mid-plane.

Poisson's equation  $\nabla^2 \Phi = 4 \pi G \rho$ relates the derivatives of the gravitational potential $\Phi$, to the local density, $\rho$.
In cylindrical coordinates Poisson's equation is
\begin{eqnarray}
4 \pi G \rho = {\partial^2 \Phi \over \partial z^2} +  {\partial^2 \Phi \over \partial r^2} + {1\over r}  {\partial \Phi \over \partial r}  .
\end{eqnarray}
Using expressions for $\Omega, \kappa$ and $\nu$ in the mid-plane, we find
\begin{eqnarray}
\rho(r,z=0) = {1 \over 4 \pi G} \left [ \nu^2 + \kappa^2 - 2\Omega^2\right].
\end{eqnarray}
The vertical resonance,
satisfies a  resonant condition $\nu = 2(\Omega- \Omega_b)$  at  a radius $r_{LR}$.
The density function
\begin{equation}
\rho_{LR}(r) = {1 \over 4 \pi G} \left[ 4 (\Omega - \Omega_b)^2 + \kappa^2 - 2 \Omega^2 \right] \label{eqn:rho_res}
\end{equation}
must be equal to the density in the mid-plane at $r_{LR}$. 
The right hand side only depends on the rotation curve in the mid-plane and the bar pattern speed.
By fitting the rotation curve measurements and an estimate for the bar pattern speed
we can compute this density function.
The rotation curve and associated functions $\Omega, \kappa$ are estimated using tangent point
velocities estimated from HI observations
 by \citet{malhotra95}, the rotation curve compiled by \citet{sofue12} or that predicted with
 the Besan\c{c}on model\footnote{http://model.obs-besancon.fr/} \citep{robin03}.
 We use a bar pattern speed of $\Omega_b = 54.2$ km/s/kpc consistent with                                                                                                                                   
$\Omega_b /\Omega_0 =1.87 \pm 0.2$ \citep{gardner10,minchev07},
where $\Omega_0$ is the angular rotation rate at the Sun's galactocentric radius.
We adopt a distance from the Sun to the galactic center of $R_0= 8.0$ kpc and
a circular velocity of $V_0 = 220$ km/s \citep{bovy12} so that $\Omega_0 = 27.5$ km/s/kpc.

\begin{figure*}
\begin{center}
\includegraphics[width=5in]{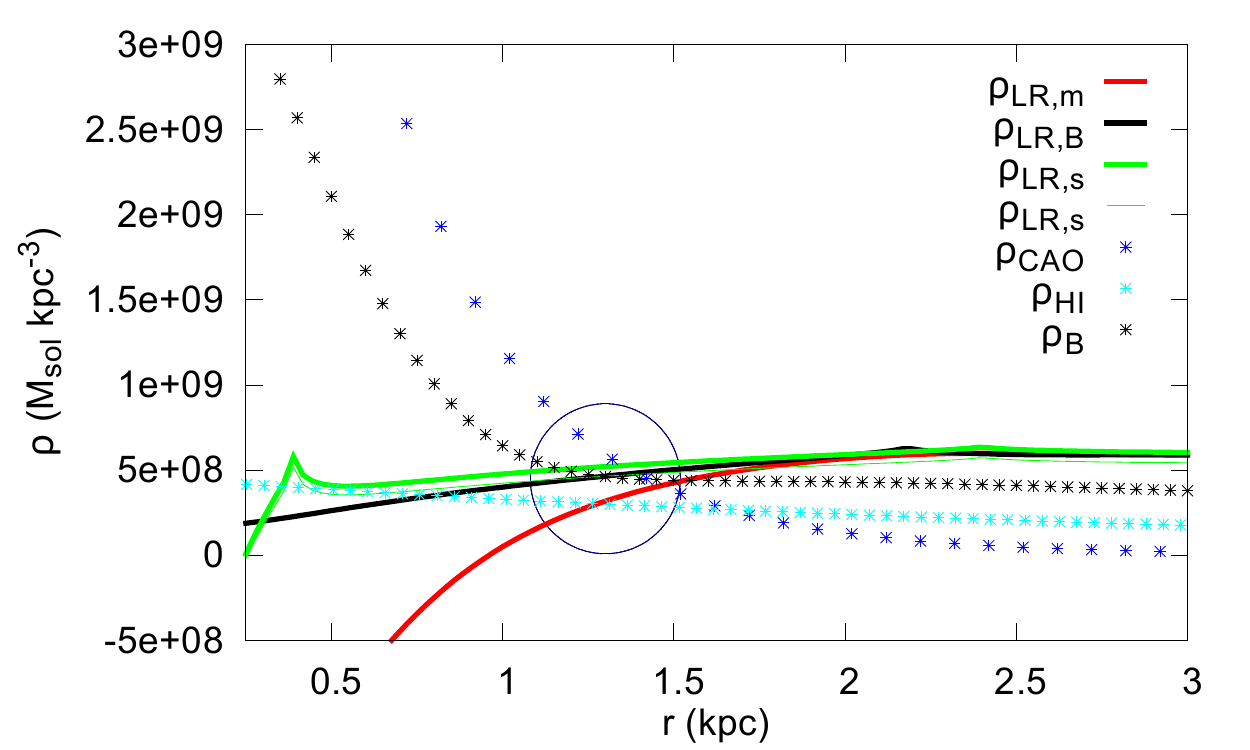} 
 \caption{
 Constraints on the Milky Way bulge.
 The lines show the mid-plane density, $\rho_{LR}(r)$, as a function of radius, with density function 
 given in equation \ref{eqn:rho_res}, computed using Poisson's equation,
  the 2:1 vertical resonance condition, and  the bar pattern speed estimated by \citet{gardner10,minchev07}.
 The red line ($\rho_{LR,m}$), and green line ($\rho_{LR,s}$),  computes this density using a rotation curve  by \citet{malhotra95} and by \citet{sofue12}, respectively.
 The  black lines is similarly computed except using the rotation curve predicted from a Besan\c{c}on model.
 The black points show the mid plane density, $\rho_B$,  for the Besan\c{c}on model.
 The blue points show the mid plane density, $\rho_{Cao}$, 
  for the azimuthally averaged bulge  E3 model by \citet{cao13}.
 The turquoise points show the mid-plane density estimated by \citet{malhotra95} from the velocity dispersion in HI.
 The region of intersection of the lines and points, shown with the navy circle,  implies 
 that the mid-plane density  is $5 \pm 1 \times 10^8 M_\odot {\rm kpc}^{-3}$
 at a radius of $1.4\pm 0.2$ kpc and that this radius is the location of the vertical resonance.
 As the lines and points approximately intersect,  the rotation curve, bar pattern speed, mid-plane density and bulge density profiles are all nearly consistent with one another.
The narrow green line shows the small effect of lowering the bar pattern speed by 5\%.
A comparison between the red, green and black lines show the sensitivity to uncertainties in the rotation curve.
The peak in the green curve at small radii is an artifact caused by a drop in the rotation velocity at very smaller radius
and the low order of polynomial used to fit the rotation curve.
   \label{fig:malfit} 
 }
\end{center}
\end{figure*}

Figure \ref{fig:malfit} shows three density functions $\rho_{LR}(r)$ computed
using equation \ref{eqn:rho_res} and predicted or observed rotation curves.  
Shown as a red line this density function is computed using the rotation
curve by \citet{malhotra95} based on  tangent point velocities measured
 between $0.2R_0$ and $R_0$.
The green line computes the density using the Milky Way rotation curve
compiled by \citet{sofue12}.  
   The black line shows the density function computed from the mid plane
rotation curve predicted from a Besan\c{c}on model.  
This model has stellar and dark matter halos, thin and thick disk components,
and ISM as described by \citet{robin03} 
  but instead of using a point mass bulge to compute the rotation curve, we use the oblate G0 bulge model by \citet{dwek95} 
  that is based on DIRBE observations, and normalized to have bulge mass $2 \times 10^{10} M_\odot$
  consistent with the Besan\c{c}on model.
 
  In Figure \ref{fig:malfit} we compare the density functions $\rho_{LR}(r)$ to estimated mid plane densities
  that are shown as points on the plot.
The azimuthally averaged mid plane 
density  for the E3 bulge model by \citet{cao13} is shown with blue points.  This has a density distribution
\begin{equation} 
\rho_{E3}(r_s) = \rho_0 K_0(r_s)
\end{equation}
where $K_0$ is the modified Bessel function of the second kind and
\begin{equation}
r_s= \left[  \left[ \left( {x \over x_0}\right)^2 + \left( {y \over y_0}\right)^2\right]^2 +  \left( {z \over z_0}\right)^4 \right]^{1\over 4}
\end{equation}
The model $E_3$ has $(x_0, y_0, z_0) = (0.67, 0.29, 0.42)$ kpc (as from their Table 1)
and $\rho_0=1.50 \times 10^{10} M_\odot$ kpc$^{-3}$. 
Figure \ref{fig:malfit} also shows the mid plane density predicted from the modified Besan\c{c}on model (shown
as black points) and as turquoise points
the mid-plane density computed from HI dispersions by \citet{malhotra95}.
The three lines and three sets of points all approximately intersect at a radius between 1.2 and 1.5 kpc.  
This radius is consistent with the radius estimated for the outer part of the X-shape
from red-clump stars by \citet{mcwilliam10} (see their Figure 6). 
The rotation curve, bar pattern speed, bulge density profile, X-shape size scale and resonance location are 
all consistent with one another.

The location of the $\rho_{LR}$  curves shown in Figure \ref{fig:malfit}  
are insensitive to the assumed bar pattern speed. 
We find that varying the bar pattern speed by 5\%) does not significantly move
the curve (see the narrow green line in Figure \ref{fig:malfit}).   
Because the relation for $\rho_{LR}$ depends on derivatives of the potential,
this function  is extremely sensitive to the type of curve fit to
the rotation curve data points and the noise and range of these points.    
This is illustrated by the difference between the density curves $\rho_{LR,m}$
and $\rho_{LR,s}$ that use two different rotation curves to generate $\Omega$ and $\kappa$.
The rotation curve by \citet{malhotra95} does not extend within 2 kpc so is purely an extrapolation within
this radius.
The mid-plane density $\rho_{HI}$ estimated from HI dispersions is extrapolated from radius $r> 2$ kpc
and so is likely an underestimate  for the mid-plane density within the bulge.
The bulge mid plane profile based on models by \citet{cao13} is somewhat high compared to the entire mid plane
density (include disk and halo components) predicted by the Besan\c{c}on model.
The Hamiltonian model we discuss here does not take into account the radial degree of freedom, however
the orbits in the X are likely to be eccentric.   To make better constraints on the mass distribution
from the observed distribution of the orbits supporting the X-shape, the orbital eccentricity 
must be understood and
taken into account.

\section{Summary and Discussion}

We have explored a Hamiltonian resonance model for X-shaped or peanut-shaped galactic bulges. 
By computing the vertical oscillation frequency
from the gravitational potential in N-body simulations, we confirm that the X- or peanut-shape is located
near the 2:1 vertical resonance with the bar, which is associated with upward and downwards facing
periodic banana-shaped orbits.
This confirms previous studies that computed periodic orbit families and attributed
the peanut-shape to stars in banana-shaped periodic orbits.
(e.g., \citealt{combes90,patsis02,martinez06}).
Here the banana shaped orbits arise in the Hamiltonian model as fixed points that correspond to
high inclination orbits in or near resonance.  They are associated with orbits that
have an  angle $\phi$, librating about 0 or $\pi$, that depends on azimuthal angle $\theta$ and
the angle associated with vertical oscillations $\theta_z$.

We examine two simulations, one contains a bar that buckles, the other exhibits
a peanut-shaped bulge even though the bar does not buckle.  
In the bar-buckling simulation, as the bulge thickens during buckling,
the vertical oscillation frequency decreases and so the resonance and peanut-shape moves to larger radius. 
In the buckle-free  simulation, the resonance and the peanut-shape
moves outward as the bar slows down.
In both cases, the location of the peanut shape is consistent with the 
 location of the resonance computed from the current mass distribution. 
The Hamiltonian model predicts a narrow resonance width in angular momentum
distance from commensurability of  $dL/L \sim 0.05$ kpc,  a region where there are no planar orbits.
As the bar slows down and disk thickens, the resonance moves outwards heating the disk.
Stars in the mid plane, just outside resonance, are pushed to high inclination when
they encounter the resonance.  They reside just inside the resonance separatrix and support
the peanut shape until the separatrix shrinks leaving them at high inclination but no longer
supporting or moving with the peanut shape.  
The peanut height predicted from the
Hamiltonian model, using the separatrix height
and coefficients measured from the simulations, is approximately consistent
with that seen in the simulations.
Using the libration frequency predicted from the Hamiltonian model, we compare the drift
rate to that defining the adiabatic limit and find that the drift rate is comparable to 
the adiabatic limit.   This implies that stars are in resonance, just long enough
to be lifted by it.

We interpret the X-shape as primarily due to the population of stars that is supporting the peanut shape
that have just recently been excited by the resonance as the resonance drifts outward.   
Stars within the resonance separatrix would either be
upwards banana shaped or downwards banana-shaped orbits.
Those recently escaping resonance, and just outside the resonance separatrix, would have morphology
similar to a sum of upwards and downwards banana-shaped orbits and spend more time aligned
with the bar and so support the peanut shape.
After stars leave resonance they no longer
support the peanut shape and so don't maintain a coherent vertical feature that is aligned  with the bar.
Stars supporting the X-shape would primarily be disk stars, the latest ones captured into resonance 
that were in the mid plane prior to their capture into resonance.  As they were all disk stars 
they should have similar metallicity and that typical of the disk just exterior to resonance.

Velocity distributions along the bar major axis in numerical simulations show that there are no stars
in orbits in the mid-plane within the vertical resonance, as expected from a drifting Hamiltonian model.    
The distribution of angular momentum versus vertical velocity shows that stars below a certain angular
momentum value associated with resonance exhibit a much larger vertical velocity dispersion than
stars outside resonance.
The velocity distributions are wide, suggesting that as the resonance is  swept through the disk,
stars are heated by it.
Division into bulge and disk might be made based on the angular momentum value
associated with the vertical resonance as stars are pushed out of the
plane via passage through the resonance.

We previously proposed that peanut shaped bulges arose from resonance capture during bar growth \citep{quillen02} and that the X-shape arose from the height of periodic orbits as a function from distance from resonance.
Here we find that that the resonance width is too narrow to account for the X-shape.
Instead the resonance is swept through the disk due to
slowing of the bar and thickening of the disk.   Because of the narrow width of the resonance, only orbits in the
vicinity of the resonance can exhibit vertical structure coherent with the bar.  Consequently
 stars recently captured into resonance or near the resonance separatrix are responsible for the X-shape.  
 The X-shape can be attributed
 to the orbital shapes of this resonant population of stars. If the resonance has drifted outward, 
 these stars lie  in the vicinity of
 the resonance separatrix at the outer boundary of the resonance.  Stars within
 the separatrix are not near periodic orbits
 but do exhibit angles $\phi$ that librate about 0 or $\pi$.  Those outside the separatrix 
 exhibit oscillating $\phi$ but spend more time with $\phi$ near $0, \pi$ than near $\pm \pi/2$.

The low order Hamiltonian model   
 provides a promising predictive dynamical framework to describe features
of peanut-shaped bulges, and relate orbital properties to the bar strength, pattern speed and
vertical oscillation frequency.    Time dependent models were only qualitatively explored here, however 
with additional calculation
they may be able to predict velocity distributions, relate the current distributions
to past evolution and reconcile the differences in theoretical framework between the vertical resonance
model and the bar buckling instability model for peanut formation. 

An estimate of the mid-plane mass density can be made in the bulge of the Milky Way
at the location of the resonance.  We find that the mid-plane density is approximately 
$5 \times 10^8 M_\odot {\rm kpc}^{-3}$ at a radius of 1.3 kpc and that this is approximately the location 
of the vertical resonance and recently discovered X-shape \citep{nataf10,mcwilliam10}.
This density is approximately consistent with the axisymmetric average at the same radius 
of the E3 bulge mode by \citet{cao13} and that of the Besan\c{c}on model.  
Thus the rotation curve, vertical resonance location, X-shape tips, and mid plane mass density, 
 are all self-consistent in the Milky Way galaxy bulge.

The largest uncertainty in estimating the location of
the resonance  is due to errors in the rotation curve in the bulge, that are exacerbated when taking derivatives.
We identify the resonance location as a function of angular momentum.  
However  the orbits supporting the X-shape
are likely to have eccentricity similar to stars just outside the vertical resonance.
That implies that these stars would be located at larger galactocentric radius near the ends
of the bar than along the bar minor axis.     
We have neglected the mean ellipticity of orbits in our computation.
To improve upon constraint based on the resonance location, the ellipticity of the orbits must be
 be taken into account.
The extreme sensitivity of the vertical resonance model to rotation curve, bar shape, strength and pattern speed, and
 disk thickness,
imply that future work will make it possible
to place  tight constraints
on the structure and evolution of the Milky Way disk and bulge. 

We explored here numerical simulations with live halos and with bars that decreased in pattern speed.
We have focused on the setting with resonance moving outward either due to disk thickening
or bar slowing.
Future studies can also study simulations (such as that by \citealt{li12}) that have steadier bars.
If the bar speeds up then the interpretation would change dramatically as in this case stars can
be captured into periodic orbits and lifted while remaining in resonance.
The Hamiltonian model has been improved and corrected in appendices \ref{ap:action} and \ref{ap:can}, 
however it does not take into
account the radial degree of freedom.  Future work is required to predict the three dimensional structure
of orbits supporting the X- or peanut-shape.

\vskip 0.5 truein
This work was in part supported by NASA grant NNX13AI27G and a travel grant from the IAU.
We thank Nanjing University and Shanghai Observatory for their generous and gracious hosting
June 2013.
We thank David Nataf, Liang Cao, Juntai Shen, and Victor Debattista for helpful comments
and communications.
\bibliographystyle{aa}

\appendix
\section{Action Angle Variables for an Axisymmetric Galaxy to low order in epicyclic amplitude and vertical
oscillation amplitude}\

\label{ap:action}

In this section we review the extension of a low order Hamiltonian theory to cover vertical oscillations
in a galactic disk.  We extend and  
 correct  the previous calculation by \citet{quillen02}.  The Hamiltonian is
expanded to low order in epicyclic and vertical action variables.  Rather than expand in 
angular momentum (as done by \citealt{cont75})  we use coefficients that are functions
of the angular momentum and we retain the angular momentum as an action variable.

The dynamics of a massless particle in a galaxy with an axisymmetric gravitational potential, $V_0(r,z)$,
can be described with the Hamiltonian 
\begin{eqnarray}
H_0(r, z,\theta; p_r, p_z, L)  = {L^2 \over 2r^2} + {p_r^2 \over 2}  
+ {p_z^2 \over 2} + V_0(r,z) 
\end{eqnarray}
where $r,z,\theta$ are cylindrical coordinates
and $p_r, p_z, L$ are associated momenta.
We can transform to new momenta and coordinates $(\theta_r,\theta_z, \theta_{new};J_r,J_z,L_{new})$
with a generating function that depends on old momenta  $(p_r,p_z,L)$ and new coordinates 
$(\theta_r,\theta_z, \theta_{new})$
\begin{eqnarray}
F_3(p_r,  p_z, L; \theta_r, \theta_z, \theta_{new} )
&=&  
{p_r^2 \over 2 \kappa(L)} \cot \theta_r
+ {p_z^2 \over 2 \nu(L)} \cot \theta_z \nonumber \\
&&- r_c(L)p_r - L \theta_{new}
\end{eqnarray}
giving a
canonical transformation that relates old coordinates and momenta to new ones
\begin{eqnarray}
r &=& r_c(L) + \sqrt{2 J_r \over \kappa(L)} \cos \theta_r \nonumber\\
p_r & = & - \sqrt{2 J_r \kappa(L)} \sin \theta_r\nonumber \\
z &=&  \sqrt{2 J_z \over \nu(L)} \cos \theta_z  \nonumber\\
p_z & = & - \sqrt{2 J_z \nu(L)} \sin \theta_z \nonumber \\
\theta &=& \theta_{new} 
- r_c'(L) \sqrt{2 J_r\kappa} \sin \theta_r \nonumber \\
&&	+ {\kappa'(L) J_r \over 2 \kappa} \sin (2\theta_r) 
	+ {\nu'(L) J_z \over 2 \nu} \sin (2\theta_z) 
	\nonumber \\
L &=& L_{new},
\end{eqnarray}
where $r_c'(L) = {\partial r_c(L) \over \partial L}$ and similarly for $\kappa'(K)$
and $\nu'(L)$.
Hereafter we do not make a distinction between $L$ and $L_{new}$ or $\theta$ and $\theta_{new}$.
The radius of a particle with angular momentum $L$ in a circular orbit is $r_c(L)$ and
\begin{equation}
L^2 = r_c^3(L) \left .{ \partial V_0  \over \partial r} \right|_{r= {r_c}(L),z=0}
\end{equation}
or 
\begin{equation}
r_c(L) = \sqrt{L \over \Omega(L)}.
\end{equation}
with $\Omega(L)$ 
the angular rotation rate of a particle in a circular orbit in the mid-plane with angular momentum $L$.
The epicyclic frequency, $\kappa(L)$, satisfies
\begin{equation}
\kappa^2(L) = 3 \Omega^2(L) + \left.{\partial^2 V_0 \over \partial r^2} \right|_{r={r_c}(L),z=0}
\end{equation}
and it may be convenient to recall
\begin{equation}
\Omega^2(L) = {1 \over r_c(L)}  \left.{\partial V_0 \over \partial r} \right|_{r={r_c}(L),z=0}.
\end{equation}
The vertical oscillation frequency, $\nu (L)$ satisfies
\begin{equation}
\nu^2(L) = \left. {\partial^2 V_0 \over \partial z^2} \right|_{r=r_c(L),z=0} 
\end{equation}

The derivatives with respect to $L$ 
\begin{eqnarray}
r_c'(L) &=& {\partial r_c(L) \over \partial L} = 
{2 L \over r_c(L)^3 \kappa(L)^2} 
= {2 \Omega(L) \over \kappa(L)} {1 \over r_c(L) \kappa(L)} \nonumber \\
\Omega'(L) &=& {1 \over r_c^2} \left[ 1 - {4 \Omega^2 \over \kappa^2} \right] \nonumber \\
\kappa'(L) &=& 
 {L \over r_c^3 \kappa^3} 
\left[ {\partial^3 V_0 \over \partial r^3} + {3 \over r_c} {\partial^2 V_0 \over\partial r}^2  - {3\over r_c^2} {\partial V_0 \over \partial r}  \right]_{r_c(L),z=0} \nonumber \\
&=& {L \over r_c^3 \kappa^3} \left. {\partial \kappa^2 \over \partial r} \right|_{r=r_c(L),z=0} \nonumber \\
\nu'(L) &=& {L \over r_c^3 \kappa^2 \nu} \left. {\partial^3 V_0 \over \partial r \partial z^2} \right|_{r=r_c(L),z=0} 
\label{eqn:derivs}
\end{eqnarray}
Here $\kappa'(L)$ is related to the $\lambda_0$ parameter used by \citet{cont75} (see his equation A10).

The variables $J_r$ and $\theta_r$ are the epicyclic action and angle 
and $J_z$ and $\theta_z$ are the action and angle for vertical oscillations.
It is sometimes convenient to discuss an epicyclic amplitude, $\sqrt{2 J_r/\kappa}$, 
or vertical oscillation amplitude,  $\sqrt{ 2 J_z/ \nu}$.
The amplitudes can be described in terms
of an orbital eccentricity, $e  = {1 \over r_c} \sqrt{2 J_r \over \kappa}$, 
or an orbital inclination, $i={1 \over r_c}\sqrt{ 2 J_z\over \nu}$.

The new Hamiltonian in the new coordinate system,
expanded to fourth order in $J_z^{1/2}$ and $J_r^{1/2}$, and assuming that $V$ is symmetric about
the mid-plane, is
\begin{eqnarray}
H_0(\theta_r,\theta_z,\theta; J_r, J_z, L )
& =& 
{L^2 \over 2 r_c^2}  +  V(r_c) 
+ \kappa J_r + \nu J_z  \nonumber
\end{eqnarray}
\begin{eqnarray}
~~~ & &+   \left( {2 J_r \over \kappa }\right)^{3\over 2} \cos^3 \theta_r \left(- {4L^2 \over 2 r_c^5} + {1 \over 6} \left. {\partial^3 V_0\over \partial r^3} \right|_{r=r_c,z=0} \right) \nonumber \\
&&+  \left( {2 J_r \over \kappa }\right)^{2} \cos^4 \theta_r \left( {5L^2 \over 2 r_c^6} + { 1\over 24} \left. {\partial^4 V_0 \over \partial r^4} \right|_{r=r_c,z=0}\right)\nonumber\\
&& +  \left( {2 J_r \over \kappa }\right)^{1\over 2}  \left( {2 J_z \over \nu }\right)\cos \theta_r \cos^2 \theta_z {1 \over 2}
    \left. { \partial^3 V_0 \over \partial r \partial^2 z} \right|_{r=r_c,z=0} \nonumber \\
&& +  \left( {2 J_r \over \kappa }\right)  \left( {2 J_z \over \nu }\right)\cos^2 \theta_r \cos^2 \theta_z {1 \over 4}
 \left. { \partial^4 V_0 \over \partial^2 r \partial^2 z} \right|_{r=r_c,z=0} \nonumber \\
 && + \left( {2 J_z \over \nu }\right)^2 \cos^4 \theta_z {1 \over 24} \left. { \partial^4 V_0 \over  \partial^4 z} \right|_{r=r_c,z=0} 
\end{eqnarray}
and our choice for functions $r_c(L), \kappa(L), \nu(L)$ has cancelled some low order terms. 

It is useful to keep in mind the trigonometric identities
\begin{eqnarray}
\cos^3 \phi & =&  {1 \over 4} {\cos 3 \phi } + {3 \over 4} \cos \phi \nonumber \\
\cos^4 \phi &=& {1\over 8} \cos 4 \phi + {1 \over 2} \cos 2 \phi + {3 \over 8} \nonumber \\
\cos^2 \phi_a \cos^2 \phi_b &=& {1\over 4} \left[ \cos (2(\phi_a+\phi_b))+ \cos (2(\phi_a-\phi_b))\right] \nonumber \\
&&+ {1\over 2} \left[\cos (2 \phi_a) + \cos (2 \phi_b) + 1 \right]
\end{eqnarray}
We can remove terms proportional to cosine functions of $\theta_z$ or $\theta_r$ by performing canonical transformations that include terms that are dependent on the Fourier components such as $\cos m \theta_r$,
or $\cos m\theta_r$ with $m=1,2,3,4$,
 as illustrated in the appendix by \citet{quillen02}.  This can be done as long as the time derivative of the angles
 are not small (there are no small divisors).  The constant coefficients in the expansion of
$ \cos^4 \theta_r$ and $\cos^4 \theta_z$ give terms proportional to $J_r^2$ and $J_z^2$.
The term with $\cos^2 \theta_r \cos^2 \theta_z$ gives a term proportional to $J _rJ_z$.
With respect to the perturbed coordinates and momenta
\begin{eqnarray}
H_0(\theta_r,\theta_z,\theta_{new}; J_r, J_z, L )
 = g_0(L)  + \kappa(L) J_r + \nu(L) J_z  \label{eqn:Ham_A} \\
\qquad \qquad \qquad + a_z(L) J_z^2 + a_r(L) J_r^2 + a_{rz}(L) J_z J_r 
\nonumber 
\end{eqnarray}
with 
\begin{eqnarray}
g_0(L)& =& {L^2 \over 2 r_c^2}  +  V(r_c) \nonumber \\
a_z(L) &=& {4 \over \nu^2} {3 \over 8} {1 \over 4!} \left. {\partial^4 V_0 \over \partial z^4} \right|_{r = r_c, z=0} 
=  {1 \over 16 \nu^2}  \left. {\partial^4 V_0 \over \partial z^4} \right|_{r = r_c, z=0} \nonumber \\
a_r(L) &=& {4 \over \kappa^2} {3 \over 8} \left( {5L^2 \over 2 r_c^4} + { 1\over 24} \left. {\partial^4 V_0 \over \partial r^4} \right|_{r=r_c,z=0}\right) \nonumber \\
&=& {15 L^2 \over 4 r_c^4 \kappa^2}  + {1 \over 16 \kappa^2}\left. {\partial^4 V_0 \over \partial r^4} \right|_{r=r_c,z=0} \nonumber \\
a_{rz}(L) &=& {1 \over 4 \nu \kappa}  \left. { \partial^4 V_0 \over  \partial^2 r \partial^2 z} \right|_{r=r_c,z=0}
\label{eqn:coeffs}
\end{eqnarray}
We correct by a factor of 2 the expression for $a_z$ from that given in equation 7 and 16 by \citet{quillen02}.
The zeroth order function $g_0(L)$ 
 is such that ${\partial g_0(L) \over \partial L} = \Omega(L)$, as expected.
The last coefficient can also be written
\begin{eqnarray}
a_{rz}(L) 
&=& {1 \over 4 \nu \kappa} \left. {\partial^2 \nu^2 \over \partial r^2} \right|_{r=r_c,z=0}
\end{eqnarray}
correcting the coefficient of
the last term in equation 7 by \citet{quillen02} that has incorrect units.

The expression for $a_r(L)$ above differs from the coefficient given in equation A32 by \citet{cont75}.
Here we have expanded to low orders in $J_r^{1/2}$ and $J_z^{1/2}$ and not
expanded about a particular $L$ value.  In other words the Hamiltonian in equation \ref{eqn:Ham_A}
contains low orders of $J_r,J_z$ but the coefficient for each term is a function of $L$. 
\citet{cont75} does not use functions of angular momentum but expands about a value for $L$ corresponding
to a particle in a circular orbit at a particular energy value.

It is convent to transform to a frame that rotates with the bar so that the new Hamiltonian is conserved.
Using a generating function $F_2(\theta_{new}, L') = (\theta_{new} - \Omega_b t)L'$
we transfer to a frame rotating with the bar finding new 
azimuthal angle in the corotating frame 
$$\theta_{cr} = \theta_{new} - \Omega_b t$$
and angular momentum unchanged $L=L'$ so we drop the prime, and new Hamiltonian
\begin{eqnarray}
K_0(\theta_r,\theta_z,\theta_{cr}; J_r, J_z, L )
 = g_0(L) - \Omega_b L  + \kappa(L) J_r  \label{eqn:Ham_Acr} \\
\qquad  + \nu(L) J_z + a_z(L) J_z^2 + a_r(L) J_r^2 + a_{rz}(L) J_z J_r 
\nonumber 
\end{eqnarray}

With the addition of a bar perturbation, $V_b$, that is time independent in the frame
rotating with the bar, the Hamiltonian is 
$K = K_0 + V_b$ and is time independent, so $K$ is a conserved quantity that is called the Jacobi
integral of motion or the Jacobi constant.
Here $V_b$ is the sum of the Fourier components (equation \ref{eqn:V_m}) and has an 
 azimuthal average of zero.  

\section{Canonical transformation to a low dimensional Hamiltonian}

\label{ap:can}

We perform a canonical transformation so that 
$\phi = \theta_z - {m\over 2}\theta_{cr} = \theta_z - {m\over 2}(\theta - \Omega_b t)$ becomes
a canonical coordinate.
We use a generating function in terms of new momenta $(J_z', J_r', I)$ and old coordinates 
$(\theta_z, \theta_r, \theta_{cr})$
\begin{eqnarray}
F_2(J_z',J_r',I; \theta_z, \theta_r, \theta_{cr})
 = \left(\theta_z - {m\over 2} \theta_{cr}  \right)J_z' + J_r' \theta_r + I \theta_{cr}
 \nonumber\\ 
\end{eqnarray}
giving new momenta and coordinates  in terms of old ones
\begin{eqnarray}
I &=& {m \over 2} J_z + L  \label{eqn:Icon} \\
\phi &=& \theta_z - {m\over 2} \theta_{cr} \label{eqn:phi_tran}
\end{eqnarray}
Momenta $J_z' = J_z$ and $J_r' = J_r$, so hereafter we drop the primes.
Coordinates $\theta$ and $\theta_r$ also remain unchanged by
the transformation.  Needed for the transformation is  ${\partial F_2 \over \partial t} = {m\over 2} \Omega_b J_z$
as it contributes to the Hamiltonian in the new variables.
The Hamiltonian (equation \ref{eqn:Ham_Acr})  in the new coordinates expanded to second order in momenta is
\begin{eqnarray}
K_0(\phi,\theta_r,\theta_{cr}; J_z, J_r, I )
= g_0(I) - \Omega_b I + \kappa(I) J_r + \delta(I) J_z +  \nonumber \\
 \qquad \qquad a_{r}(I) J_r^2 + a_{cz}(I) J_z^2 + a_{crz}(I) J_z J_r  \nonumber \\
 \label{eqn:K_ham}
\end{eqnarray}
with 
\begin{eqnarray}
\delta(I) &=&  \nu(I) - {m\over 2}(\Omega(I) - \Omega_b) \\
a_{cz}(I) &=& a_z(I) + {m^2 \over 8}{\partial \Omega \over \partial L}  - {m \over 2} {\partial \nu \over \partial L} \nonumber \\ 
&=&  {1 \over 16\nu^2}{\partial V_0^4 \over \partial z^4} + {m^2 \over 8 r_c^2} \left[ 1 - {4 \Omega^2 \over \kappa^2} \right] \nonumber \\
&& 
 \qquad - {m\over 2} {\Omega \over \nu} {1 \over r_c \kappa^2} {\partial \nu^2 \over \partial r}  \label{eqn:acz} \\
a_{crz}(I) &=& a_{rz}(I) - {m\over 2} {\partial \kappa (I)\over \partial L}  \nonumber \\
&=&
{1 \over 4 \nu \kappa}  { \partial^4 V_0 \over  \partial^2 r \partial^2 z}
-{m\over 2}  {\Omega \over r_c \kappa^3}  {\partial \kappa^2 \over \partial r}  \label{eqn:acrz}
\end{eqnarray}
where we have used equations \ref{eqn:derivs} for derivatives with respect to $L$.

With the addition of a perturbation that depends only on $\phi$, the momentum $I$ (equation \ref{eqn:Icon}) 
is conserved.
Ignoring the epicyclic variations, the Hamiltonian is reduced to a single dimension
and becomes that given in equation \ref{eqn:Ham} with $a = a_{cz}$ and $\delta$ shown above
as above.
\citet{quillen02} did not compute the second and third terms in equation \ref{eqn:acz} however each of 
these  two terms can dominate the first one.  
The second and third term have opposite signs, reducing the strength of $a_{cz}$.

\section{Shift in resonance location due to the $J_r J_z$ cross term}
\label{ap:shift}

The Hamiltonian contains a term proportional to $J_r J_z$.   
As Hamilton's equation $\dot \theta_z  = {\partial H \over \partial J_z}$ the $J_r J_z$ term causes
a shift in the frequency $\dot \theta_z$ and so a shift in the location of the vertical resonance.
Using the expression for $J_r = e^2 r_c^2 \kappa/2$ and eccentricity $e = {1 \over r_c} \sqrt{2 J_r \over \kappa}$
we can estimate  a frequency shift, $\omega = a_{crz} J_r$,  that depends on eccentricity
\begin{eqnarray}
\delta_{\omega} = 
{\omega \over \nu} = {a_{crz} J_r \over \nu} =  {e^2 }  \left[
 {1 \over 8} {r_c^2 \over \nu^2} {\partial^2 \nu^2 \over \partial r^2} 
 - {m \over 4} {\Omega\over \nu} {r_c \over \kappa^2}
 {\partial \kappa^2 \over \partial r} \right],
\end{eqnarray}
using equation \ref{eqn:acrz}.
Assuming that both $\kappa$ and $\nu$ are approximately inversely proportional to $r$,
we find that the second term dominates and 
the frequency shift should be positive.
Using the approximation $\kappa \propto r^{-1}$, $\nu \sim 2 \Omega$ and for $m=4$
we estimate
\begin{equation}
\delta_\omega \sim e^2 .
\end{equation}
For an eccentricity of $e\sim 0.3$ the frequency shift could be of size 0.1.
Because the frequency shift depends on the square of the eccentricity, even within the bar,
the shift cannot be large.

When taking into account the $J_r J_z$ term in the Hamiltonian, 
the distance to the vertical Lindblad resonance is given by the frequency
\begin{equation}
\delta = \nu(1 + \delta_\omega) - 2(\Omega - \Omega_b) ,
\end{equation}
and the resonance is located where
\begin{equation}
\Omega -{ \nu \over 2} \left(1 + \delta_\omega \right) \sim \Omega_b.
\end{equation}
We expect the  shift $\delta_\omega$ to be positive and so it should increase the effect of $\nu$
in the negative term in the above equation and so would 
move the resonance slightly inward.

\section{Radial degree of freedom and orbital eccentricity}
\label{ap:radial}

Outside of Lindblad resonances a low order approximation can be used to estimate orbital eccentricity.
Neglecting the vertical degree of freedom and to first order in the action variable $J_r$, 
the unperturbed Hamiltonian (based on equation \ref{eqn:Ham_Acr})
is 
\begin{eqnarray}
K_0(J_r, L; \theta_{r}, \theta_{cr}) &\approx& \kappa(L) J_r + g_0(L) -\Omega_b L. 
\end{eqnarray}
To this we add a perturbation term that is associated with an $m$-th Lindblad resonance, 
$ \epsilon_m J_r^{1/2} \cos ( \theta_r - m(\theta - \Omega_b t))$ where $\epsilon_b$ can
be estimated from a bar potential's Fourier components
\begin{equation}
\epsilon_m \approx \sqrt{2 \over \kappa} \left[  \frac{1}{2} \frac{ \partial C_m }{ \partial m} + \frac{m \Omega C_m }{ r \kappa} \right] \label{eqn:epsm}
\end{equation}
 \citep{cont75},
 where $C_m$ is the strength of the $m$-th Fourier component (equation \ref{eqn:V_m}).
This gives a total Hamiltonian (in the rotating frame)
\begin{eqnarray}
K_0(J_r, L; \theta_{r}, \theta_{cr}) &\approx& \kappa(L) J_r + g_0(L) -\Omega_b L \nonumber \\
&&+ \epsilon_m J_r^{1/2} \cos ( \theta_r - m\theta_{cr}). 
\end{eqnarray}
 
Following a canonical transformation or using Hamilton's equations, 
we can estimate  $J_r$ at a fixed point \begin{equation}
J_{r,fixed} = \left({\epsilon_m \over 2 \Delta}\right)^2 
\end{equation}
 where $\Delta = \kappa - m(\Omega - \Omega_b)$ is
the distance to resonance.   
Taking $m=2$, relevant for the 2:1 Lindblad resonance and using equation \ref{eqn:epsm} for $\epsilon_2$, 
we estimate
\begin{equation}
{J_{r,fixed} } \approx 2   r^2 \Omega
\left({\Omega \over \kappa}\right)^3 \left( {C_2 \over r^2 \Omega^2}\right)^2
\left({\Omega \over \Delta} \right)^2
 \label{eqn:rfixed}
\end{equation}
Fixed points for $m=2$ 
corresponding to oval periodic orbits in the plane
that are aligned with the bar.

We can relate this
to the eccentricity of the periodic orbit with $e= r_c^{-1} \sqrt{2J_{r,fixed}/\kappa}$ giving
\begin{equation}
e_{periodic} =  
\left({\Omega \over \kappa}\right)^2 \left( {C_2 \over r^2 \Omega^2}\right)
\left({\Omega \over \Delta} \right)
\end{equation}
giving a relation between the bar's Fourier components and the eccentricity of periodic orbits.
The relation diverges near the Lindblad resonance where $\Delta \to 0$, 
but this is an artifact of the low order (in $J_r$) of the  Hamiltonian model.
As long as the Lindblad resonance is not coincident with the vertical Lindblad resonance,
the approximation is not divergent.
Where $J_r \ne J_{r,fixed}$ orbits oscillate about $J_{r,fixed}$, so $e_{periodic}$ can be 
used to estimate the eccentricity of the orbital distribution.

\end{document}